\documentclass[reprint,superscriptaddress,aps,prd]{revtex4-2}
\usepackage{amsmath,amssymb,mathrsfs,multirow}
\usepackage{graphicx}
\usepackage{array}[=2016-10-06]
\usepackage{dcolumn}
\usepackage{bm,slashed}
\usepackage[colorlinks=true,linkcolor=blue,urlcolor=blue,citecolor=blue]{hyperref}
\usepackage[compat=1.1.0]{tikz-feynman}
\usepackage[utf8]{inputenc}
\usepackage{subfig}
\usepackage{orcidlink}

\begin{document}
\title{Probing anomalous quartic gauge couplings via vector boson scattering at the same-sign muon collider}

\author{Lopamudra Mukherjee\orcidlink{0000-0001-8765-7563}}
\email{lopamudra.physics@gmail.com}
\affiliation{Department of Physics, University of Calcutta, 92, Acharya Prafulla Chandra Road, Kolkata 700009, India}

\author{Abhik Sarkar\orcidlink{0000-0003-1449-2934}}
\email{sarkar.abhik@iitg.ac.in}
\affiliation{Department of Physics, Indian Institute of Technology Guwahati, North Guwahati, 781039, India}

\begin{abstract}
{The measurement of quartic gauge couplings (QGCs) provides a crucial test of the non-Abelian gauge structure of the Standard Model and offers sensitivity to new physics effects. In this work, we explore the potential of the proposed multi-TeV same-sign muon collider, $\mu$TRISTAN, to probe anomalous quartic gauge couplings (aQGCs) through vector boson scattering (VBS) processes. Owing to the same-sign initial state, s-channel contributions are absent, rendering VBS as the dominant production mode and thereby significantly enhancing the sensitivity to aQGCs.
Using dimension-8 Standard Model Effective Field Theory (SMEFT) operators, we classify the relevant operator sets contributing to charged and neutral QGCs, and confront them with existing bounds from the LHC. A detailed collider analysis is performed across multiple final states: $2V2\nu$, $V\gamma \ell \nu$, $2V\ell \nu$, $2\gamma 2\ell$, and $2V2\ell$ ($V=$ reconstructed $W$ and $Z$ boson), applying optimized selection strategies. We present the projected sensitivities at the $\mu$TRISTAN with center-of-mass energies 2 TeV and 6 TeV, with integrated luminosities of 1 ab$^{-1}$ and 10 ab$^{-1}$, and demonstrate significant improvements over current experimental limits from the LHC. Our results establish $\mu$TRISTAN as a powerful probe of electroweak symmetry breaking dynamics and aQGCs in a model-independent framework.}
\end{abstract}
\maketitle
\flushbottom
\section{Introduction}
\label{sec:introduction}
The $SU(2)_L \times U(1)_Y$ non-Abelian gauge structure of the standard model (SM) predicts a set of self-interactions among the electroweak gauge bosons known as the triple gauge couplings (TGC) and quartic gauge couplings (QGC). Among these, the QGCs, involving the simultaneous interaction of four gauge bosons, are essential to maintaining unitarity in high-energy vector boson scattering (VBS) processes and testing the internal consistency of the SM and the electroweak symmetry breaking (EWSB). Precise measurements of QGCs therefore offer a crucial test of the electroweak sector and is a powerful probe of physics beyond the Standard Model (BSM). Many well-known extensions of the SM, including composite Higgs models, extra-dimensional theories, and scenarios with strongly coupled dynamics, predict modifications to quartic gauge couplings~\cite{Fichet:2013ola,Gupta:2011be}. The deviations from the SM predictions can be parametrized in a model-independent way using the framework of Effective Field Theory (EFT) using higher-dimensional operators, particularly dimension-8 terms. Using proper choice of basis, EFT operators can be constructed which contribute to the anomalous quartic gauge couplings (aQGCs) without affecting triple gauge boson couplings (TGCs) at leading order \cite{Degrande:2013kka, Eboli:2016kko}.

VBS processes help us probe the aQGCs directly since the quartic gauge interactions appear at leading order. New physics may significantly enhance the VBS cross-sections and alter the high-energy tails of kinematic distributions. The ATLAS and CMS experiments at the Large Hadron Collider (LHC) have already provided significant constraints on anomalous QGCs through measurements of same-sign $W^\pm W^\pm jj$, $WZjj$, $ZZjj$ and $Z\gamma jj$ final states \cite{ATLAS:2014jzl,CMS:2014mra,ATLAS:2016bkj,ATLAS:2016nmw,ATLAS:2016snd,CMS:2016gct,CMS:2017rin,CMS:2017zmo,CMS:2017fhs,CMS:2019uys,CMS:2019qfk,CMS:2020gfh,CMS:2020ypo,CMS:2020fqz,CMS:2021gme,ATLAS:2022nru,ATLAS:2023sua,CMS:2025dbm}. These analyses have established the first systematic bounds on dimension-8 operator coefficients and demonstrated the feasibility of precision studies of QGCs in hadron collider environments. However, current constraints remain limited by statistical uncertainties and the available energy reach, leaving substantial room for improvements at future facilities.

In this regard, the future lepton colliders are expected to dramatically extend the sensitivity to quartic gauge interactions. The characteristic energy growth of dimension-8 operator contributions implies that high-energy colliders are particularly well suited to probing QGCs, potentially pushing sensitivity to scales far beyond the direct production reach of new particles. In this work, we investigate the current and projected constraints on dimension-8 operators contributing to QGCs within the Standard Model Effective Field Theory (SMEFT) framework. Such interactions have been extensively studied in the context of opposite-sign muon colliders~\cite{Abbott:2022jqq,Yang:2022fhw,Gutierrez-Rodriguez:2025wcy} operating at high energies. However, the $\mu^+\mu^-$ initial state leads to a profusion of annihilation-driven $s$-channel SM processes, which act as significant backgrounds to VBS, particularly at moderate energies, thus necessitating higher center-of-mass (CM) energy, $\sqrt{s}$ to suppress these contributions. In contrast, the $\mu^+\mu^+$ configuration provides a cleaner environment, with the absence of $s$-channel annihilation topologies, thereby reducing background contamination and rendering VBS the dominant production mode for probing such interactions.

We probe aQGCs at the proposed $\mu$TRISTAN facility~\cite{Hamada:2022mua} and assess its sensitivity relative to existing LHC bounds. $\mu$TRISTAN proposes to collide the ultra-cold and highly focused $\mu^+$ beams that have been developed by the J-PARC muon $g-2$ experiment, accelerate them up to 1 TeV energy and collide them with the 30 GeV $e^-$ beam from the existing TRISTAN storage ring, thereby achieving a CM energy, $\sqrt{s} = 346$ GeV and expected integrated luminosity $\mathfrak{L}_{\rm int} = 1~\rm{ab}^{-1}$. However, owing to its relatively low CM energy, this asymmetric collider lacks sufficient kinematic reach to probe momentum-dependent dimension-8 aQGC structures. Alternatively, this facility could operate in the $\mu^+\mu^+$ mode at $\sqrt{s}=2$ TeV, offering a high-energy leptonic environment capable of probing TeV-scale physics within currently achievable accelerator technologies, and thereby providing a clean avenue to test such effective vertices. Future upgrades in beam intensity and repetition rate are expected to extend the reach to integrated luminosities of $\mathfrak{L}_{\rm int}=10~\rm{ab}^{-1}$ and CM energies of $\sqrt{s}=6$ and $10$ TeV in the same-sign collision mode. This would establish $\mu$TRISTAN as a powerful platform for both precision measurements~\cite{Hamada:2022uyn, Chen:2024tqh} and broad searches for physics beyond the Standard Model~\cite{Bossi:2020yne, Das:2022mmh,Dev:2023nha,Fridell:2023gjx,Fukuda:2023yui,Lichtenstein:2023iut,Das:2024gfg,Ding:2024zaj,Calibbi:2024rcm,Das:2024kyk,Sarkar:2025bgo,Barducci:2025kuq}. 

The paper is organized as follows: In Section~\ref{sec:aqgcmodel}, we introduce the EFT framework employed to describe aQGCs, summarize the existing constraints from LHC studies and discuss the theoretical constraints on the operators. In Section~\ref{sec:collider}, we present the collider analysis for the $\mu$TRISTAN 2 TeV run, adopting a signal region-based strategy. In Section~\ref{sec:sensitivity}, we derive constraints on the aQGC effective couplings using a multi-observable likelihood-based approach and discuss the theoretical validity of the projected limits. Finally, we summarize our results in Section~\ref{sec:summary}.
\section{Effective framework}
\label{sec:aqgcmodel}
The theoretical framework employed in this work is based on SMEFT which provides a systematic and model-independent description of potential deviations from the SM at energies below a new physics scale $\Lambda$.
The charged quartic gauge couplings (cQGCs): $WWWW$, $WW\gamma \gamma$, $WW \gamma Z$ and $WWZZ$, are present in the SM and the anomalous contributions are generated at dimension-6 SMEFT. The neutral quartic gauge couplings (nQGCs): $\gamma \gamma \gamma \gamma$, $Z\gamma \gamma \gamma$, $ZZ\gamma \gamma$, $ZZZ\gamma$ and $ZZZZ$, are absent in the SM and first appear at dimension-8. To maintain uniformity with existing studies, we adopt the dimension-8 operator basis used in \cite{Eboli:2016kko, Abbott:2022jqq,Yang:2022fhw,Gutierrez-Rodriguez:2025wcy}. For completeness, the impact of dimension-6 operators on cQGC processes is discussed in Appendix~\ref{appA}. We parametrize the effective Lagrangian as
\begin{widetext}
\begin{equation}
\label{eq:LSMEFT}
\mathcal{L}_{\rm SMEFT}= \mathcal{L}_{SM} \;+\;\sum_{i} \frac{f_{S, i}}{\Lambda^4}\,\mathcal{O}_{S, i} \;+\;\sum_{j} \frac{f_{M, j}}{\Lambda^4}\,\mathcal{O}_{M, j} \;+\; \sum_{k}\frac{f_{T, k}}{\Lambda^4}\,\mathcal{O}_{T, k}
\end{equation}
\end{widetext}
where the operators are organized into the scalar (S), mixed (M) and tensor (T) classes and the corresponding Wilson coefficients $f_{S,i}$, $f_{M,i}$, and $f_{T,i}$ encode the effects of new physics at scale $\Lambda$.
After removing redundancies, the three classes of dimension-8 operators are listed below. It should be noted that these set of operators contribute to QGCs without disturbing the TGC vertices.

\paragraph*{\textbf{Class A}} These operators only involve the covariant derivatives of the Higgs doublet.
\begin{eqnarray}
\mathcal{O}_{S, 0}&=&[(D_\mu H)^\dagger (D_\nu H)]\times [(D^\mu H)^\dagger (D^\nu H)], \nonumber\\
\mathcal{O}_{S, 1}&=&[(D_\mu H)^\dagger (D^\mu H)]\times [(D_\nu H)^\dagger (D^\nu H)],  \nonumber\\
\mathcal{O}_{S, 2}&=&[(D_\mu H)^\dagger (D_\nu H)]\times [(D^\nu H)^\dagger (D^\mu H)]. \nonumber
\end{eqnarray}

\paragraph*{\textbf{Class B}}These operators involve Higgs field and the field tensors of the gauge bosons.
\begin{eqnarray}
\mathcal{O}_{M, 0}&=&[W^{I}_{\mu\nu} W^{I\mu\nu}]\times [(D_\beta H)^\dagger (D^\beta H)],  \nonumber\\
\mathcal{O}_{M, 1}&=&[W^{I}_{\mu\nu} W^{I\nu\beta}]\times [(D_\beta H)^\dagger (D^\mu H)],  \nonumber\\
\mathcal{O}_{M, 7}&=&[(D_\mu H)^\dagger \tau^{I} \tau^{J} (D^\nu H)] \times [W^{I}_{\beta\nu} W^{J\beta\mu}]. \nonumber\\
\mathcal{O}_{M, 2}&=&[B_{\mu\nu} B^{\mu\nu}]\times [(D_\beta H)^\dagger (D^\beta H)],  \nonumber\\
\mathcal{O}_{M, 3}&=&[B_{\mu\nu} B^{\nu\beta}]\times [(D_\beta H)^\dagger (D^\mu H)],  \nonumber\\
\mathcal{O}_{M, 4}&=&[(D_\mu H)^\dagger \tau^{I} (D^\mu H)]\times [W^{I}_{\beta\nu} B^{\beta\nu}],  \nonumber\\
\mathcal{O}_{M, 5}&=&[(D_\mu H)^\dagger \tau^{I} (D^\nu H)]\times [W^{I}_{\beta\nu} B^{\beta\mu}],  \nonumber
\end{eqnarray}

\paragraph*{\textbf{Class C}} These operators are constructed purely out of the gauge boson field tensors.
\begin{eqnarray}
\mathcal{O}_{T, 0}&=&[W^{I}_{\mu\nu} W^{I\mu\nu}]\times[W^{J}_{\alpha\beta}W^{J\alpha\beta}],  \nonumber\\
\mathcal{O}_{T, 1}&=&[W^{I}_{\alpha\nu} W^{I\mu\beta}]\times[W^{J}_{\mu\beta}W^{J\alpha\nu}],  \nonumber\\
\mathcal{O}_{T, 2}&=&[W^{I}_{\alpha\mu} W^{I\mu\beta}]\times[W^{J}_{\beta\nu}W^{J\nu\alpha}],  \nonumber\\
\mathcal{O}_{T, 5}&=&[W^{I}_{\mu\nu} W^{I\mu\nu}]\times [B_{\alpha\beta}B^{\alpha\beta}],  \nonumber\\
\mathcal{O}_{T, 6}&=&[W^{I}_{\alpha\nu} W^{I\mu\beta}]\times [B_{\mu\beta}B^{\alpha\nu}],  \nonumber\\
\mathcal{O}_{T, 7}&=&[W^{I}_{\alpha\mu} W^{I\mu\beta}]\times [B_{\beta\nu}B^{\nu\alpha}],  \nonumber\\
\mathcal{O}_{T, 8}&=&[B_{\mu\nu} B^{\mu\nu}] \times [B_{\alpha\beta}B^{\alpha\beta}],  \nonumber\\
\mathcal{O}_{T, 9}&=& [B_{\alpha\mu} B^{\mu\beta}] \times [B_{\beta\nu} B^{\nu\alpha}]. \nonumber
\end{eqnarray}
In Table~\ref{tab2}, we further organize the operators into subclasses to streamline the analysis. For each subclass, we list the specific vertices that are modified or induced. The subclass $\mathrm{A}_{0}$ contains all scalar operators, while the subclasses $\mathrm{B}$ and $\mathrm{C}$ correspond to mixed and tensor-type operators, respectively.
\begin{table*}[htb!]
\centering
\begin{tabular}{
>{\centering\arraybackslash}p{2cm}
>{\arraybackslash}p{3.0cm}
>{\centering\arraybackslash}p{2.5cm}
>{\centering\arraybackslash}p{2.5cm}
>{\centering\arraybackslash}p{2.5cm}
>{\centering\arraybackslash}p{2.5cm}
}
\hline \hline
Subclass & Operators & $WWWW$ & $WWZZ$ & $WWZ\gamma$ & $WW\gamma\gamma$ \\
\hline \hline
$\rm A_0$ & $\mathcal{O}_{S,0}$, $\mathcal{O}_{S,1}$, $\mathcal{O}_{S,2}$ & $\checkmark$ & $\checkmark$ & $\times$ & $\times$ \\ \hline
$\rm B_1$ & $\mathcal{O}_{M,0}$, $\mathcal{O}_{M,1}$, $\mathcal{O}_{M,7}$ & $\checkmark$ & $\checkmark$ & $\checkmark$ & $\checkmark$ \\
$\rm B_2$ & $\mathcal{O}_{M,2}$, $\mathcal{O}_{M,3}$ & $\times$ & $\checkmark$ & $\checkmark$ & $\checkmark$ \\
$\rm B_3$ & $\mathcal{O}_{M,4}$, $\mathcal{O}_{M,5}$ & $\times$ & $\checkmark$ & $\checkmark$ & $\checkmark$ \\ \hline
$\rm C_1$ & $\mathcal{O}_{T,0}$, $\mathcal{O}_{T,1}$, $\mathcal{O}_{T,2}$ & $\checkmark$ & $\checkmark$ & $\checkmark$ & $\checkmark$ \\
$\rm C_2$ & $\mathcal{O}_{T,5}$, $\mathcal{O}_{T,6}$, $\mathcal{O}_{T,7}$ & $\times$ & $\checkmark$ & $\checkmark$ & $\checkmark$ \\
$\rm C_3$ & $\mathcal{O}_{T,8}$, $\mathcal{O}_{T,9}$ & $\times$ & $\times$ & $\times$ & $\times$ \\
\hline \hline
& & & & & \\
\end{tabular}
\centering
\begin{tabular}{
>{\centering\arraybackslash}p{1.95cm}
>{\arraybackslash}p{2.9cm}
>{\centering\arraybackslash}p{2.0cm}
>{\centering\arraybackslash}p{2.0cm}
>{\centering\arraybackslash}p{2.0cm}
>{\centering\arraybackslash}p{2.0cm}
>{\centering\arraybackslash}p{2.0cm}
}
\hline \hline
Subclass & Operators  & $ZZZZ$ & $ZZZ\gamma$ & $ZZ\gamma \gamma$ & $Z\gamma\gamma\gamma$ & $\gamma\gamma\gamma\gamma$ \\
\hline \hline
$\rm A_0$ & $\mathcal{O}_{S,0}$, $\mathcal{O}_{S,1}$, $\mathcal{O}_{S,2}$ & $\checkmark$ & $\times$ & $\times$ & $\times$ & $\times$ \\ \hline
$\rm B_1$ & $\mathcal{O}_{M,0}$, $\mathcal{O}_{M,1}$, $\mathcal{O}_{M,7}$  & $\checkmark$ & $\checkmark$ & $\checkmark$ & $\times$ & $\times$ \\
$\rm B_2$ & $\mathcal{O}_{M,2}$, $\mathcal{O}_{M,3}$ & $\checkmark$ & $\checkmark$ & $\checkmark$ & $\times$ & $\times$ \\
$\rm B_3$ & $\mathcal{O}_{M,4}$, $\mathcal{O}_{M,5}$ & $\checkmark$ & $\checkmark$ & $\checkmark$ & $\times$ & $\times$ \\ \hline
$\rm C_1$ & $\mathcal{O}_{T,0}$, $\mathcal{O}_{T,1}$, $\mathcal{O}_{T,2}$ & $\checkmark$ & $\checkmark$ & $\checkmark$ & $\checkmark$ & $\checkmark$ \\
$\rm C_2$ & $\mathcal{O}_{T,5}$, $\mathcal{O}_{T,6}$, $\mathcal{O}_{T,7}$ & $\checkmark$ & $\checkmark$ & $\checkmark$ & $\checkmark$ & $\checkmark$ \\
$\rm C_3$ & $\mathcal{O}_{T,8}$, $\mathcal{O}_{T,9}$ & $\checkmark$ & $\checkmark$ & $\checkmark$ & $\checkmark$ & $\checkmark$ \\
\hline \hline
\end{tabular}
\caption{The cQGCs (\textit{upper}) and nQGCs (\textit{lower}) modified by the listed dimension-8 operators are shown with $\checkmark$.}
\label{tab2}
\end{table*}
\subsection{Existing LHC constraints}
\label{sec:constraints}
Existing constraints on aQGCs include a combination of theoretical consistency requirements and experimental measurements. On the theory side, perturbative unitarity imposes important bounds on the Wilson coefficients of dimension-8 operators, as the energy growth of scattering amplitudes due to additional contributions from new physics can lead to a breakdown of unitarity at sufficiently high energies \cite{Cornwall:1974km, Passarino:1990hk}. This restricts the allowed parameter space for the coefficients $f_{S,i}/\Lambda^{4}$, $f_{M,j}/\Lambda^{4}$, and $f_{T,k}/\Lambda^{4}$, or necessitates the implementation of appropriate unitarization procedures~\cite{Almeida:2020ylr}. The unitarity constraints are further supplemented by positivity constraints which ensure the mathematical consistency of the EFT by enforcing causality and analyticity in the underlying UV completion \cite{Zhang:2018shp,Bi:2019phv,Yamashita:2020gtt}. Complementing these theoretical constraints are experimental measurements by the ATLAS and CMS Collaborations at the LHC. The most stringent experimental constraints till date have been obtained through measurements of VBS and triboson production processes (such as $W\gamma jj$, $Z\gamma jj$, and $Z\gamma\gamma$) which have been used to set 95\% C.L. limits on Wilson coefficients for scalar (S), mixed (M), and tensor (T) type operators \cite{ATLAS:2014jzl,CMS:2014mra,ATLAS:2016bkj,ATLAS:2016snd,CMS:2016gct,ATLAS:2016nmw,CMS:2017fhs,CMS:2017rin,CMS:2017zmo,ATLAS:2017vqm,CMS:2019uys,CMS:2019qfk,CMS:2020gfh,CMS:2020ypo,CMS:2020fqz,CMS:2021gme,ATLAS:2022nru,ATLAS:2023sua,CMS:2025dbm}. Table \ref{tab:QGC_LHC_combined} lists the most updated experimental bound on the operators from the 13 TeV run of the ATLAS and CMS where the tightest bound is highlighted in bold for each case. 

\begin{table*}[htb!]
\centering
\begin{tabular}{
>{\centering\arraybackslash}p{1.5cm}
>{\centering\arraybackslash}p{4.5cm}
>{\centering\arraybackslash}p{4.5cm}
>{\centering\arraybackslash}p{4.5cm}
}
\hline \hline
\multirow{2}*{WCs} 
& ATLAS (13 TeV)~\cite{ATLAS:2023sua} 
& ATLAS (13 TeV)~\cite{ATLAS:2022nru} 
& CMS (13 TeV)~\cite{CMS:2025dbm} \\
& $W^\pm W^\pm jj$ (139 fb$^{-1}$) & $Z(\nu\bar\nu)\gamma jj$ (139 fb$^{-1}$) & WV+ZV (138 fb$^{-1}$) \\
\hline \hline
$f_{S,0}/\Lambda^4$ & $[-5.9,\; 5.9]\times10^{0}$ & $-$ & $\mathbf{[-2.9,\; 3.0]\times10^{0}}$ \\
$f_{S,1}/\Lambda^4$ & $[-2.4,\; 2.4]\times10^{0}$ & $-$ & $\mathbf{[-4.0,\; 4.0]\times10^{0}}$ \\
$f_{S,2}/\Lambda^4$ & $[-5.9,\; 5.9]\times10^{0}$ & $-$ & $\mathbf{[-4.0,\; 4.1]\times10^{0}}$ \\
\hline\hline
$f_{M,0}/\Lambda^4$ & $[-4.1,\; 4.1]\times10^{0}$ & $[-4.6,\; 4.6]\times10^{0}$ & $\mathbf{[-5.4,\; 5.3]\times10^{-1}}$ \\
$f_{M,1}/\Lambda^4$ & $[-6.8,\; 7.0]\times10^{0}$ & $[-7.7,\; 7.7]\times10^{0}$ & $\mathbf{[-1.6,\; 1.6]\times10^{0}}$ \\
$f_{M,7}/\Lambda^4$ & $[-9.8,\; 9.5]\times10^{0}$ & $-$ & $\mathbf{[-2.6,\; 2.6]\times10^{0}}$ \\ \hline
$f_{M,2}/\Lambda^4$ & $-$ & $[-1.9,\; 1.9]\times10^{0}$ & $\mathbf{[-7.0,\; 7.0]\times10^{-1}}$ \\
$f_{M,3}/\Lambda^4$ & $-$ & $-$ & $\mathbf{[-2.6,\; 2.6]\times10^{0}}$ \\ \hline
$f_{M,4}/\Lambda^4$ & $-$ & $-$ & $\mathbf{[-1.5,\; 1.5]\times10^{0}}$ \\
$f_{M,5}/\Lambda^4$ & $-$ & $-$ & $\mathbf{[-2.1,\; 2.1]\times10^{0}}$ \\
\hline \hline
$f_{T,0}/\Lambda^4$ & $[-3.6,\; 3.6]\times10^{-1}$ & $[-9.4,\; 8.4]\times10^{-2}$ & $\mathbf{[-9.2,\; 7.8]\times10^{-2}}$ \\
$f_{T,1}/\Lambda^4$ & $[-1.7,\; 1.9]\times10^{-1}$ & $-$ & $\mathbf{[-8.6,\; 9.4]\times10^{-2}}$ \\
$f_{T,2}/\Lambda^4$ & $[-6.3,\; 7.4]\times10^{-1}$ & $-$ & $\mathbf{[-2.1,\; 2.1]\times10^{-1}}$ \\ \hline
$f_{T,5}/\Lambda^4$ & $-$ & $\mathbf{[-8.8,\; 9.9]\times10^{-2}}$ & $[-2.6,\; 2.4]\times10^{-1}$ \\
$f_{T,6}/\Lambda^4$ & $-$ & $-$ & $\mathbf{[-5.0,\; 4.8]\times10^{-1}}$ \\
$f_{T,7}/\Lambda^4$ & $-$ & $-$ & $\mathbf{[-8.5,\; 8.0]\times10^{-1}}$ \\ \hline
$f_{T,8}/\Lambda^4$ & $-$ & $\mathbf{[-5.9,\; 5.9]\times10^{-2}}$ & $[-5.3,\; 5.3]\times10^{-1}$ \\
$f_{T,9}/\Lambda^4$ & $-$ & $\mathbf{[-1.3,\; 1.3]\times10^{-1}}$ & $[-1.2,\; 1.2]\times10^{0}$ \\
\hline \hline
\end{tabular}
\caption{Observed 95\% C.L. intervals (TeV$^{-4}$) on dimension-8 quartic gauge couplings from LHC measurements. The tightest bound per operator is highlighted in bold.}
\label{tab:QGC_LHC_combined}
\end{table*}

\subsection{Muon collider projections}
\label{sec:mu-projection}
Several studies in the context of $\mu^{+}\mu^{-}$ colliders have explicitly investigated aQGCs across different production channels. In Ref.~\cite{Gutierrez-Rodriguez:2025wcy}, nQGCs are probed using $\mathcal{O}_{T,k}$-type operators that induce $Z\gamma\gamma\gamma$, $ZZ\gamma\gamma$, and $ZZZ\gamma$ interactions via the process $\mu^{+}\mu^{-} \to \mu^{+}\mu^{-}Z\gamma$ at a CM energy of $\sqrt{s} = 10$~TeV with an integrated luminosity of $\mathfrak{L}_{\rm int} = 10~\mathrm{ab}^{-1}$. The resulting constraints improve by nearly an order of magnitude over the bounds on $f_{T,k}/\Lambda^{4}$ as listed in Table~\ref{tab:QGC_LHC_combined}. A similar analysis is presented in Ref.~\cite{Yang:2022fhw}, where a subset of operators from all three classes (S, M, and T) is considered. This study focuses on vector boson scattering processes, specifically $W^{+}W^{-} \to W^{+}W^{-}$, initiated by $W$ bosons at a $\mu^{+}\mu^{-}$ collider with $\sqrt{s} = 30$~TeV and integrated luminosities of $\mathfrak{L}_{\rm int} = 10$ and $90~\mathrm{ab}^{-1}$. At such high energies and luminosities, the sensitivity to the most constrained operator, $f_{T,2}/\Lambda^{4}$, reaches the level of $\mathcal{O}(10^{-5})$. In a related study, Ref.~\cite{Abbott:2022jqq} also examines a subset of operators from all three categories at $\mu^{+}\mu^{-}$ colliders with $\sqrt{s} = 6$, $10$, and $30$~TeV, and integrated luminosities of $\mathfrak{L}_{\rm int} = 4$ and $10~\mathrm{ab}^{-1}$. The analysis focuses on the channels $\mu^{+}\mu^{-} \to WW + \nu\nu/\mu\mu$ and $\mu^{+}\mu^{-} \to WW/ZZ + \nu\nu$, achieving sensitivities of $\mathcal{O}(10^{-5})$ for the most constrained operator, $f_{T,0}/\Lambda^{4}$. The projected 95\% C.L. limits are summarized in Table~\ref{tab:QGC_muC_combined}.

\begin{table*}[htb!]
\centering
\begin{tabular}{
>{\centering\arraybackslash}p{1.5cm}
>{\centering\arraybackslash}p{4.5cm}
>{\centering\arraybackslash}p{4.5cm}
>{\centering\arraybackslash}p{4.5cm}
}
\hline \hline
\multirow{2}*{WCs} 
& $\mu^{+}\mu^{-} \to \mu^{+}\mu^{-}Z\gamma$
& $W^+W^- \to W^+W^-$ 
& $\mu^{+}\mu^{-} \to VV+\mu\mu/\nu\nu$ \\
& (10 TeV, 10 ab$^{-1}$)~\cite{Gutierrez-Rodriguez:2025wcy} & (30 TeV , 90 ab$^{-1}$)~\cite{Yang:2022fhw} & (30 TeV, 10 ab$^{-1}$)~\cite{Abbott:2022jqq} \\
\hline \hline
$f_{S,0}/\Lambda^4$ & $-$ & $0.5^{+2.1}_{-0.3}\times 10^{-2}$ & ${[-4.6,\; 4.5]\times10^{-3}}$ \\
$f_{S,1}/\Lambda^4$ & $-$ & $3.1^{+0.1}_{-3.0}\times 10^{-2}$ & ${[-2.5,\; 2.5]\times10^{-3}}$ \\
$f_{S,2}/\Lambda^4$ & $-$ & $0.6^{+1.9}_{-0.4}\times 10^{-2}$ & ${[-2.5,\; 2.5]\times10^{-3}}$ \\
\hline\hline
$f_{M,0}/\Lambda^4$ & $-$ & $2.6^{+0.1}_{-0.0}\times10^{-4}$ & ${[-4.6,\; 4.6]\times10^{-4}}$ \\
$f_{M,1}/\Lambda^4$ & $-$ & $6.2^{+0.1}_{-0.0}\times10^{-4}$ & ${[-1.2,\; 1.1]\times10^{-3}}$ \\
$f_{M,7}/\Lambda^4$ & $-$ & $1.1^{+0.0}_{-0.0}\times 10^{-3}$ & ${[-2.1,\; 2.2]\times10^{-3}}$ \\ \hline
$f_{M,2}/\Lambda^4$ & $-$ & $-$ & $-$ \\
$f_{M,3}/\Lambda^4$ & $-$ & $-$ & $-$ \\ \hline
$f_{M,4}/\Lambda^4$ & $-$ & $-$ & $-$ \\
$f_{M,5}/\Lambda^4$ & $-$ & $-$ & $-$ \\
\hline \hline
$f_{T,0}/\Lambda^4$ & $[-7.6,\; 7.4]\times10^{-2}$ & $2.5^{+0.0}_{-0.0} \times10^{-5}$ & ${[-3.1,\; 1.8]\times10^{-5}}$ \\
$f_{T,1}/\Lambda^4$ & $[-7.6,\; 7.4]\times10^{-2}$ & $3.4^{+0.0}_{-0.0}\times10^{-5}$ & ${[-5.0,\; 3.1]\times10^{-5}}$ \\
$f_{T,2}/\Lambda^4$ & $[-1.2,\; 1.3]\times10^{-1}$ & $5.8^{+0.0}_{-0.0}\times10^{-5}$ & ${[-9.1,\; 4.2]\times10^{-5}}$ \\ \hline
$f_{T,5}/\Lambda^4$ & $[-3.2,\; 3.0]\times10^{-2}$ & $-$ & $-$ \\
$f_{T,6}/\Lambda^4$ & $[-3.4,\; 3.5]\times10^{-2}$ & $-$ & ${[-2.7,\; 2.4]\times10^{-5}}$ \\
$f_{T,7}/\Lambda^4$ & $[-7.9,\; 7.8]\times10^{-2}$ & $-$ & ${[-3.4,\; 2.8]\times10^{-5}}$ \\ \hline
$f_{T,8}/\Lambda^4$ & $[-8.6,\; 8.6]\times10^{-3}$ & $-$ & $-$ \\
$f_{T,9}/\Lambda^4$ & $[-1.7,\; 1.7]\times10^{-2}$ & $-$ & $-$ \\
\hline \hline
\end{tabular}
\caption{Projected 95\% C.L. intervals (TeV$^{-4}$) on dimension-8 quartic gauge couplings from future $\mu^{+}\mu^{-}$ collider studies, involving different processes. Here, $V=W,Z$.}
\label{tab:QGC_muC_combined}
\end{table*}

\subsection{Theoretical constraints and EFT validity}
\label{sec:theory_constraints}
The dimension-8 parametrization used in Eq.~\eqref{eq:LSMEFT} is valid only within a finite energy range. Since the aQGC operators considered here
generate amplitudes that grow with the hard scattering energy, the projected collider sensitivities must be interpreted together with constraints from the EFT cutoff, perturbative unitarity, and positivity.

As shown in Eq.~\eqref{eq:LSMEFT}, the coefficient of a generic dimension-8 operator is given by $c_{X,i} \equiv \left({f_{X,i}}/{\Lambda^4}\right)$. For a one-coefficient interpretation with a  dimensionless Wilson coefficient of order unity, the EFT scale associated with this interaction
is estimated as
\begin{equation}
    \Lambda_{X,i}=|c_{X,i}|^{-1/4}.
    \label{eq:LambdaEFT}
\end{equation}
More generally, the corresponding cut-off scale is $\Lambda_{X,i} = (|f_{X,i}|/|c_{X,i}|)^{1/4}$. The relevant energy regime for the EFT validity is the hard VBS scale $ Q_{\rm VBS}\equiv \sqrt{\hat{s}_{VV}}$, which may be identified with the final-state diboson invariant mass $m_{VV}$ at the partonic level. This scale should be distinguished from the full muon-collider CM energy $\sqrt{s}$. 

Perturbative unitarity gives an additional restriction. The dimension-8 aQGC amplitudes scale as $c_{X,i}Q_{\rm VBS}^4$, and the corresponding partial-wave
eigenvalues can violate unitarity at sufficiently large energies. Following
Ref.~\cite{Almeida:2020ylr}, the one-coefficient partial-wave unitarity condition may be written as
\begin{equation}
    |c_{X,i}| < \frac{A_{X,i}}{Q_{\rm VBS}^4}\,,
\end{equation}
or equivalently in terms of the unitarity scale,
\begin{equation}
    Q_{\rm VBS} < Q^{U}_{X,i} =
    \left(\frac{A_{X,i}}{|c_{X,i}|}\right)^{1/4}.
    \label{eq:unitarity_scale}
\end{equation}
Here $A_{X,i}$ is obtained from the largest eigenvalue of the coupled $VV,Vh,hh\to VV,Vh,hh$ partial-wave scattering matrix for the given operator and is listed in Table~\ref{tab:aqgc_theory_constraints}. Combining the EFT cut-off estimate and the unitarity bound, a conservative EFT validity condition is given by
\begin{equation}
    Q_{\rm VBS} < Q^{\rm valid}_{X,i}
    \equiv
    \min\left(\Lambda_{X,i},Q^{U}_{X,i}\right).
    \label{eq:Qvalid}
\end{equation}
This condition will be used below to assess the theoretical interpretation of the projected sensitivities at the $\mu$TRISTAN collider.

Finally, the allowed Wilson coefficients are further restricted by positivity bounds which necessitate certain linear combinations of these coefficients must be positive. These follow from general assumptions on the ultraviolet completion, such as Lorentz invariance, analyticity, causality, and unitarity. In the one-coefficient scenarios considered in this work, positivity imposes sign restrictions on several dimension-8 aQGC coefficients, while some isolated one-operator directions are not compatible with positivity and require correlated nonzero coefficients in a global EFT interpretation~\cite{Bi:2019phv}. The numerical impact of these theoretical requirements on our projected limits will be discussed later.
\section{Collider analysis}
\label{sec:collider}
\begin{figure}[htb!]
    \centering
    \includegraphics[width = 0.5\linewidth]{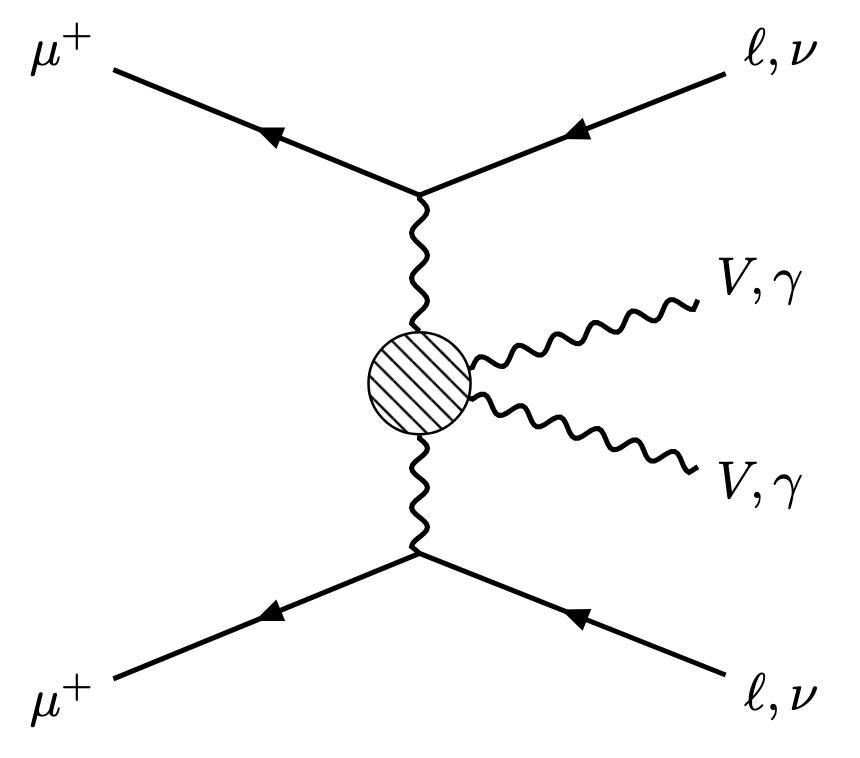}
    \caption{Feynman diagrams corresponding to diboson production via vector boson scattering at the $\mu$TRISTAN. Here, $V = W, Z$. The blob includes both SM and EFT operator contributions.}
    \label{fig:muTc}
\end{figure}

In this section, we present the collider analysis strategy employed to probe aQGCs at the $\mu^+\mu^+$ mode of the $\mu$TRISTAN collider. Our analysis focuses on VBS-initiated topologies, where the effects of higher dimensional operators become increasingly significant at high energies. A similar analysis can, in principle, be performed in the $\mu^{+}e^{-}$ mode of $\mu$TRISTAN. However, the higher CM energy achievable in the $\mu^{+}\mu^{+}$ mode provides a more favorable environment for probing aQGCs. Moreover, in VBS-initiated processes, the $\mu^{+}e^{-}$ mode primarily offers better sensitivity to cQGCs, whereas the $\mu^{+}\mu^{+}$ mode allows simultaneous access to both cQGCs as well as nQGCs. This makes the $\mu^{+}\mu^{+}$ configuration a more powerful and comprehensive probe of aQGCs. We consider multiple final states involving reconstructed weak bosons, photons, leptons, and missing transverse energy (MET), which together provide complementary sensitivity to different operator subclasses introduced in Section~\ref{sec:aqgcmodel}. A schematic representation of the VBS topology relevant to our analysis is shown in Figure~\ref{fig:muTc}. This Feynman diagram captures the characteristic structure where incoming same-sign anti-muons radiate electroweak gauge bosons that subsequently undergo scattering, leading to multi-boson final states. The blob denotes the combined contribution from both the SM interactions and the EFT operators.
The aQGC model implementation is done using \texttt{FeynRules}~\cite{Alloul:2013bka}. The generated UFO~\cite{Degrande:2011ua} model is fed to \texttt{MG5\_aMC}~\cite{Alwall:2011uj}, where, event samples for both the SM and SMEFT contributions are generated at parton level and subsequently passed through \texttt{Pythia8}~\cite{Bierlich:2022pfr} for parton showering. The detector response is modeled using {\tt Delphes3}~\cite{deFavereau:2013fsa}, employing a default detector card appropriate for a generic high-energy lepton collider analysis. Jets are reconstructed using the anti-$k_T$ algorithm using \texttt{FastJet3}~\cite{Cacciari:2011ma} within {\tt Delphes3} framework, and standard object identification criteria are applied for leptons and photons.
\subsection{Signal region categorization}
\label{sec:events}
In order to access multiple subclasses of operators, we divide the possible final states from VBS into different signal regions. The identification of signal regions is based on the multiplicity of reconstructed objects in the final state, namely jets ($j$), leptons ($\ell$), and photons ($\gamma$). These selections are designed to isolate distinct VBS topologies while maintaining sensitivity to different classes of operators. For the reconstruction of hadronically decaying weak bosons, we select events containing pairs of hard jets and impose an invariant mass requirement:
\begin{equation}
    \left| M^{V}_{jj} - M_{Z} \right| < 30~\mathrm{GeV}\,,
\end{equation}
which efficiently captures both $W$ and $Z$ bosons due to the finite detector resolution. In events with two reconstructed bosons, jet pairing is performed by minimizing the difference between the invariant masses of jet pairs and the nominal weak boson mass, followed by comparing with each possible system. The strategy can be outlines as follows:
\begin{widetext}
\begin{equation}
\begin{split}
    &\delta M_{j_1 j_2; j_3 j_4} = \sqrt{\left(\Delta M^{V}_{j_1 j_2}\right)^{2} + \left(\Delta M^{V}_{j_3 j_4}\right)^{2}} \quad \longrightarrow \quad
    \begin{cases} \Delta M^{V}_{j_1 j_2} = \left| M^{V}_{j_1 j_2} - M_{Z} \right| \\ \Delta M^{V}_{j_3 j_4} = \left| M^{V}_{j_3 j_4} - M_{Z} \right| \\ {\rm Ensure\,:}\;M^{V}_{j_1 j_2} > M^{V}_{j_3 j_4} \end{cases},  \\
    &{\rm If}\;\delta M_{j_1 j_2; j_3 j_4} = {\rm min}\begin{cases}  \delta M_{j_1 j_2; j_3 j_4}\\ \delta M_{j_1 j_3; j_2 j_4}\\ \delta M_{j_1 j_4; j_2 j_3} \end{cases} \quad \longrightarrow \quad \begin{cases} V_{1} \Rightarrow {\rm Reconstructed\;from\;}j_{1},j_{2}\\ V_{2} \Rightarrow {\rm Reconstructed\;from\;}j_{3},j_{4} \end{cases}.
\end{split}
\end{equation}
\end{widetext}
Based on object multiplicities, we define the following signal regions:
\begin{eqnarray}
    \text{Signal Region}\quad& 2V2\nu \;:\;& (N_{j},N_{\ell},N_{\gamma}) = (4, 0, 0)\,,\nonumber\\
    \text{Signal Region}\quad& V\gamma\ell\nu \;:\;& (N_{j},N_{\ell},N_{\gamma}) = (2, 1, 1)\,,\nonumber\\
    \text{Signal Region}\quad& 2V\ell\nu \;:\;& (N_{j},N_{\ell},N_{\gamma}) = (4, 1, 0)\,,\\
    \text{Signal Region}\quad& 2\gamma2\ell \;:\;& (N_{j},N_{\ell},N_{\gamma}) = (0, 2, 2)\,,\nonumber\\
    \text{Signal Region}\quad& 2V2\ell \;:\;& (N_{j},N_{\ell},N_{\gamma}) = (4, 2, 0)\,.\nonumber
\end{eqnarray}
For event simulation, we include all relevant SM contributions to each final state, ensuring that not only the VBS topology but also additional contributions, such as radiative processes, are fully taken into account. For the invisible final state, in addition to the above selection criteria, MET requirements are imposed to suppress possible visible backgrounds, that can mimic the accompanying visible final states. Specifically, we require
\begin{itemize}
    \itemsep-0em
    \item $\mathrm{MET} > 10$~GeV for single neutrino final states,
    \item $\mathrm{MET} > 25$~GeV for double neutrino final states.
\end{itemize}
These signal regions collectively probe different cQGC and nQGC couplings, as summarized in Table~\ref{tab:table4}. The complementarity among the channels allows for a comprehensive exploration of the EFT parameter space.
\begin{table*}[htb!]
    \centering
    \begin{tabular}{
>{\centering\arraybackslash}p{2.5cm}
>{\centering\arraybackslash}p{2.0cm}
>{\centering\arraybackslash}p{2.0cm}
>{\centering\arraybackslash}p{2.0cm}
>{\centering\arraybackslash}p{2.0cm}
>{\arraybackslash}p{4.5cm}
    }
    \hline \hline
    Signal Region & $WWWW$ & $WWZZ$ & $WWZ\gamma$ & $WW\gamma\gamma$ & Operator Subclasses$\;\;\;\;\;\;$ \\
    \hline \hline
    $2V 2\nu$ & $\checkmark$ & $\times$ & $\times$ & $\times$ & $\rm A_0, B_1, C_1$ \\
    $V\gamma\ell\nu$ & $\times$ & $\times$ & $\checkmark$ & $\checkmark$ & $\rm B_1, B_2, B_3, C_1, C_2$ \\
    $2V\ell\nu$ & $\times$ & $\checkmark$ & $\checkmark$ & $\times$ & $\rm A_0, B_1, B_2, B_3, C_1, C_2$ \\
    \hline \hline
    & & & & & \\
    \end{tabular}
    \centering
    \begin{tabular}{
>{\centering\arraybackslash}p{2.4cm}
>{\centering\arraybackslash}p{1.6cm}
>{\centering\arraybackslash}p{1.6cm}
>{\centering\arraybackslash}p{1.6cm}
>{\centering\arraybackslash}p{1.6cm}
>{\centering\arraybackslash}p{1.6cm}
>{\arraybackslash}p{4.45cm}
    }
    \hline \hline
    Signal Region & $ZZZZ$ & $ZZZ\gamma$ & $ZZ\gamma \gamma$ & $Z\gamma\gamma\gamma$ & $\gamma\gamma\gamma\gamma$ & $\;\;$Operator Subclasses$\;\;\;\;$ \\
    \hline \hline
    $2\gamma 2\ell$ & $\times$ & $\times$ & $\checkmark$ & $\checkmark$ & $\checkmark$ & $\;\;\rm B_1, B_2, B_3, C_1, C_2, C_3$ \\
    $2V 2\ell$ & $\checkmark$ & $\checkmark$ & $\checkmark$ & $\times$ & $\times$ & $\;\;\rm A_0, B_1, B_2, B_3, C_1, C_2, C_3$ \\
    \hline \hline
    \end{tabular}
    \caption{Signal and respective cQGC (\textit{upper}) and nQGC (\textit{lower}) vertices at the $\mu^+\mu^+$ collider.}
    \label{tab:table4}
\end{table*}
In Table~\ref{tab:ncross}, we present the event yields in each signal region for different collider configurations, considering only SM contributions. For $\sqrt{s} = 2$~TeV, the $V\gamma\ell\nu$ signal region yields the highest event count, followed by the $2V2\nu$ channel. In contrast, at $\sqrt{s} = 6$~TeV, the $2V2\nu$ signal region becomes dominant, while $V\gamma\ell\nu$ provides a subleading contribution. This shift in the relative event rates can be attributed to the increased boost of the vector bosons at higher energies. As a result, the decay products of the bosons, particularly jets, become increasingly collimated and may merge into a single reconstructed jet. Consequently, final states requiring resolved jet pairs are comparatively suppressed at higher energies. We note that in such boosted regimes, a dedicated fat-jet and jet substructure-based analysis would provide a more efficient reconstruction strategy. However, incorporating such techniques is beyond the scope of the present study.
\begin{table*}[htb!]
    \centering
    \begin{tabular}{
>{\centering\arraybackslash}p{3.75cm}
>{\centering\arraybackslash}p{2.25cm}
>{\centering\arraybackslash}p{2.25cm}
>{\centering\arraybackslash}p{2.25cm}
>{\centering\arraybackslash}p{2.25cm}
>{\centering\arraybackslash}p{2.25cm}
    }
    \hline \hline
    \multirow{2}*{Setup $(\sqrt{s}, \mathfrak{L}_{\rm int})$} & \multicolumn{5}{c}{Event counts in different signal regions} \\ \cline{2-6}
     & $2V2\nu$ & $V\gamma\ell\nu$ & $2V\ell\nu$ & $2\gamma 2\ell$ & $2V2\ell$ \\
    \hline \hline
    $(2\text{ TeV},\;\,1\text{ ab}^{-1})$ & $8.1\times10^{3}$ & $2.2\times10^{5}$ & $5.0\times10^{3}$ & $6.4\times10^{3}$ & $1.5\times10^{3}$ \\
    $(2\text{ TeV},10\text{ ab}^{-1})$ & $8.1\times10^{4}$ & $2.2\times10^{6}$ & $5.0\times10^{4}$ & $6.4\times10^{4}$ & $1.5\times10^{4}$ \\
    $(6\text{ TeV},\;\,1\text{ ab}^{-1})$ & $2.4\times10^{4}$ & $1.0\times10^{4}$ & $4.7\times10^{3}$ & $1.0\times10^{3}$ & $5.5\times10^{2}$ \\
    $(6\text{ TeV},10\text{ ab}^{-1})$ & $2.4\times10^{5}$ & $1.0\times10^{5}$ & $4.7\times10^{4}$ & $1.0\times10^{4}$ & $5.5\times10^{3}$  \\
    \hline \hline
    \end{tabular}
    \caption{The SM event counts for each signal region for different CM energy, $\sqrt{s}$ (TeV) and integrated luminosity, $\mathfrak{L}_{\rm int}$ (ab$^{-1}$) combinations.}
    \label{tab:ncross}
\end{table*}
\section{Projected sensitivity}
\label{sec:sensitivity}
In this section, we present the projected sensitivity to the effective operator coefficients, $f_{X,i}/\Lambda^{4}$, using a binned likelihood analysis. By leveraging the shape information of suitably chosen differential distributions, we go beyond total rate measurements and extract stronger constraints. This approach captures not only the overall event yield but also the distinct kinematic features of signal and background processes, thereby enhancing the sensitivity to new physics. We outline the statistical framework used to quantify these projections below. We estimate the projected sensitivity to the operator coefficients by constructing a binned likelihood using a differential observable, say $\varphi$. The expected number of events in the ${\tt r}^{\rm th}$ bin of the observable $\varphi^{(k)}$ is given by:
\begin{equation}
    \mu^{(k)}_{\tt r} \left(f_{X,i}/\Lambda^{4}\right) = \mathfrak{L}_{\rm int} \times \sigma^{(k)}_{\tt r} \left(f_{X,i}/\Lambda^{4}\right)\,,
\end{equation}
where $\mathfrak{L}_{\rm int}$ is the integrated luminosity, $\sigma^{(k)}_{\tt r}$ is the cross-section of the ${\tt r}^{\rm th}$ bin of the observable $\varphi^{(k)}$, which depends on the operator coefficient, $f_{X,i}/\Lambda^{4}$. The corresponding likelihood function \cite{ParticleDataGroup:2024cfk} is defined as a product of Poisson probabilities over all bins:
\begin{equation}
    \mathscr{L} \left(f_{X,i}/\Lambda^{4}\right)  = \prod_{{\tt r},k} \frac{\left[\mu^{(k)}_{\tt r} \left(f_{X,i}/\Lambda^{4}\right) \right]^{n_{\tt r}}e^{-\mu^{(k)}_{\tt r} \left(f_{X,i}/\Lambda^{4}\right)}}{n_{\tt r}!}\,,
\end{equation}
where, $n_{\tt r}$ is the observed number of events in the ${\tt r}^{\rm th}$ bin of the observable $\varphi^{(k)}$. Since no observed data exist for future colliders, the observed event counts are assumed to equal to the SM yields. So, the observed events are taken as the background expectation:
\begin{equation}
    n_{\tt r} = \mathfrak{L}_{\rm int} \times \sigma^{(k)}_{\tt r} \left(0\right)\,.
\end{equation}
The log-likelihood is then given by:
\begin{equation}
\begin{split}
\log{\mathscr{L} \left(f_{X,i}/\Lambda^{4}\right)} &= \sum_{{\tt r},k} \Big[n^{(k)}_{\tt r} \log{\mu^{(k)}_{\tt r} \left(f_{X,i}/\Lambda^{4}\right)} \\ &- \mu^{(k)}_{\tt r} \left(f_{X,i}/\Lambda^{4}\right) \Big] + {\rm constant}\,.
\end{split}
\end{equation}
To quantify the sensitivity, we define the profile likelihood ratio:
\begin{equation} \label{eq:profile}
\begin{split}
    \lambda \left(f_{X,i}/\Lambda^{4}\right) &= \frac{\mathscr{L} \left(f_{X,i}/\Lambda^{4}\right)}{\mathscr{L} \left(\hat{f}_{X,i}/\Lambda^{4}\right)}\,, \\
    \mathcal{Q} \left(f_{X,i}/\Lambda^{4}\right) &= -2 \log{\lambda \left(f_{X,i}/\Lambda^{4}\right)}\,,
\end{split}
\end{equation}
where, $\hat{f}_{X,i}/\Lambda^{4}$ denotes the value of the operator coefficient that maximizes the likelihood. Since we are evaluating the sensitivity to new physics against the SM, we perform the analysis around the null hypothesis, i.e., $\hat{f}_{X,i}/\Lambda^{4} = 0$.

The test statistic $\mathcal{Q}$ defined in Eq.~\eqref{eq:profile} is used to derive projected bounds on the operator coefficient $f_{X,i}/\Lambda^{4}$ at a given confidence level (C.L.). According to Wilks' theorem~\cite{Wilks:1938dza}, in the asymptotic limit, the distribution of $\mathcal{Q}$ approaches a chi-squared ($\chi^2$) distribution with degrees of freedom equal to the number of parameters being tested. It is important to note that in the low-statistics regime, $\mathcal{Q}$ may deviate from the ideal $\chi^2$ behavior, in which case the true distribution should be estimated using Monte Carlo simulations. However, such simulations are computationally demanding. For projection studies focused on assessing the reach of future colliders, the use of Wilks' theorem provides a reasonable and commonly adopted approximation. Using the Wilks' theorem, we translate the value of $\mathcal{Q}$ into confidence intervals on $f_{X,i}/\Lambda^{4}$. For instance, the critical values of the test statistic $\mathcal{Q}$ corresponding to the 68\% and 95\% C.L. are given by:
\begin{equation}
\begin{split}
    \mathcal{Q} \left(f_{X,i}/\Lambda^{4}\right) \leq \chi^2_{1,\,68\%} &= 1.00\,, \\
    \mathcal{Q} \left(f_{X,i}/\Lambda^{4}\right) \leq \chi^2_{1,\,95\%} &= 3.84\,, \\
\end{split}
\end{equation}
where $\chi^2_{f,\,p\%}$ denotes the $p^{\rm th}$ percentile of the chi-squared distribution with $f$ degrees of freedom. These thresholds define the regions of parameter space that are consistent with the background-only hypothesis at the corresponding confidence levels. For different signal regions we use different observables ($\varphi^{(k)}$):
\begin{equation}
\begin{split}
    2V2\nu &\;:\; \left\{\Delta \phi_{VV},\; \Delta \eta_{VV},\; \phi_{V_1},\; \eta_{V_1} \right\}\,, \\
    V\gamma\ell\nu &\;:\; \left\{\Delta \phi_{V\gamma},\; \Delta \eta_{V\gamma},\; \phi_{V_1},\; \eta_{V_1} \right\}\,, \\
    2V\ell\nu &\;:\; \left\{\Delta \phi_{VV},\; \Delta \eta_{VV},\; \phi_{V_1},\; \eta_{V_1} \right\}\,, \\
    2\gamma 2\ell &\;:\; \left\{\Delta \phi_{\gamma\gamma},\; \Delta \eta_{\gamma\gamma},\; \phi_{\gamma_1},\; \eta_{\gamma_1} \right\}\,, \\
    2V2\ell &\;:\; \left\{\Delta \phi_{VV},\; \Delta \eta_{VV},\; \phi_{V_1},\; \eta_{V_1} \right\}\,.\\
\end{split}
\end{equation}
In the following subsections, we present the limits on the operators of different subclasses.
\subsection{Operators of subclass A}
\begin{figure*}[htb!]
    \centering
    \includegraphics[width=0.35\linewidth]{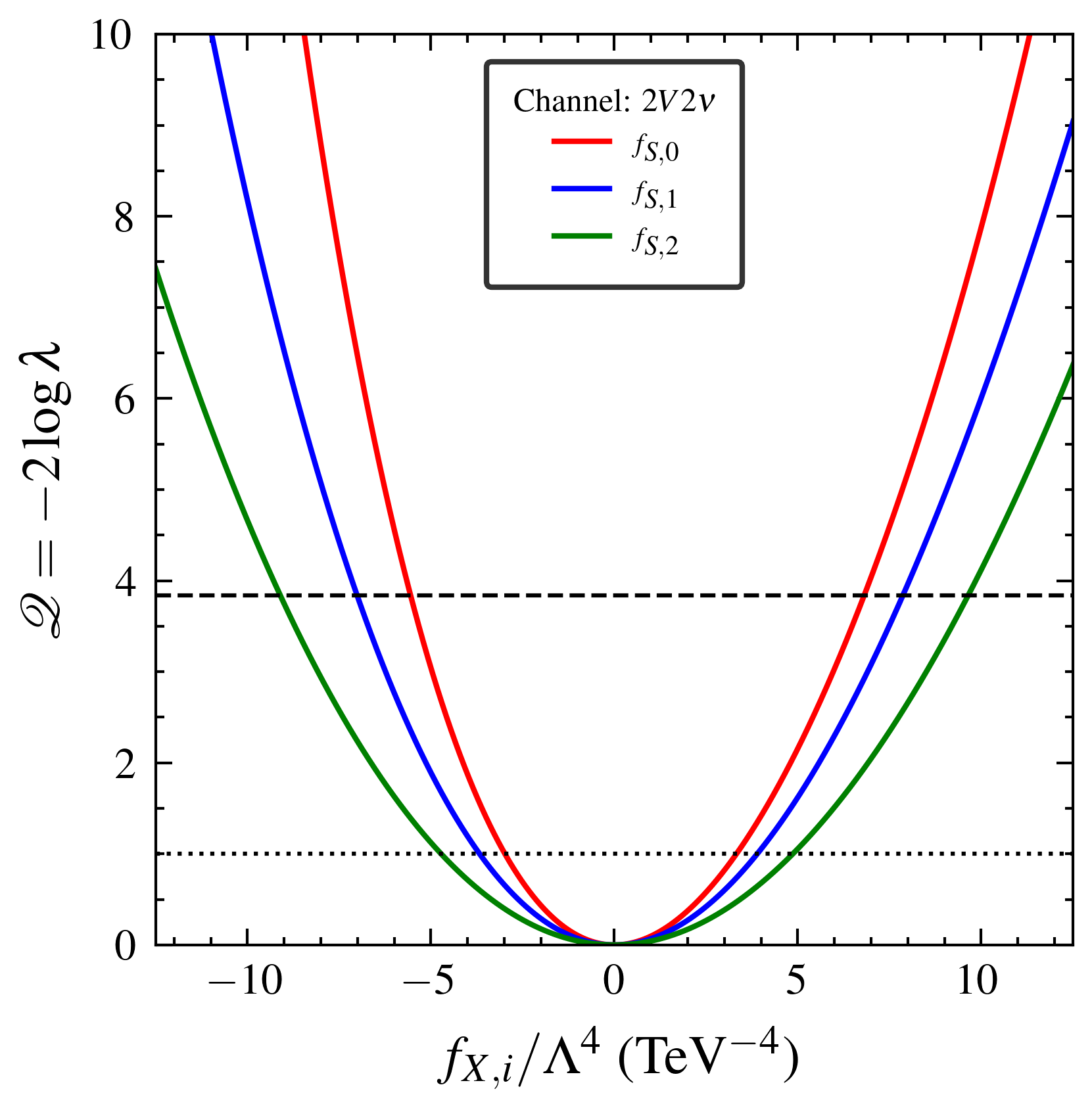}
    \includegraphics[width=0.35\linewidth]{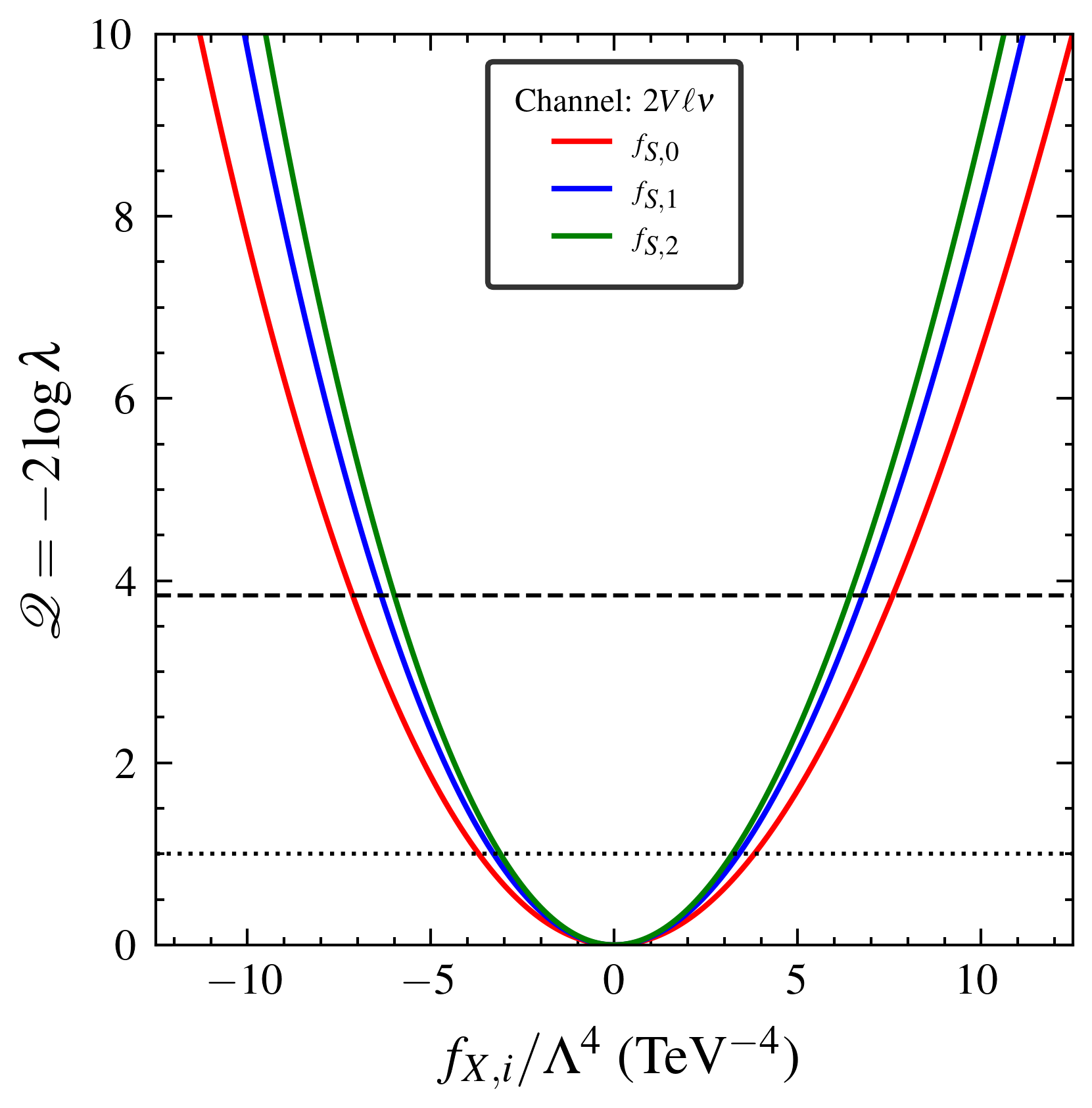} \\
    \includegraphics[width=0.35\linewidth]{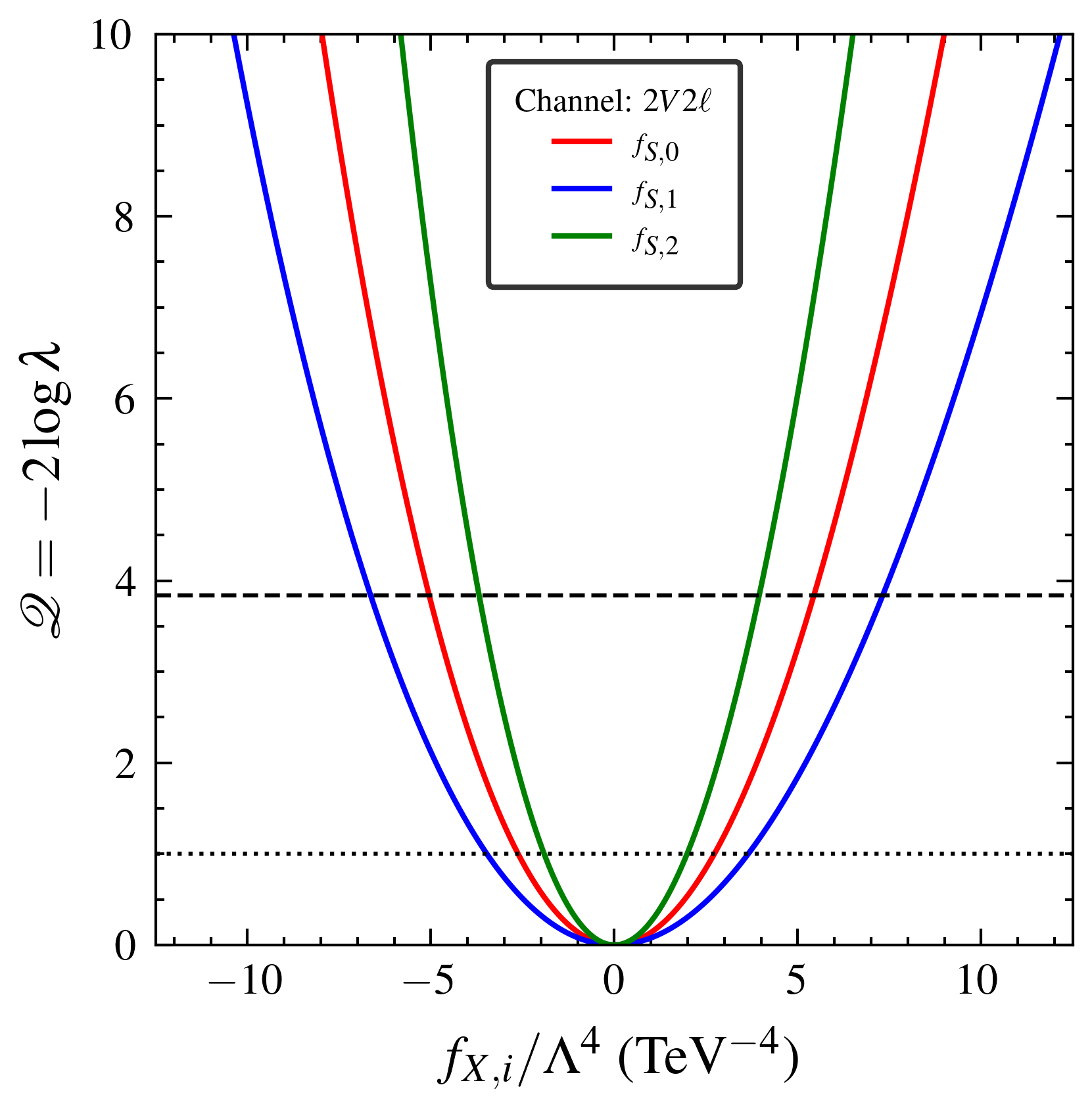}
    \includegraphics[width=0.35\linewidth]{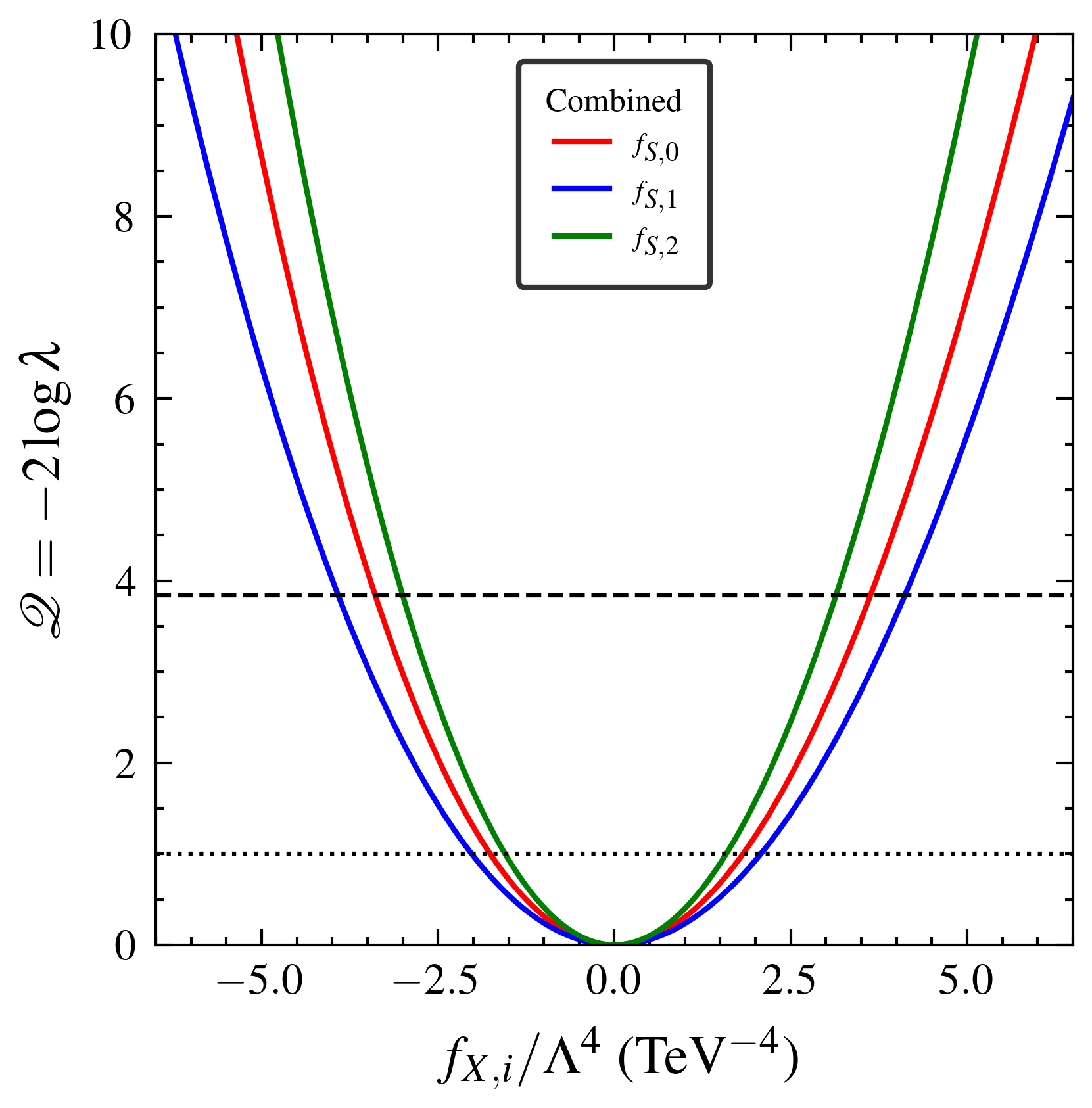} 
    \caption{95\% C.L. sensitivity plots on aQGCs of type $f_{S,i}/\Lambda^{4}$ (in TeV$^{-4}$) from different signal regions, for $\mu$TRISTAN $\sqrt{s} = 2$ TeV, with $\mathfrak{L}_{\rm int} = 1~\mathrm{ab}^{-1}$.}
    \label{fig:s1}
\end{figure*}

We, first, start with the analysis of subclass $\mathrm{A}_{0}$ operators, which correspond to the scalar class of dimension-8 operators involving only covariant derivatives of the Higgs doublet. These operators predominantly modify charged electroweak quartic interactions such as $WWWW$ and $WWZZ$, and therefore mainly contribute to final states with multiple reconstructed weak bosons. The 95\% C.L. limits on $f_{S,i}/\Lambda^{4}$ (in TeV$^{-4}$) from different signal regions, for $\mu$TRISTAN $\sqrt{s} = 2$ TeV, with $\mathfrak{L}_{\rm int} = 1~\mathrm{ab}^{-1}$, are shown in Table~\ref{tab:setA}.
\begin{table*}[htb!]
    \centering
    \begin{tabular}{>{\centering\arraybackslash}p{2.5cm}
                    >{\centering\arraybackslash}p{1.0cm}
                    >{\centering\arraybackslash}p{3.5cm}
                    >{\centering\arraybackslash}p{1.0cm}
                    >{\centering\arraybackslash}p{3.5cm}
                    >{\centering\arraybackslash}p{1.0cm}
                    >{\centering\arraybackslash}p{3.5cm}}
    \hline \hline
    \multicolumn{7}{c}{Operators: $\rm A_{0}$} \\ \hline \hline
    \multirow{2}*{$f_{S,0}/\Lambda^{4}$} & $2V2\nu$ & $[-5.6,\; 6.8]\times10^{0}$ & $V\gamma\ell\nu$ & $-$ & $2V\ell\nu$ & $[-7.1,\; 7.6]\times10^{0}$ \\
    & $2\gamma2\ell$ & $-$ & $2V2\ell$ & $[-5.0,\; 5.4]\times10^{0}$ & \textbf{$\boldsymbol{\rightarrow}$} & $\boldsymbol{[-3.4,\; 3.6]\times10^{0}}$ \\ \hline
    \multirow{2}*{$f_{S,1}/\Lambda^{4}$} & $2V2\nu$ & $[-7.0,\; 7.9]\times10^{0}$ & $V\gamma\ell\nu$ & $-$ & $2V\ell\nu$ & $[-6.4,\; 6.8]\times10^{0}$ \\
    & $2\gamma2\ell$ & $-$ & $2V2\ell$ & $[-6.6,\; 7.3]\times10^{0}$ & \textbf{$\boldsymbol{\rightarrow}$} & $\boldsymbol{[-3.9,\; 4.1]\times10^{0}}$ \\  \hline \multirow{2}*{$f_{S,2}/\Lambda^{4}$} & $2V2\nu$ & $[-9.1,\; 9.6]\times10^{0}$ & $V\gamma\ell\nu$ & $-$ & $2V\ell\nu$ & $[-6.0,\; 6.4]\times10^{0}$ \\
    & $2\gamma2\ell$ & $-$ & $2V2\ell$ & $[-3.7,\; 3.9]\times10^{0}$ & \textbf{$\boldsymbol{\rightarrow}$} & $\boldsymbol{[-3.0,\; 3.1]\times10^{0}}$ \\ 
    \hline \hline
    \end{tabular}
    \caption{95\% C.L. sensitivity constraints on aQGCs of type $f_{S,i}/\Lambda^{4}$ (in TeV$^{-4}$) from different signal regions, for $\mu$TRISTAN $\sqrt{s} = 2$ TeV, with $\mathfrak{L}_{\rm int} = 1~\mathrm{ab}^{-1}$. $\boldsymbol{\rightarrow}$ refers to as the combined limit from all the channels.}
    \label{tab:setA}
\end{table*}
As seen from Table~\ref{tab:setA}, the sensitivity to these operators arises primarily from the $2V2\nu$, $2V\ell\nu$, and $2V2\ell$ signal regions, while channels involving photons do not contribute. Among these, the $2V2\ell$ channel provides the strongest individual constraints due to its clean final state and reduced combinatorial ambiguities. The combined limits, obtained by statistically combining all relevant signal regions, show a noticeable improvement over individual channels. This highlights the complementarity between different final states in constraining the parameter space. The resulting bounds remain at $\mathcal{O}(1)$ in units of TeV$^{-4}$, comparative to the LHC bounds. The corresponding likelihood profiles for individual and combined channels are shown in Figure~\ref{fig:s1}, illustrating the relative constraining power of each signal region.
\subsection{Operators of subclass B}
\begin{figure*}[htb!]
    \centering
    \includegraphics[width=0.35\linewidth]{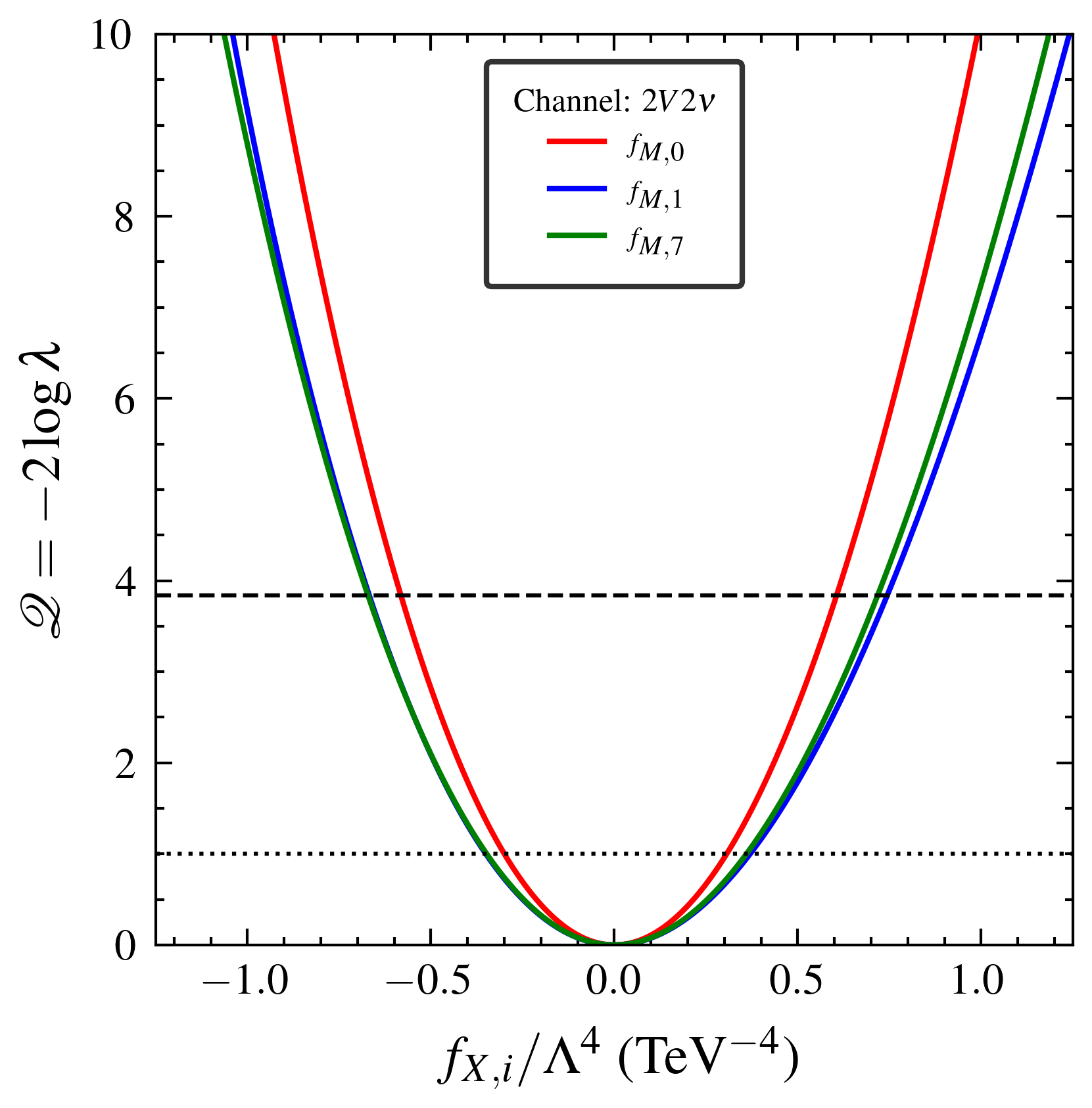}
    \includegraphics[width=0.35\linewidth]{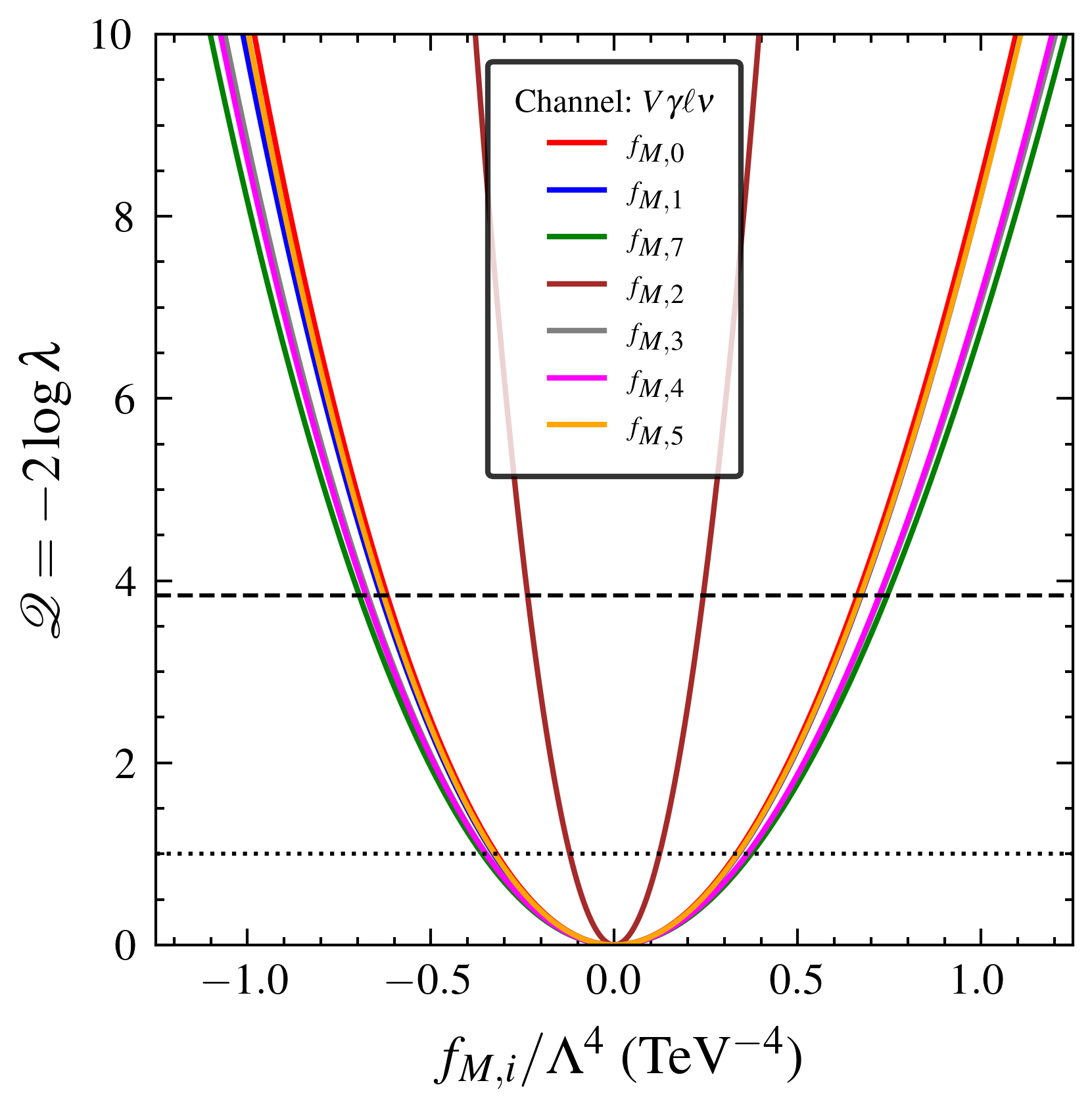}
    \includegraphics[width=0.35\linewidth]{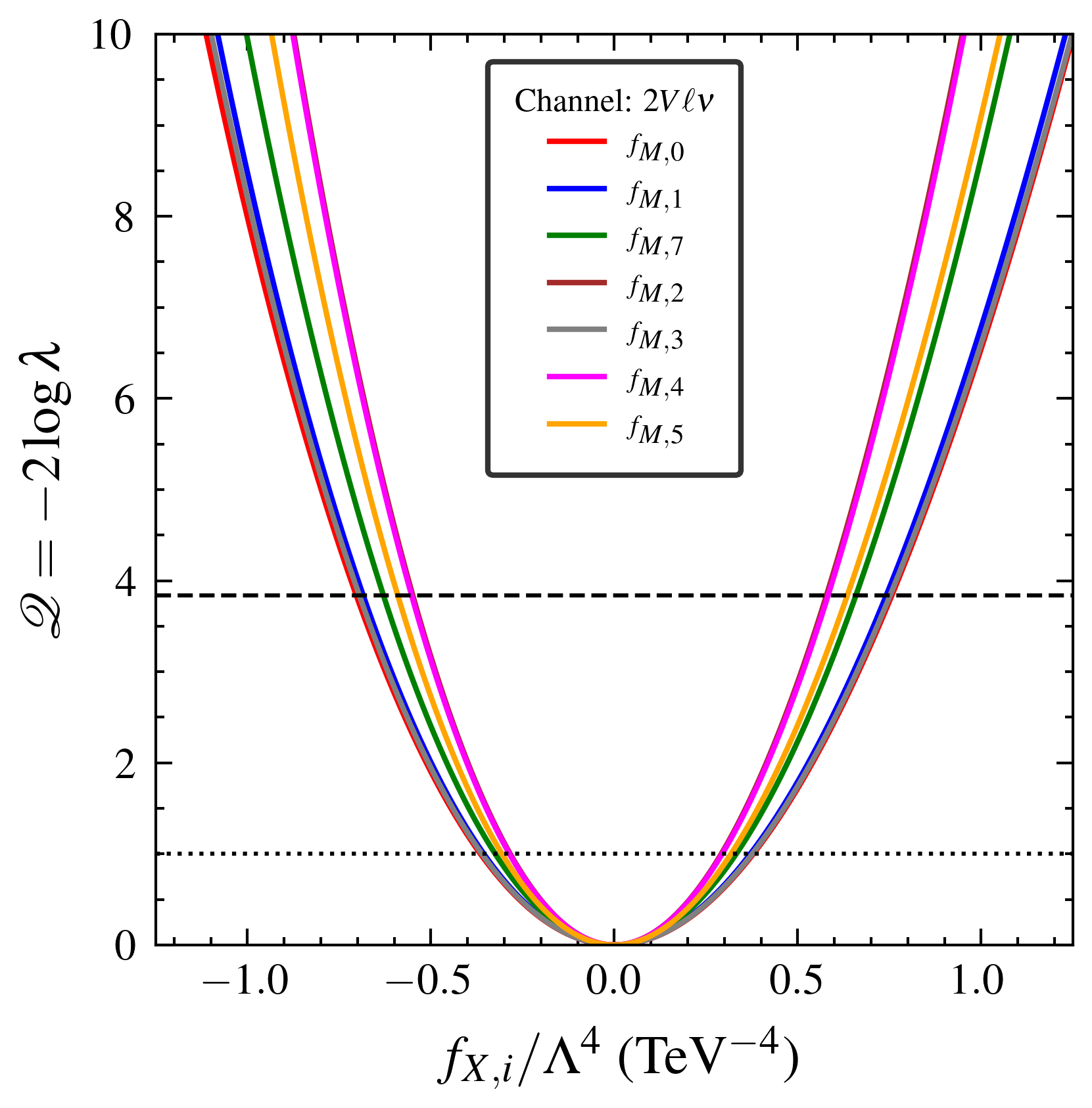}
    \includegraphics[width=0.35\linewidth]{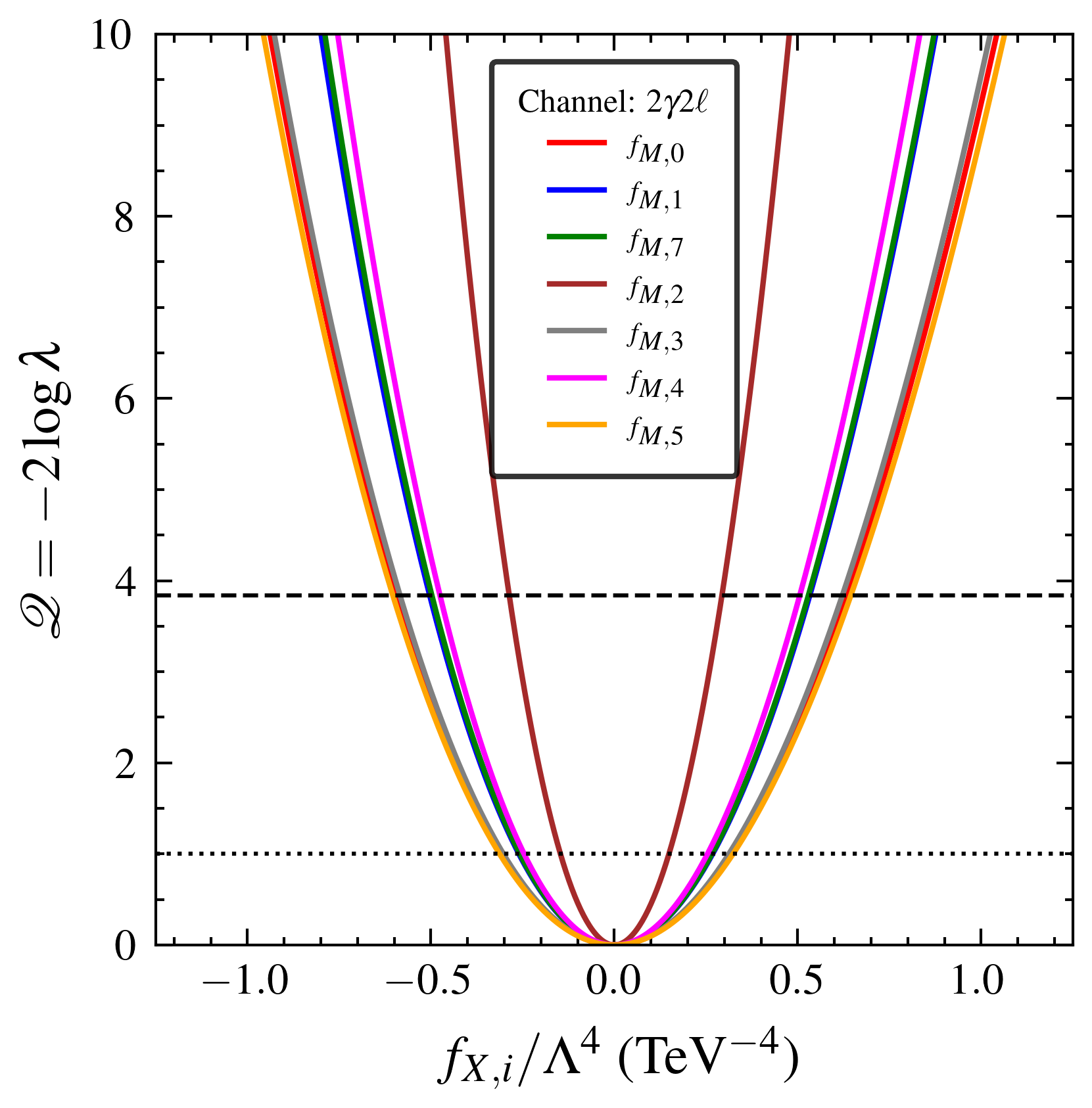}
    \includegraphics[width=0.35\linewidth]{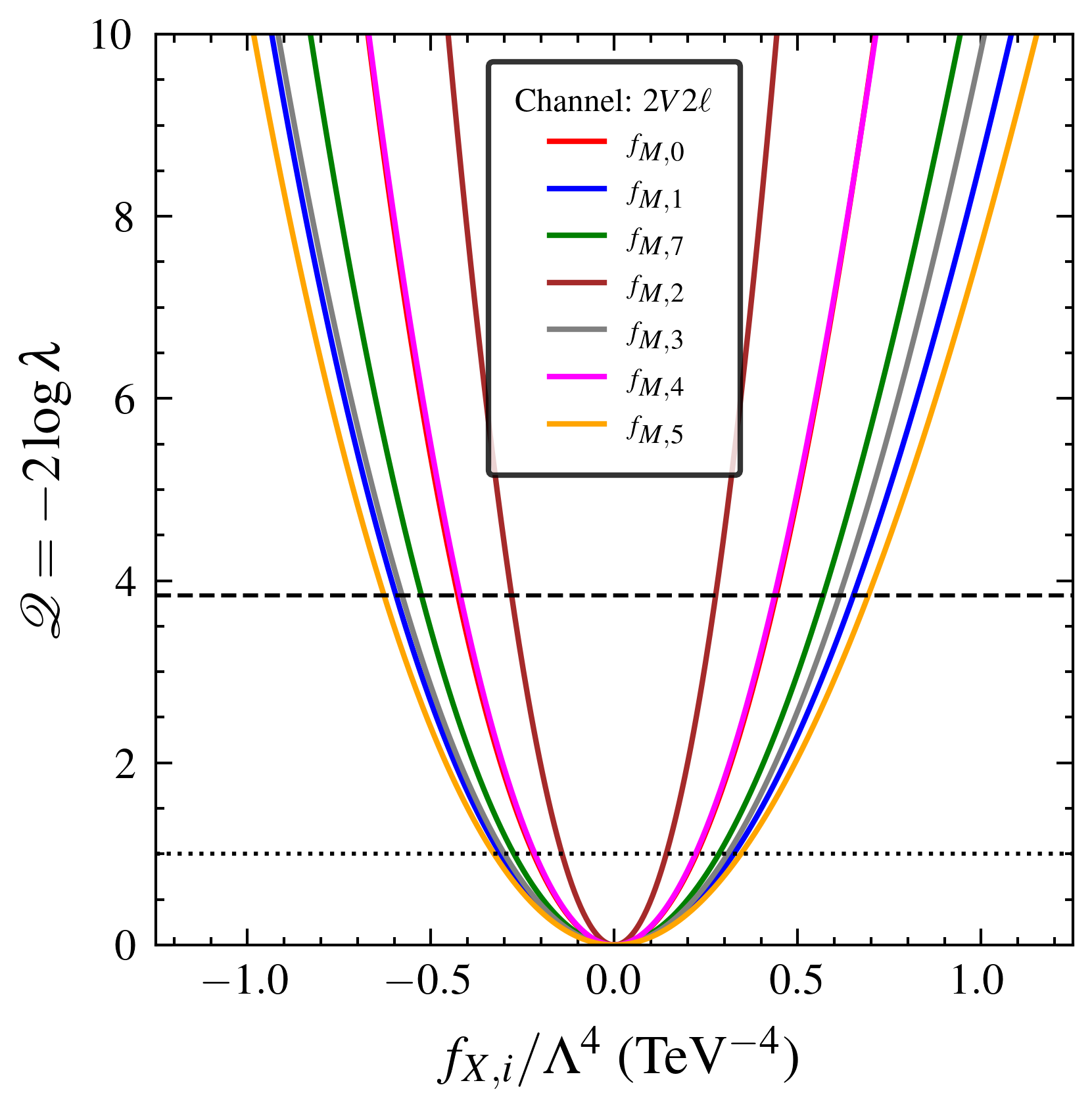}
    \includegraphics[width=0.35\linewidth]{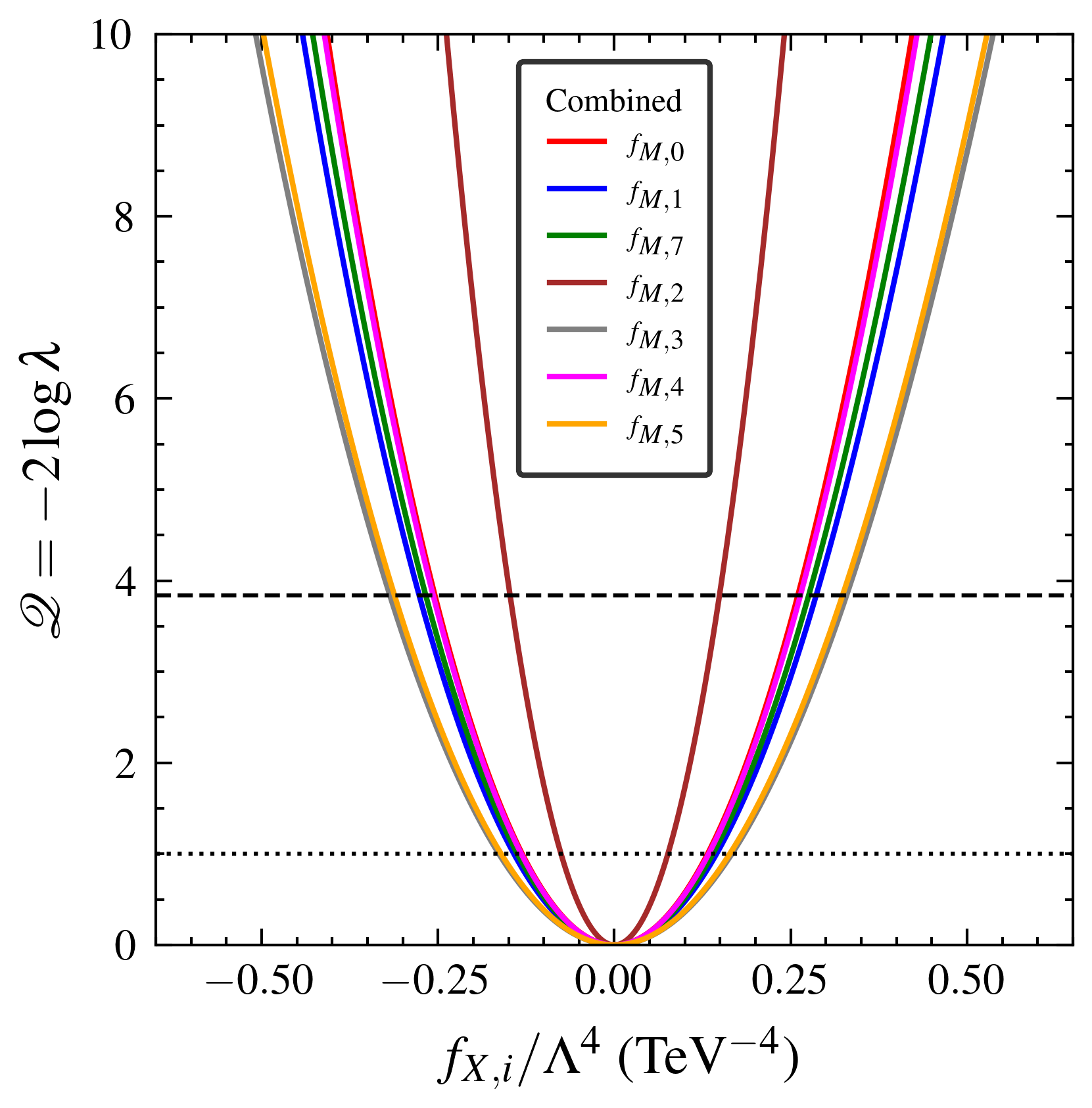}
    \caption{95\% C.L. sensitivity plots on aQGCs of type $f_{M,i}/\Lambda^{4}$ (in TeV$^{-4}$) from different signal regions, for $\mu$TRISTAN $\sqrt{s} = 2$ TeV, with $\mathfrak{L}_{\rm int} = 1~\mathrm{ab}^{-1}$.}
    \label{fig:s2}
\end{figure*}

Next, we consider subclass $\mathrm{B}$ operators, which belong to the mixed category i.e. scalar and tensor combination, involving both Higgs fields and field strength tensors. These operators induce a broader range of quartic interactions, including $WW\gamma\gamma$ and $WWZ\gamma$, thereby contributing to both pure bosonic and photon associated final states. The 95\% C.L. limits on $f_{M,i}/\Lambda^{4}$ (in TeV$^{-4}$) from different signal regions, for $\mu$TRISTAN $\sqrt{s} = 2$ TeV, with $\mathfrak{L}_{\rm int} = 1~\mathrm{ab}^{-1}$, are shown in Table~\ref{tab:setB}.
\begin{table*}[htb!]
    \centering
    \begin{tabular}{>{\centering\arraybackslash}p{2.5cm}
                    >{\centering\arraybackslash}p{1.0cm}
                    >{\centering\arraybackslash}p{3.5cm}
                    >{\centering\arraybackslash}p{1.0cm}
                    >{\centering\arraybackslash}p{3.5cm}
                    >{\centering\arraybackslash}p{1.0cm}
                    >{\centering\arraybackslash}p{3.5cm}}
    \hline \hline
    \multicolumn{7}{c}{Operators: $\rm B_{1}$} \\ \hline \hline
    \multirow{2}*{$f_{M,0}/\Lambda^{4}$} & $2V2\nu$ & $[-5.8,\; 6.0]\times10^{-1}$ & $V\gamma\ell\nu$ & $[-6.2,\; 6.6]\times10^{-1}$ & $2V\ell\nu$ & $[-7.0,\; 7.6]\times10^{-1}$ \\
    & $2\gamma2\ell$ & $[-5.9,\; 6.3]\times10^{-1}$ & $2V2\ell$ & $[-4.3,\; 4.4]\times10^{-1}$ & \textbf{$\boldsymbol{\rightarrow}$} & $\boldsymbol{[-2.5,\; 2.6]\times10^{-1}}$ \\ \hline
    \multirow{2}*{$f_{M,1}/\Lambda^{4}$} & $2V2\nu$ & $[-6.7,\; 7.4]\times10^{-1}$ & $V\gamma\ell\nu$ & $[-6.4,\; 6.7]\times10^{-1}$ & $2V\ell\nu$ & $[-6.8,\; 7.4]\times10^{-1}$ \\
    & $2\gamma2\ell$ & $[-5.0,\; 5.3]\times10^{-1}$ & $2V2\ell$ & $[-6.0,\; 6.5]\times10^{-1}$ & \textbf{$\boldsymbol{\rightarrow}$} & $\boldsymbol{[-2.8,\; 2.9]\times10^{-1}}$ \\  \hline
    \multirow{2}*{$f_{M,7}/\Lambda^{4}$} & $2V2\nu$ & $[-6.7,\; 7.2]\times10^{-1}$ & $V\gamma\ell\nu$ & $[-7.0,\; 7.4]\times10^{-1}$ & $2V\ell\nu$ & $[-6.3,\; 6.6]\times10^{-1}$ \\
    & $2\gamma2\ell$ & $[-5.0,\; 5.3]\times10^{-1}$ & $2V2\ell$ & $[-5.3,\; 5.7]\times10^{-1}$ & \textbf{$\boldsymbol{\rightarrow}$} & $\boldsymbol{[-2.7,\; 2.8]\times10^{-1}}$ \\ 
    \hline \hline
    \multicolumn{7}{c}{Operators: $\rm B_{2}$} \\ \hline \hline
    \multirow{2}*{$f_{M,2}/\Lambda^{4}$} & $2V2\nu$ & $-$ & $V\gamma\ell\nu$ & $[-2.4,\; 2.4]\times10^{-1}$ & $2V\ell\nu$ & $[-5.5,\; 5.8]\times10^{-1}$ \\
    & $2\gamma2\ell$ & $[-2.9,\; 2.9]\times10^{-1}$ & $2V2\ell$ & $[-2.8,\; 2.8]\times10^{-1}$ & \textbf{$\boldsymbol{\rightarrow}$} & $\boldsymbol{[-1.5,\; 1.5]\times10^{-1}}$ \\ \hline
    \multirow{2}*{$f_{M,3}/\Lambda^{4}$} & $2V2\nu$ & $-$ & $V\gamma\ell\nu$ & $[-6.7,\; 7.2]\times10^{-1}$ & $2V\ell\nu$ & $[-7.0,\; 7.5]\times10^{-1}$ \\
    & $2\gamma2\ell$ & $[-5.8,\; 6.2]\times10^{-1}$ & $2V2\ell$ & $[-5.8,\; 6.1]\times10^{-1}$ & \textbf{$\boldsymbol{\rightarrow}$} & $\boldsymbol{[-3.2,\; 3.3]\times10^{-1}}$ \\
    \hline \hline
    \multicolumn{7}{c}{Operators: $\rm B_{3}$} \\ \hline \hline
    \multirow{2}*{$f_{M,4}/\Lambda^{4}$} & $2V2\nu$ & $-$ & $V\gamma\ell\nu$ & $[-6.8,\; 7.2]\times10^{-1}$ & $2V\ell\nu$ & $[-5.5,\; 5.8]\times10^{-1}$ \\
    & $2\gamma2\ell$ & $[-4.8,\; 5.0]\times10^{-1}$ & $2V2\ell$ & $[-4.2,\; 4.4]\times10^{-1}$ & \textbf{$\boldsymbol{\rightarrow}$} & $\boldsymbol{[-2.6,\; 2.6]\times10^{-1}}$ \\ \hline
    \multirow{2}*{$f_{M,5}/\Lambda^{4}$} & $2V2\nu$ & $-$ & $V\gamma\ell\nu$ & $[-6.3,\; 6.7]\times10^{-1}$ & $2V\ell\nu$ & $[-5.9,\; 6.4]\times10^{-1}$ \\
    & $2\gamma2\ell$ & $[-6.0,\; 6.4]\times10^{-1}$ & $2V2\ell$ & $[-6.3,\; 6.9]\times10^{-1}$ & \textbf{$\boldsymbol{\rightarrow}$} & $\boldsymbol{[-3.1,\; 3.2]\times10^{-1}}$ \\
    \hline \hline    
    \end{tabular}
    \caption{95\% C.L. sensitivity constraints on aQGCs of type $f_{M,i}/\Lambda^{4}$ (in TeV$^{-4}$) from different signal regions, for $\mu$TRISTAN $\sqrt{s} = 2$ TeV, with $\mathfrak{L}_{\rm int} = 1~\mathrm{ab}^{-1}$. $\boldsymbol{\rightarrow}$ refers to as the combined limit from all the channels.}
    \label{tab:setB}
\end{table*}
From Table~\ref{tab:setB}, it is evident that all signal regions, including the photon associated regions $V\gamma\ell\nu$ and $2\gamma2\ell$, play an important role in constraining these operators. In particular, photon rich final states significantly enhance sensitivity due to the higher kinematic access and direct coupling of these operators to electroweak field strengths, which induce momentum dependent Lorentz structures. The interplay between different signal regions leads to comparable sensitivities across multiple channels. The combined limits show a substantial improvement over individual constraints, reaching $\mathcal{O}(10^{-1})$ in TeV$^{-4}$. This improvement reflects both the structure of these operators as well as the increased statistics from combining multiple final states. The sensitivity projections and likelihood distributions are illustrated in Figure~\ref{fig:s2}, where the impact of each signal region and their combination can be clearly observed.
\subsection{Operators of subclass C}
\begin{figure*}[htb!]
    \centering
    \includegraphics[width=0.35\linewidth]{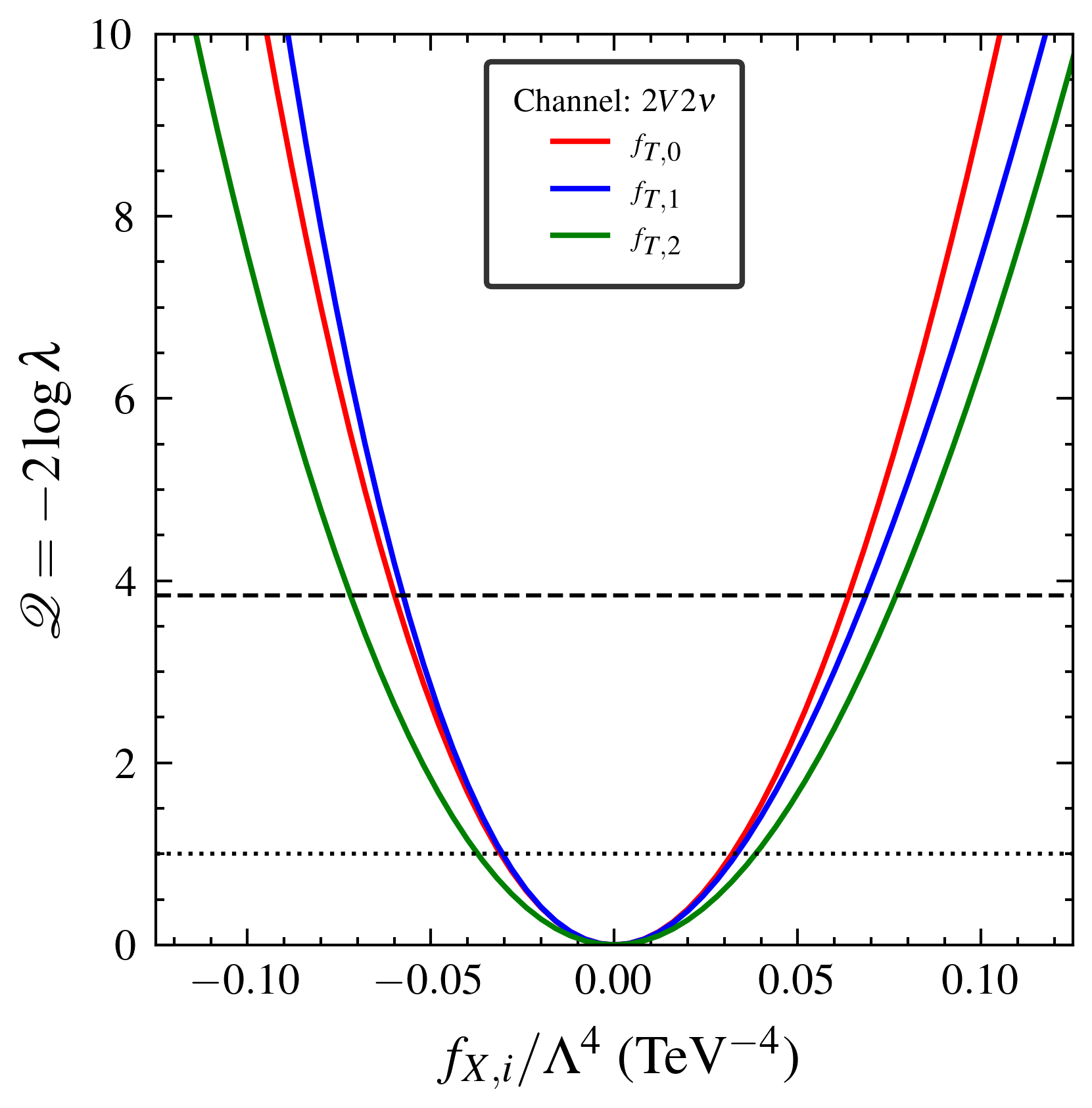}
    \includegraphics[width=0.35\linewidth]{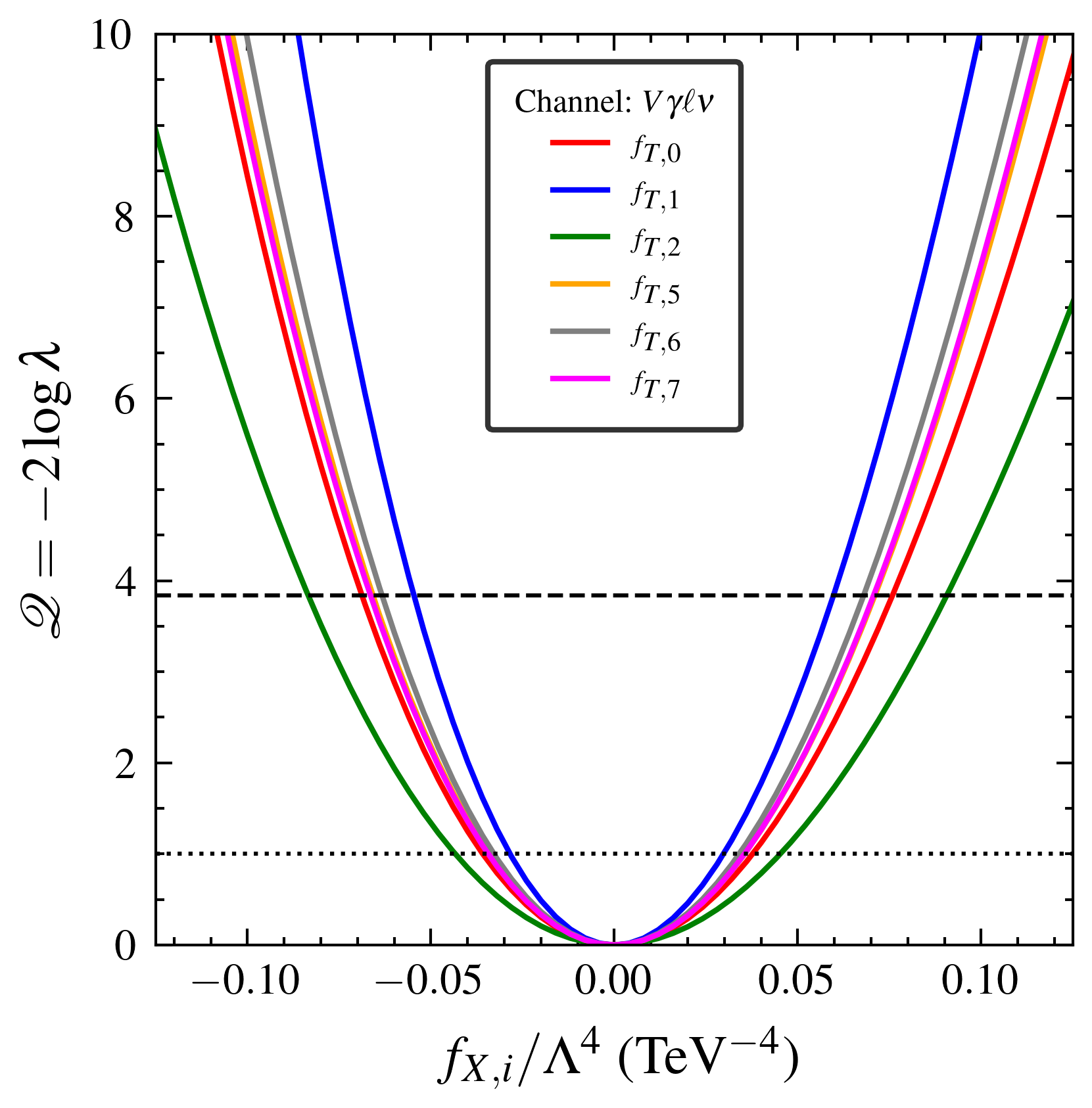}
    \includegraphics[width=0.35\linewidth]{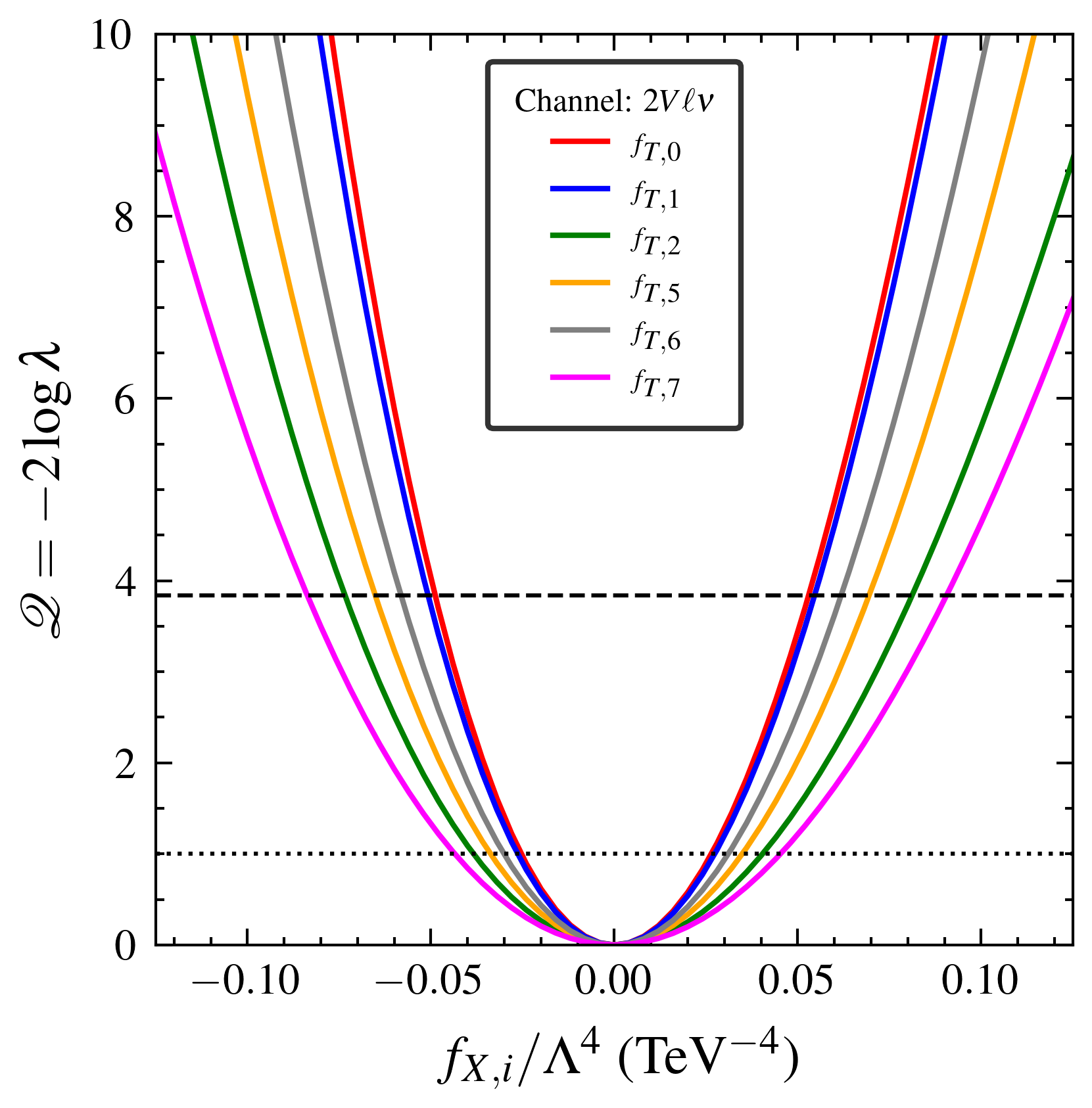}
    \includegraphics[width=0.35\linewidth]{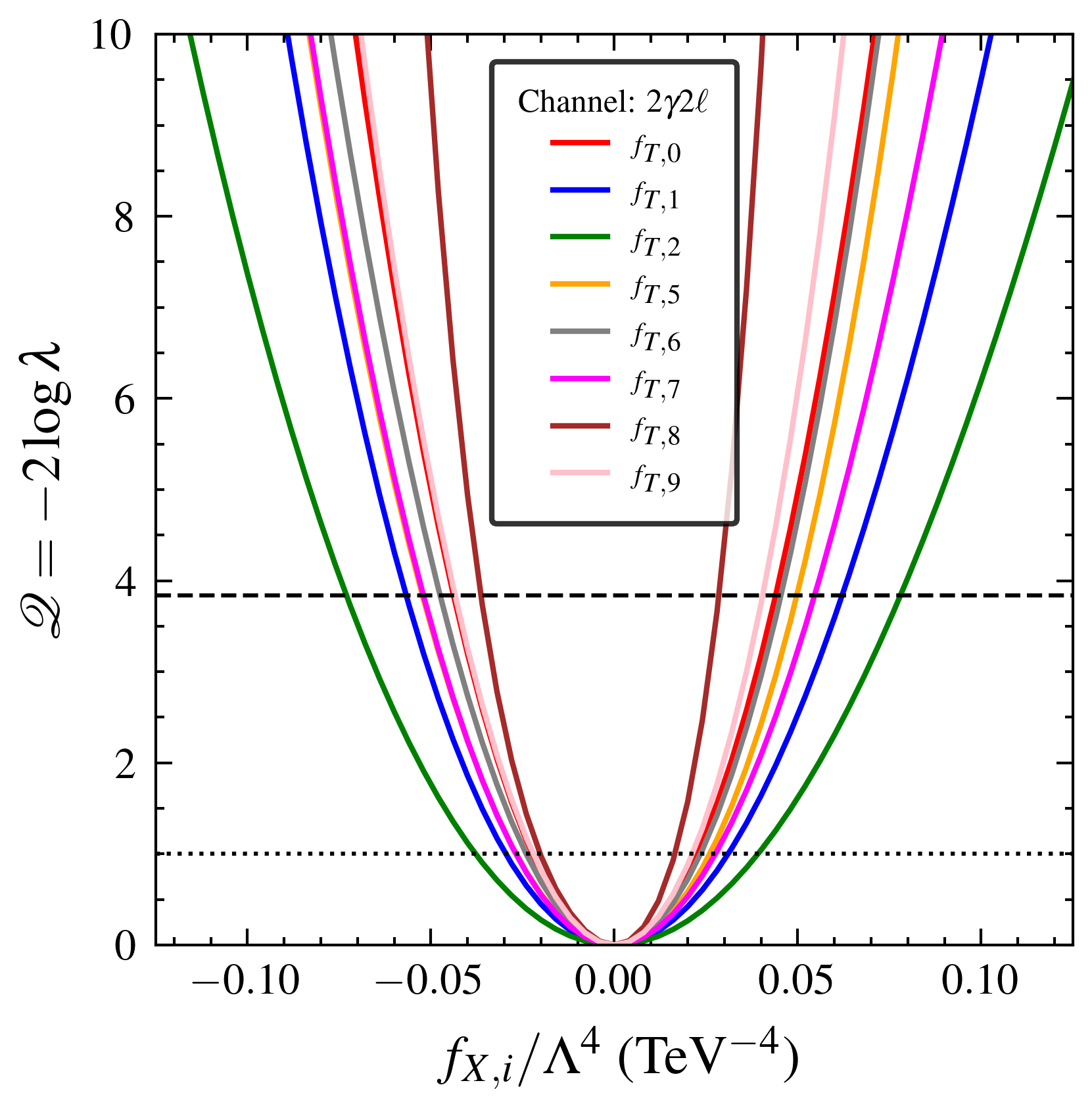}
    \includegraphics[width=0.35\linewidth]{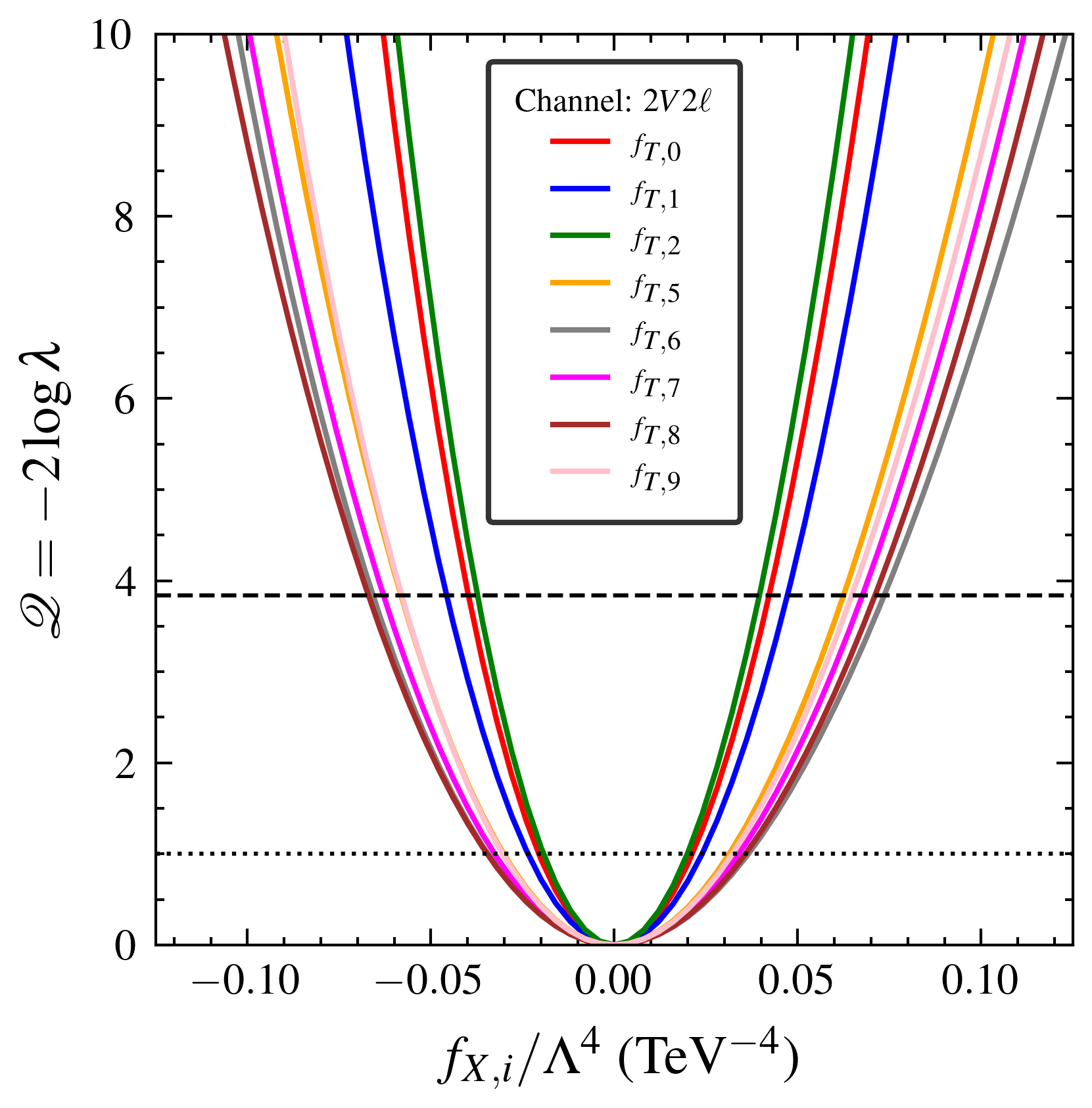}
    \includegraphics[width=0.35\linewidth]{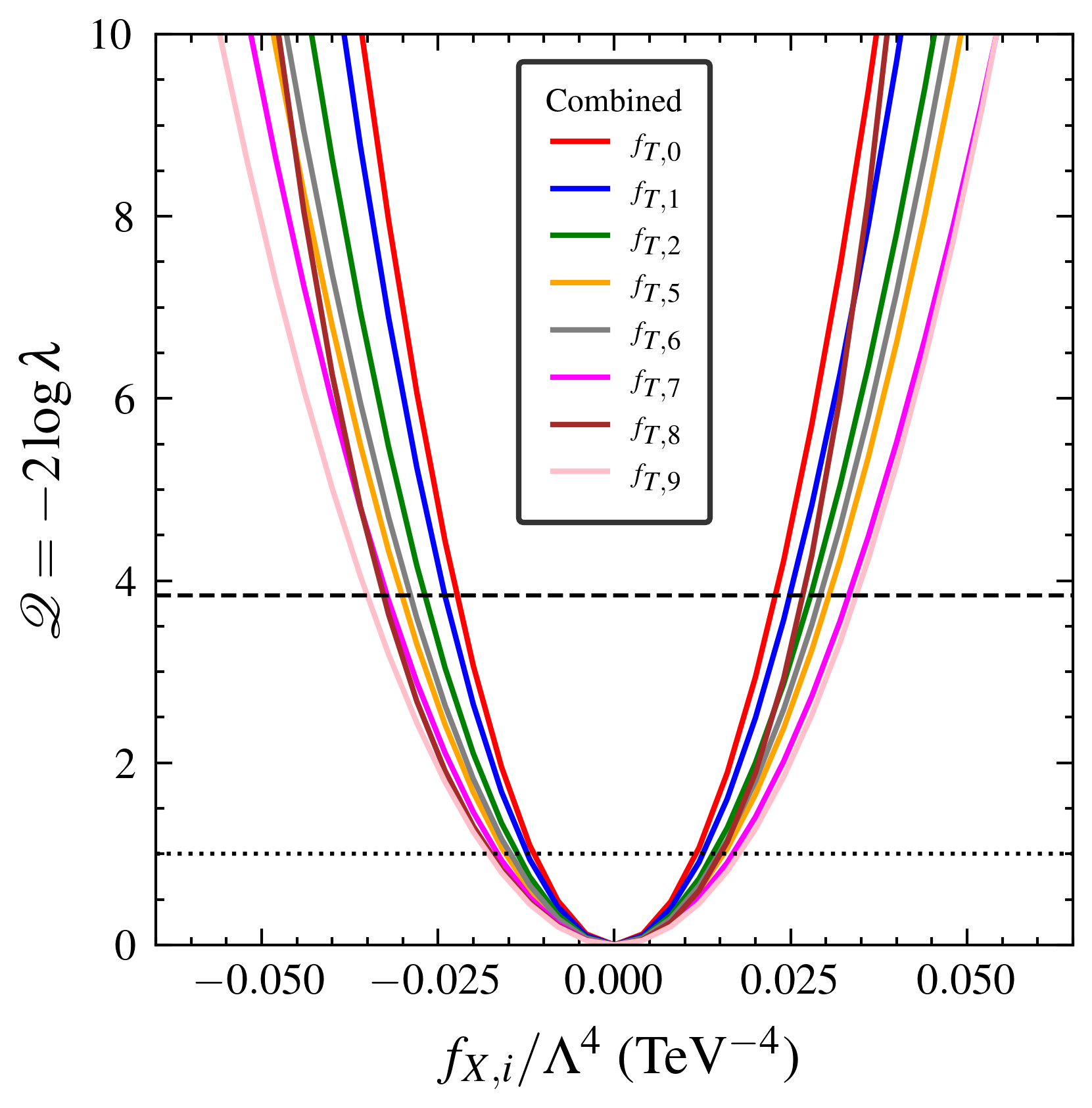}
    \caption{95\% C.L. sensitivity plots on aQGCs of type $f_{T,i}/\Lambda^{4}$ (in TeV$^{-4}$) from different signal regions, for $\mu$TRISTAN $\sqrt{s} = 2$ TeV, with $\mathfrak{L}_{\rm int} = 1~\mathrm{ab}^{-1}$.}
    \label{fig:s3}
\end{figure*}

Finally, we analyze subclass $\mathrm{C}$ operators, corresponding to the pure tensor class constructed from gauge field strength tensors. These operators generate the most pronounced energy growing effects and contribute to both cQGC and nQGC couplings, including purely neutral quartic gauge interactions absent in the SM at tree level. The 95\% C.L. limits on $f_{T,i}/\Lambda^{4}$ (in TeV$^{-4}$) from different signal regions, for $\mu$TRISTAN $\sqrt{s} = 2$ TeV, with $\mathfrak{L}_{\rm int} = 1~\mathrm{ab}^{-1}$, are shown in Table~\ref{tab:setC}.
\begin{table*}[htb!]
    \centering
    \begin{tabular}{>{\centering\arraybackslash}p{2.5cm}
                    >{\centering\arraybackslash}p{1.0cm}
                    >{\centering\arraybackslash}p{3.5cm}
                    >{\centering\arraybackslash}p{1.0cm}
                    >{\centering\arraybackslash}p{3.5cm}
                    >{\centering\arraybackslash}p{1.0cm}
                    >{\centering\arraybackslash}p{3.5cm}}
    \hline \hline
    \multicolumn{7}{c}{Operators: $\rm C_{1}$} \\ \hline \hline
    \multirow{2}*{$f_{T,0}/\Lambda^{4}$} & $2V2\nu$ & $[-6.0,\; 6.4]\times10^{-2}$ & $V\gamma\ell\nu$ & $[-6.9,\; 7.6]\times10^{-2}$ & $2V\ell\nu$ & $[-4.9,\; 5.3]\times10^{-2}$ \\
    & $2\gamma2\ell$ & $[-4.4,\; 4.4]\times10^{-2}$ & $2V2\ell$ & $[-4.0,\; 4.2]\times10^{-2}$ & \textbf{$\boldsymbol{\rightarrow}$} & $\boldsymbol{[-2.2,\; 2.3]\times10^{-2}}$ \\ \hline
    \multirow{2}*{$f_{T,1}/\Lambda^{4}$} & $2V2\nu$ & $[-5.8,\; 6.8]\times10^{-2}$ & $V\gamma\ell\nu$ & $[-5.5,\; 6.0]\times10^{-2}$ & $2V\ell\nu$ & $[-5.1,\; 5.5]\times10^{-2}$ \\
    & $2\gamma2\ell$ & $[-5.7,\; 6.2]\times10^{-2}$ & $2V2\ell$ & $[-4.6,\; 4.7]\times10^{-2}$ & \textbf{$\boldsymbol{\rightarrow}$} & $\boldsymbol{[-2.4,\; 2.5]\times10^{-2}}$ \\  \hline
    \multirow{2}*{$f_{T,2}/\Lambda^{4}$} & $2V2\nu$ & $[-7.2,\; 7.7]\times10^{-2}$ & $V\gamma\ell\nu$ & $[-8.4,\; 9.1]\times10^{-2}$ & $2V\ell\nu$ & $[-7.3,\; 8.1]\times10^{-2}$ \\
    & $2\gamma2\ell$ & $[-7.3,\; 7.8]\times10^{-2}$ & $2V2\ell$ & $[-3.7,\; 4.0]\times10^{-2}$ & \textbf{$\boldsymbol{\rightarrow}$} & $\boldsymbol{[-2.7,\; 2.8]\times10^{-2}}$ \\ 
    \hline \hline
    \multicolumn{7}{c}{Operators: $\rm C_{2}$} \\ \hline \hline
    \multirow{2}*{$f_{T,5}/\Lambda^{4}$} & $2V2\nu$ & $-$ & $V\gamma\ell\nu$ & $[-6.6,\; 7.1]\times10^{-2}$ & $2V\ell\nu$ & $[-6.5,\; 6.9]\times10^{-2}$ \\
    & $2\gamma2\ell$ & $[-5.2,\; 5.0]\times10^{-2}$ & $2V2\ell$ & $[-5.8,\; 6.2]\times10^{-2}$ & \textbf{$\boldsymbol{\rightarrow}$} & $\boldsymbol{[-3.0,\; 3.0]\times10^{-2}}$ \\ \hline
    \multirow{2}*{$f_{T,6}/\Lambda^{4}$} & $2V2\nu$ & $-$ & $V\gamma\ell\nu$ & $[-6.3,\; 6.8]\times10^{-2}$ & $2V\ell\nu$ & $[-5.8,\; 6.2]\times10^{-2}$ \\
    & $2\gamma2\ell$ & $[-4.8,\; 4.5]\times10^{-2}$ & $2V2\ell$ & $[-6.6,\; 7.4]\times10^{-2}$ & \textbf{$\boldsymbol{\rightarrow}$} & $\boldsymbol{[-2.9,\; 2.9]\times10^{-2}}$ \\ \hline
    \multirow{2}*{$f_{T,7}/\Lambda^{4}$} & $2V2\nu$ & $-$ & $V\gamma\ell\nu$ & $[-6.6,\; 7.1]\times10^{-2}$ & $2V\ell\nu$ & $[-8.4,\; 9.1]\times10^{-2}$ \\
    & $2\gamma2\ell$ & $[-5.2,\; 5.5]\times10^{-2}$ & $2V2\ell$ & $[-6.3,\; 6.8]\times10^{-2}$ & \textbf{$\boldsymbol{\rightarrow}$} & $\boldsymbol{[-3.2,\; 3.3]\times10^{-2}}$ \\
    \hline \hline
    \multicolumn{7}{c}{Operators: $\rm C_{3}$} \\ \hline \hline
    \multirow{2}*{$f_{T,8}/\Lambda^{4}$} & $2V2\nu$ & $-$ & $V\gamma\ell\nu$ & $-$ & $2V\ell\nu$ & $-$ \\
    & $2\gamma2\ell$ & $[-3.6,\; 2.8]\times10^{-2}$ & $2V2\ell$ & $[-6.7,\; 7.1]\times10^{-2}$ & \textbf{$\boldsymbol{\rightarrow}$} & $\boldsymbol{[-3.3,\; 2.7]\times10^{-2}}$ \\ \hline
    \multirow{2}*{$f_{T,9}/\Lambda^{4}$} & $2V2\nu$ & $-$ & $V\gamma\ell\nu$ & $-$ & $2V\ell\nu$ & $-$ \\
    & $2\gamma2\ell$ & $[-4.3,\; 4.0]\times10^{-2}$ & $2V2\ell$ & $[-5.8,\; 6.5]\times10^{-2}$ & \textbf{$\boldsymbol{\rightarrow}$} & $\boldsymbol{[-3.5,\; 3.4]\times10^{-2}}$ \\
    \hline \hline    
    \end{tabular}
    \caption{95\% C.L. sensitivity constraints on aQGCs of type $f_{T,i}/\Lambda^{4}$ (in TeV$^{-4}$) from different signal regions, for $\mu$TRISTAN $\sqrt{s} = 2$ TeV, with $\mathfrak{L}_{\rm int} = 1~\mathrm{ab}^{-1}$. $\boldsymbol{\rightarrow}$ refers to as the combined limit from all the channels.}
    \label{tab:setC}
\end{table*}
As shown in Table~\ref{tab:setC}, the sensitivity to these operators is significantly stronger compared to subclasses $\mathrm{A}$ and $\mathrm{B}$, with constraints reaching $\mathcal{O}(10^{-2})-\mathcal{O}(10^{-3})$ in TeV$^{-4}$. All signal regions contribute meaningfully, with the photon rich and multiboson final states playing a particularly important role due to the structure of the operators. The combined limits benefit greatly from the inclusion of all channels, especially those sensitive to neutral quartic interactions such as $2\gamma2\ell$. The enhanced reach reflects the strong energy dependence of tensor operators, making them especially well-suited for probing at high energy lepton colliders, which can access large CM energies at the parton level. The corresponding likelihood profiles are presented in Figure~\ref{fig:s3}, demonstrating the superior sensitivity for this class of operators and the importance of combining multiple channels.

\subsection{Future upgrades}
\label{sec:future}
In this section, we explore the projected sensitivity to aQGCs under future upgrade scenarios of the $\mu$TRISTAN facility. In particular, we consider increases in both the CM energy and the integrated luminosity, which play a crucial role in enhancing the reach of higher dimensional operators within the EFT framework. The results for three representative configurations, namely $(\sqrt{s}, \mathcal{L}) = (2~\mathrm{TeV},\,10~\mathrm{ab}^{-1})$, $(6~\mathrm{TeV},\,1~\mathrm{ab}^{-1})$, and $(6~\mathrm{TeV},\,10~\mathrm{ab}^{-1})$, are summarized in Table~\ref{tab:future_MT}. These projections are obtained using the same multi-observable likelihood framework described in Section~\ref{sec:sensitivity}, combining information from all relevant signal regions. A clear hierarchy in sensitivity improvement is observed as the collider configurations i.e. CM energy and the integrated luminosity, are upgraded. Increasing the integrated luminosity at fixed energy, for instance from $\mathfrak{L}_{\rm int} = 1~\mathrm{ab}^{-1}$ to $\mathfrak{L}_{\rm int} = 10~\mathrm{ab}^{-1}$ at $\sqrt{s} = 2$~TeV, leads to a moderate improvement in the bounds, typically by a factor of $\sim 2-3$. This behavior is consistent with the expected statistical scaling, as the sensitivity in the statistics dominated regime improves approximately as $1/\sqrt{\mathfrak{L}_{\rm int}}$. In contrast, increasing the CM energy has a far more pronounced impact on the sensitivity. Moving from $\sqrt{s} = 2$~TeV to $6$~TeV results in an order of magnitude or more improvement in the constraints in case of most operators. This enhancement arises primarily from the intrinsic energy dependence of dimension-8 operator contributions, which grow rapidly with energy and dominate over the SM background in the high energy tails of kinematic distributions. As a result, higher energy configurations significantly extend the reach of the SMEFT parameter space. This trend is particularly striking for tensor operators of subclass $\mathrm{C}$, operators of purely tensor type, where the projected sensitivities reach the level of $\mathcal{O}(10^{-4})-\mathcal{O}(10^{-5})$~TeV$^{-4}$ at $\sqrt{s}=6~\mathrm{TeV}$ with $\mathfrak{L}_{\rm int} = 10~\mathrm{ab}^{-1})$. Such precision represents a significant improvement of several orders of magnitude over current LHC bounds (see Table~\ref{tab:QGC_LHC_combined}), and comparable to analysis at $\mu^{+}\mu^{-}$ colliders~\cite{Yang:2022fhw,Gutierrez-Rodriguez:2025wcy}, which require way larger CM energy or the integrated luminosity, in comparison. Mixed operators of subclass $\mathrm{B}$ also exhibit substantial gains, reaching sensitivities at the $\mathcal{O}(10^{-3})$ level, while scalar operators of subclass $\mathrm{A}$ improve to the $\mathcal{O}(10^{-2})$ regime. It is also worth noting that the relative improvement is not uniform across all operators. Operators contributing to photon rich final states, such as those involving $WW\gamma\gamma$ and $Z\gamma\gamma\gamma$ vertices, benefit significantly from higher energies due to both enhanced production rates and much cleaner experimental signatures. Similarly, operators inducing purely nQGCs, which are absent at leading order in the SM, experience reduced background contamination and thus yield stronger constraints. This highlights the importance of pursuing both high energy and high luminosity upgrades in future collider programs. In particular, the strong energy scaling of dimension-8 operators makes high energy lepton colliders uniquely powerful probes of aQGCs, surpassing the capabilities of current hadron colliders. Overall, the future upgrade scenarios of $\mu$TRISTAN demonstrate a remarkable potential to probe aQGCs with unprecedented precision, opening a new window into the electroweak sector and possible BSM physics.
\begin{table*}[htb!]
\centering
\begin{tabular}{
>{\centering\arraybackslash}p{2.25cm}
>{\centering\arraybackslash}p{4.25cm}
>{\centering\arraybackslash}p{4.25cm}
>{\centering\arraybackslash}p{4.25cm}
}
\hline \hline
\multirow{2}*{WCs} & \multicolumn{3}{c}{Future $\mu$TRISTAN configurations} \\ \cline{2-4}
& $2\text{ TeV},\;10\text{ ab}^{-1}$ & $6\text{ TeV},\;1\text{ ab}^{-1}$ & $6\text{ TeV},\;10\text{ ab}^{-1}$ \\
\hline \hline
$f_{S,0}/\Lambda^4$ & $[-1.1,\;1.1]\times10^{0}$ & $[-2.0,\;2.1]\times10^{-1}$ & $[-6.4,\;6.5]\times10^{-2}$ \\
$f_{S,1}/\Lambda^4$ & $[-1.2,\;1.3]\times10^{0}$ & $[-2.9,\;3.0]\times10^{-1}$ & $[-9.2,\;9.4]\times10^{-2}$ \\
$f_{S,2}/\Lambda^4$ & $[-9.6,\;9.8]\times10^{-1}$ & $[-2.9,\;3.0]\times10^{-1}$ & $[-9.3,\;9.4]\times10^{-2}$ \\
\hline \hline
$f_{M,0}/\Lambda^4$ & $[-8.1,\;8.2]\times10^{-2}$ & $[-1.1,\;1.1]\times10^{-2}$ & $[-3.4,\;3.4]\times10^{-3}$ \\
$f_{M,1}/\Lambda^4$ & $[-8.8,\;8.9]\times10^{-2}$ & $[-2.3,\;2.4]\times10^{-2}$ & $[-7.3,\;7.4]\times10^{-3}$ \\
$f_{M,7}/\Lambda^4$ & $[-1.0,\;1.0]\times10^{-1}$ & $[-2.0,\;2.1]\times10^{-2}$ & $[-6.5,\;6.5]\times10^{-3}$ \\ \hline
$f_{M,2}/\Lambda^4$ & $[-8.5,\;8.6]\times10^{-2}$ & $[-1.0,\;1.0]\times10^{-2}$ & $[-3.1,\;3.2]\times10^{-3}$ \\
$f_{M,3}/\Lambda^4$ & $[-4.7,\;4.7]\times10^{-2}$ & $[-2.8,\;2.9]\times10^{-2}$ & $[-8.9,\;9.0]\times10^{-3}$ \\ \hline
$f_{M,4}/\Lambda^4$ & $[-1.0,\;1.0]\times10^{-1}$ & $[-1.7,\;1.7]\times10^{-2}$ & $[-5.4,\;5.4]\times10^{-3}$ \\
$f_{M,5}/\Lambda^4$ & $[-8.2,\;8.3]\times10^{-2}$ & $[-2.3,\;2.3]\times10^{-2}$ & $[-7.3,\;7.3]\times10^{-3}$ \\
\hline \hline
$f_{T,0}/\Lambda^4$ & $[-6.9,\;7.0]\times10^{-3}$ & $[-5.4,\;5.4]\times10^{-4}$ & $[-1.5,\;1.5]\times10^{-4}$ \\
$f_{T,1}/\Lambda^4$ & $[-7.6,\;7.7]\times10^{-3}$ & $[-4.3,\;4.3]\times10^{-4}$ & $[-9.4,\;9.4]\times10^{-5}$ \\
$f_{T,2}/\Lambda^4$ & $[-8.5,\;8.6]\times10^{-3}$ & $[-1.6,\;1.6]\times10^{-3}$ & $[-4.9,\;4.9]\times10^{-4}$ \\ \hline
$f_{T,5}/\Lambda^4$ & $[-9.4,\;9.4]\times10^{-3}$ & $[-6.2,\;6.2]\times10^{-4}$ & $[-1.9,\;1.9]\times10^{-4}$ \\
$f_{T,6}/\Lambda^4$ & $[-9.0,\;9.1]\times10^{-3}$ & $[-9.6,\;9.6]\times10^{-4}$ & $[-2.9,\;2.9]\times10^{-4}$ \\
$f_{T,7}/\Lambda^4$ & $[-1.0,\;1.0]\times10^{-2}$ & $[-1.9,\;2.0]\times10^{-3}$ & $[-6.2,\;6.2]\times10^{-4}$ \\ \hline
$f_{T,8}/\Lambda^4$ & $[-1.1,\;1.0]\times10^{-2}$ & $[-2.5,\;2.4]\times10^{-4}$ & $[-3.4,\;3.3]\times10^{-5}$ \\
$f_{T,9}/\Lambda^4$ & $[-1.1,\;1.1]\times10^{-2}$ & $[-5.1,\;5.0]\times10^{-4}$ & $[-1.3,\;1.3]\times10^{-4}$ \\
\hline \hline
\end{tabular}
\caption{Combined 95\% C.L. intervals for different future upgrades of $\mu$TRISTAN.}
\label{tab:future_MT}
\end{table*}

\subsection{Theoretical validity of the projected limits}
\label{sec:theory_interpretation_results}
\begin{figure*}[htb!]
    \centering   \includegraphics[width=0.8\linewidth]{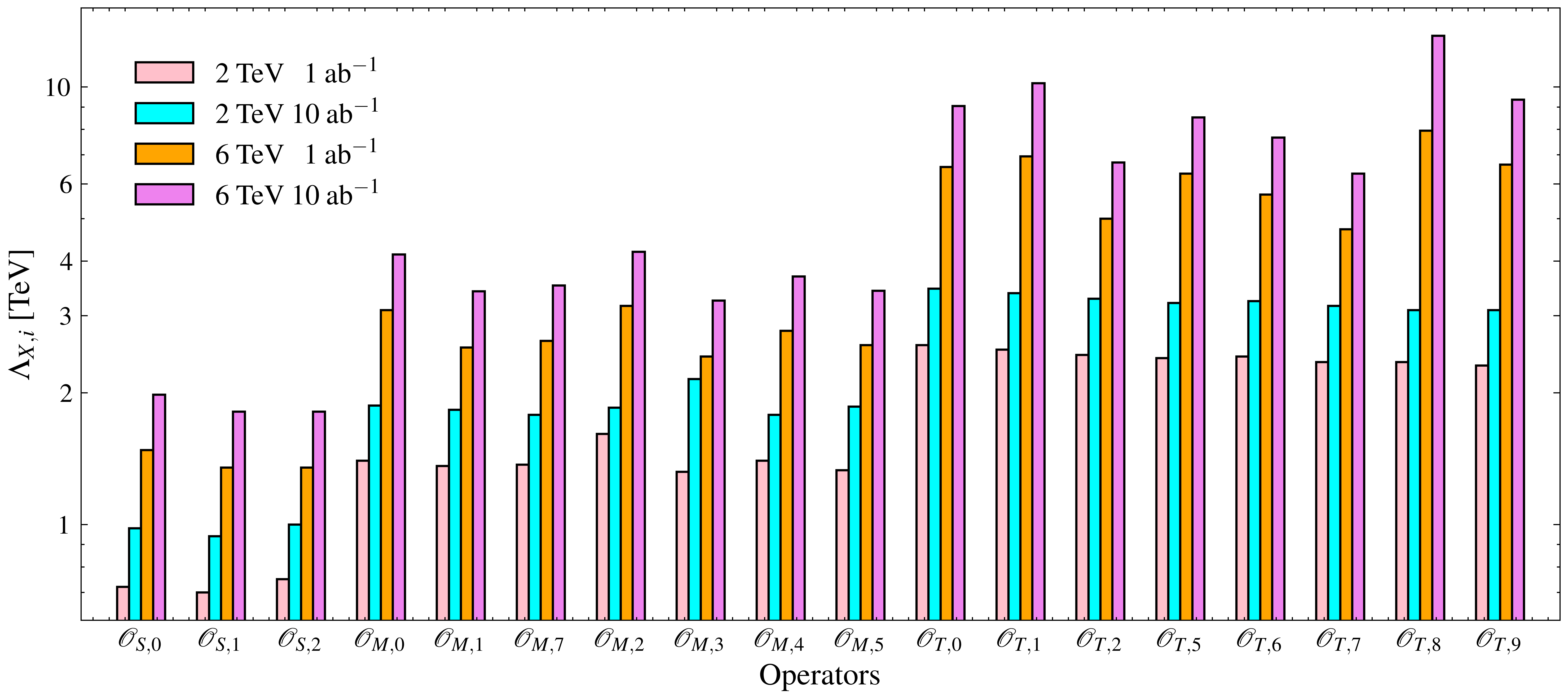}
    \caption{Limits on $\Lambda_{X,i}$ from different $\mu$TRISTAN configurations derived from limits on $f_{X,i}/\Lambda^{4}$, with $|f_{X,i}|=1$.}
    \label{fig:eftval}
\end{figure*}

In this section we combine the projected one-coefficient sensitivities obtained above with the EFT-validity criteria introduced in Section~\ref{sec:theory_constraints}. For each operator we assign $|c_{X,i}|$ to the larger absolute end-point of the combined $95\%$ C.L. sensitivity highlighted in Tables~\ref{tab:setA}, \ref{tab:setB}, and \ref{tab:setC} to estimate the EFT validity scale $\Lambda_{X,i}$ and the perturbative unitarity scale $Q^{U}_{X,i}$ according to Eqs.~\eqref{eq:LambdaEFT}, and \eqref{eq:unitarity_scale} where the constants $A_{X,i}$ are taken from the one-operator partial-wave unitarity bounds of Ref.~\cite{Almeida:2020ylr}. The resulting values for the $2~{\rm TeV}$, $1~{\rm ab}^{-1}$ configuration are shown in Table~\ref{tab:aqgc_theory_constraints}. A conservative sufficient condition for a fully perturbative interpretation over the entire kinematical regions is $Q^{\rm valid}_{X,i} \equiv \min\left(\Lambda_{X,i},Q^{U}_{X,i}\right) > \sqrt{s} = 2~\rm{TeV}$. The $\Lambda_{X,i}$ estimates for the other three $\mu$TRISTAN configurations, namely $\sqrt{s}=2~{\rm TeV}$ with $\mathfrak{L}_{\rm int}=10~{\rm ab}^{-1}$, $\sqrt{s}=6~{\rm TeV}$ with $\mathfrak{L}_{\rm int}=1~{\rm ab}^{-1}$, and $\sqrt{s}=6~{\rm TeV}$ with $\mathfrak{L}_{\rm int}=10~{\rm ab}^{-1}$, are presented in Fig~\ref{fig:eftval} for comparison. A clear hierarchy is observed among the three classes of operators. The \textit{Subclass A (Scalar)} operators correspond to the lowest sensitivity scales, remaining below approximately $2~{\rm TeV}$ even for the most optimistic collider configuration. The \textit{Subclass B (Mixed)} operators exhibit intermediate sensitivity scales, reaching about $4~{\rm TeV}$ at $\sqrt{s}=6~{\rm TeV}$ with an integrated luminosity of $10~{\rm ab}^{-1}$. In contrast, the \textit{Subclass C (Tensor)} operators consistently yield the highest sensitivity scales, with several operators attaining values above $10~{\rm TeV}$ for the same collider configuration. The sensitivity scales increase systematically with both the CM energy and the integrated luminosity. While increasing the integrated luminosity from $1~{\rm ab}^{-1}$ to $10~{\rm ab}^{-1}$ at fixed CM energy leads to a significant improvement, the enhancement due to increasing the CM energy from $2~{\rm TeV}$ to $6~{\rm TeV}$ is considerably more significant. This trend reflects the stronger energy dependence of the amplitudes induced by dimension-8 operators, making high-energy collisions particularly sensitive to aQGC structures. These sensitivity scales form the basis of the comparison with the corresponding theoretical validity bounds presented in Sec.~\ref{sec:sensitivity}.

Increasing the luminosity or collider energy improves the coefficient reach and therefore raises the inferred scale \(\Lambda_{X,i}\).  The improvement is most consequential for the tensor operators, whereas the scalar and mixed classes generally remain below the conservative
full-phase-space validity criterion. 

It is important to note that, a value of $\Lambda_{X,i}$ below the nominal collider energy reflects the limited statistical sensitivity of a particular collider configuration, rather than a breakdown of the EFT description. In such a case, the measurement simply does not provide sufficient precision to constrain the corresponding interaction at a higher characteristic scale. This behaviour is illustrated directly in Fig.~\ref{fig:eftval}. As the statistical precision improves with increasing integrated luminosity, the inferred
scales \(\Lambda_{X,i}\) move to higher values. Increasing the CM energy provides an even larger improvement because of the strong energy dependence of the dimension-8 contributions.

\begin{table*}[htb!]
\centering
\begin{tabular}{
>{\centering\arraybackslash}p{1.5cm}
>{\centering\arraybackslash}p{1.5cm}
>{\centering\arraybackslash}p{1.5cm}
>{\centering\arraybackslash}p{1.5cm}
>{\centering\arraybackslash}p{1.5cm}
>{\centering\arraybackslash}p{7.0cm}
}
\hline \hline
Class
& Operator
& $\Lambda_{X,i}~[{\rm TeV}]$
& $A_{X,i}$ \cite{Almeida:2020ylr}
& $Q^U_{X,i}~[{\rm TeV}]$
& Positivity constraint~\cite{Bi:2019phv}\\
\hline \hline

\multirow{3}*{$\rm A_0$}
& $\mathcal{O}_{S,0}$
& $0.73$
& $32\pi$
& $2.30$
& $c_{S,0}\geq 0$ \\

& $\mathcal{O}_{S,1}$
& $0.70$
& $96\pi/7$
& $1.80$
& $c_{S,1}\geq 0$ \\

& $\mathcal{O}_{S,2}$
& $0.75$
& $96\pi/5$
& $2.10$
& $c_{S,2}\geq 0$ \\
\hline \hline

\multirow{3}*{$\rm B_1$}
& $\mathcal{O}_{M,0}$
& $1.40$
& $32\pi/\sqrt{6}$
& $3.54$
& isolated $c_{M,0}\neq 0$ not allowed \\

& $\mathcal{O}_{M,1}$
& $1.36$
& $128\pi/\sqrt{6}$
& $4.88$
& $c_{M,1}\leq 0$ \\

& $\mathcal{O}_{M,7}$
& $1.37$
& $256\pi/\sqrt{6}$
& $5.85$
& $c_{M,7}\geq 0$ \\
\hline

\multirow{2}*{$\rm B_2$}
& $\mathcal{O}_{M,2}$
& $1.61$
& $16\pi/\sqrt{2}$
& $3.92$
& isolated $c_{M,2}\neq 0$ not allowed \\

& $\mathcal{O}_{M,3}$
& $1.32$
& $64\pi/\sqrt{2}$
& $4.56$
& $c_{M,3}\leq 0$ \\
\hline

\multirow{2}*{$\rm B_3$}
& $\mathcal{O}_{M,4}$
& $1.40$
& $32\pi$
& $4.43$
& isolated $c_{M,4}\neq 0$ not allowed \\

& $\mathcal{O}_{M,5}$
& $1.33$
& $64\pi$
& $5.01$
& isolated $c_{M,5}\neq 0$ not allowed \\
\hline \hline

\multirow{3}*{$\rm C_1$}
& $\mathcal{O}_{T,0}$
& $2.57$
& $12\pi/5$
& $4.26$
& $c_{T,0}\geq 0$ \\

& $\mathcal{O}_{T,1}$
& $2.51$
& $24\pi/5$
& $4.96$
& $c_{T,1}\geq 0$ \\

& $\mathcal{O}_{T,2}$
& $2.44$
& $96\pi/13$
& $5.37$
& $c_{T,2}\geq 0$ \\
\hline

\multirow{3}*{$\rm C_2$}
& $\mathcal{O}_{T,5}$
& $2.40$
& $8\pi/\sqrt{3}$
& $4.69$
& isolated $c_{T,5}\neq 0$ not allowed \\

& $\mathcal{O}_{T,6}$
& $2.42$
& $48\pi/7$
& $5.22$
& $c_{T,6}\geq 0$ \\

& $\mathcal{O}_{T,7}$
& $2.35$
& $32\pi/\sqrt{3}$
& $6.48$
& isolated $c_{T,7}\neq 0$ not allowed \\
\hline

\multirow{2}*{$\rm C_3$}
& $\mathcal{O}_{T,8}$
& $2.35$
& $3\pi/2$
& $3.46$
& $c_{T,8}\geq 0$ \\

& $\mathcal{O}_{T,9}$
& $2.31$
& $24\pi/7$
& $4.19$
& $c_{T,9}\geq 0$ \\
\hline \hline
\end{tabular}
\caption{EFT cutoff and perturbative-unitarity scales corresponding to the combined
95\% C.L. limits at $\sqrt{s}=2~{\rm TeV}$ and $\mathfrak{L}_{\rm int}=1~{\rm ab}^{-1}$ assuming $|f_{X,i}|=1$. The last column
lists the positivity conditions for the one-coefficient scenarios.}
\label{tab:aqgc_theory_constraints}
\end{table*}

In addition to the EFT validity and unitarity requirements, positivity bounds provide an independent theoretical consistency condition. The one-coefficient positivity constraints for the operators considered in this work are summarized in the last column of Table~\ref{tab:aqgc_theory_constraints}, following Ref.~\cite{Bi:2019phv}. For several operators, positivity requires a definite sign of the Wilson coefficient, while for others an isolated non-zero coefficient is not compatible with the positivity conditions, implying that additional operators must be simultaneously present. These bounds are complementary to the EFT validity criteria discussed above and should be imposed when interpreting the projected sensitivities.

\section{Summary}
\label{sec:summary}
In this work, we have performed a comprehensive investigation of the potential of the proposed $\mu$TRISTAN collider to probe aQGCs using the $\mu^+\mu^+$ collision mode. Using the SMEFT framework, we systematically classified dimension-8 operators across scalar (Class A), mixed (Class B), and tensor (Class C) subclasses and analyzed their impact on on VBS processes. Our detailed collider analysis, spanning multiple signal regions ($2V2\nu, V\gamma \ell\nu, 2V\ell\nu, 2\gamma2\ell,$ and $2V2\ell$), highlights the following conclusions:
\begin{itemize}
    \item \textbf{Sensitivity at 2 TeV:} At the initial $\sqrt{s}=2$ TeV stage with $1\text{ ab}^{-1}$, $\mu$TRISTAN achieves sensitivities to scalar operators (\textit{Subclass $A$}) at the $\mathcal{O}(1)\text{ TeV}^{-4}$ level, as detailed in Table~\ref{tab:setA} and Figure~\ref{fig:s1}. Similarly, for \textit{Subclass $B$} and \textit{Subclass $C$}, the projected limits are shown in Tables~\ref{tab:setB} and~\ref{tab:setC}, and Figures~\ref{fig:s2} and~\ref{fig:s3}, respectively. These results are already competitive with the most recent constraints from the 13 TeV LHC run shown in Table~\ref{tab:QGC_LHC_combined}.
    \item \textbf{Impact of Energy Upgrades:} As expected, increasing the CM energy provides a significantly more pronounced sensitivity gain than increasing integrated luminosity alone. Transitioning from 2 TeV to 6 TeV results in an order of magnitude or more improvement for most operators due to the intrinsic energy dependence of dimension-8 contributions while increasing luminosity from $1\text{ ab}^{-1}$ to $10\text{ ab}^{-1}$ at 2 TeV provides a statistical improvement of roughly a factor of 2–3 only.
    \item \textbf{Operator Class Performance:} As summarized in Table~\ref{tab:future_MT}, the projected bounds reach unprecedented precision at 6 TeV with $10\text{ ab}^{-1}$:
    \begin{enumerate}
        \item \textit{Subclass A (Scalar)}: These operators improve existing constraints to the $\mathcal{O}(10^{-2})$ $\text{ TeV}^{-4}$ level for the 6 TeV configuration.
        \item \textit{Subclass B (Mixed)}: Projections for these operators, which benefit from photon-rich final states, reach the $\mathcal{O}(10^{-3})\text{ TeV}^{-4}$ regime.
        \item \textit{Subclass C (Tensor)}: These operators exhibit the strongest energy growth, reaching sensitivities of $\mathcal{O}(10^{-4})-\mathcal{O}(10^{-5})\text{ TeV}^{-4}$. For example, the limit for $f_{T,0}/\Lambda^4$ improves from the current of the LHC's $[-9.2, 7.8] \times 10^{-2} \text{ TeV}^{-4}$ to a projected $[-1.5, 1.5] \times 10^{-4} \text{ TeV}^{-4}$.
    \end{enumerate}
    \item \textbf{Multichannel Gains:} The combination of all signal regions significantly enhances the reach of the SMEFT parameter space, especially for operators inducing neutral quartic interactions that are absent in the SM.
    \item \textbf{Theoretical Validity:} Figure~\ref{fig:eftval} presents the characteristic scales $\Lambda_{X,i}$ corresponding to the projected sensitivities for the four proposed $\mu$TRISTAN configurations, obtained using $|f_{X,i}|=1$. At the initial \(2~{\rm TeV},1~{\rm ab}^{-1}\) stage, the scalar operators have \(\Lambda_{X,i}<1~{\rm TeV}\), and therefore their limits should be regarded as a limitation of the measurement's
    constraining power rather than as evidence for an inconsistency of the EFT framework. The mixed operators are more informative but remain close to the boundary of a conservative EFT interpretation at \(2~{\rm TeV}\). The tensor operators are the most robust already at the initial stage, with inferred scales above \(2~{\rm TeV}\), especially for the neutral-QGC-sensitive subclass \(C_3\).

The upgraded configurations improve this interpretation substantially.
Increasing the luminosity to \(10~{\rm ab}^{-1}\) at \(2~{\rm TeV}\) mainly
strengthens the tensor limits and makes the mixed sector less marginal, while
the \(6~{\rm TeV}\) runs provide a much larger gain because of the energy
growth of the dimension-8 amplitudes. In particular, at
\(6~{\rm TeV},10~{\rm ab}^{-1}\), the inferred scales reach roughly
\(2~{\rm TeV}\) for scalar operators, \(4~{\rm TeV}\) for mixed operators,
and exceed \(6~{\rm TeV}\) for tensor operators, with several tensor
directions extending above \(10~{\rm TeV}\). Thus, the higher-energy and
high-luminosity runs not only improve the projected numerical sensitivities,
but also enlarge the region where the EFT interpretation is perturbatively
reliable. 

In addition to the EFT cutoff and perturbative unitarity requirements, the one-coefficient limits must also be interpreted in light of positivity
bounds. As summarized in Table~\ref{tab:aqgc_theory_constraints}, positivity imposes sign restrictions on several single-operator directions, while some isolated mixed and tensor
operators require correlated nonzero coefficients in a global
UV-consistent EFT interpretation.
\end{itemize}
In general, our results establish the $\mu^+\mu^+$ collider as a uniquely powerful probe of electroweak symmetry breaking dynamics. The ability to access high CM energies at a leptonic platform allows for precision measurements of aQGCs that surpass the current and projected capabilities of hadron colliders.

\acknowledgments
LM would like to acknowledge ANRF,
Government of India, for financial support through the Project ANRF/CRG/2025/000133.

\appendix
\section{Dimension-6 SMEFT effects on aQGCs}
\label{appA}
As discussed in Sec.~\ref{sec:aqgcmodel}, aQGCs, particularly cQGCs, can also arise from dimension-6 SMEFT operators. Pure gauge field-strength operators contributing to cQGCs are given by
\begin{equation}
\begin{split}
    \frac{f_{W3}}{\Lambda^2}\,\mathcal{O}_{W3},& \quad \mathcal{O}_{W3}: \epsilon^{IJK} W^{I}_{\mu\nu} W^{J\mu}_{\rho} W^{K\rho\nu}\,, \\
    \frac{f_{\widetilde{W}3}}{\Lambda^2}\,\mathcal{O}_{\widetilde{W}3},& \quad \mathcal{O}_{\widetilde{W}3}: \epsilon^{IJK} \widetilde{W}^{I}_{\mu\nu} W^{J\mu}_{\rho} W^{K\rho\nu}\,,
\end{split}
\end{equation}
where $\widetilde{W}^{I}_{\mu\nu}$ denotes the dual field-strength tensor, and $\mathcal{O}_{\widetilde{W}3}$ is CP-odd. In addition, Higgs-gauge operators can also induce cQGCs,
\begin{equation}
\begin{split}
    \frac{f_{HW}}{\Lambda^2}\,\mathcal{O}_{HW},& \quad \mathcal{O}_{HW}: (H^\dagger H)\,W^{I}_{\mu\nu}W^{I\mu\nu}\,, \\
    \frac{f_{H\widetilde{W}}}{\Lambda^2}\,\mathcal{O}_{H\widetilde{W}},& \quad \mathcal{O}_{H\widetilde{W}}: (H^\dagger H)\,\widetilde{W}^{I}_{\mu\nu}W^{I\mu\nu}\,,
\end{split}
\end{equation}
where $\mathcal{O}_{H\widetilde{W}}$ is likewise CP-odd. Unlike the genuine dimension-8 operators considered in this work, these dimension-6 operators simultaneously modify both TGCs and QGCs. Consequently, for most collider processes, including the VBS channels studied here, the sensitivity is primarily driven by the TGC contributions, while the genuine QGC effects remain subdominant.

\begin{figure*}[htb!]
    \centering
    \includegraphics[width=0.3\linewidth]{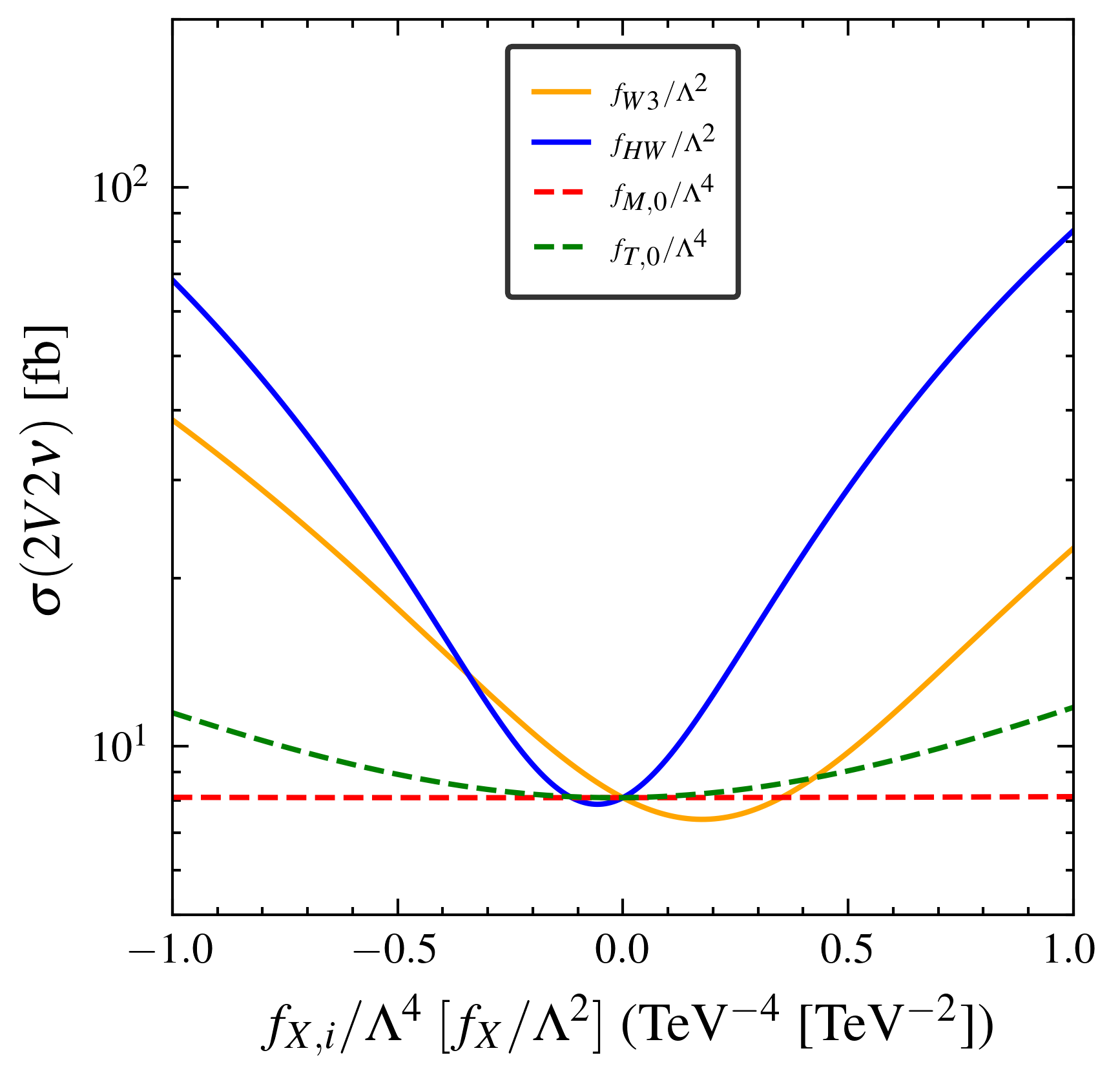}
    \includegraphics[width=0.3\linewidth]{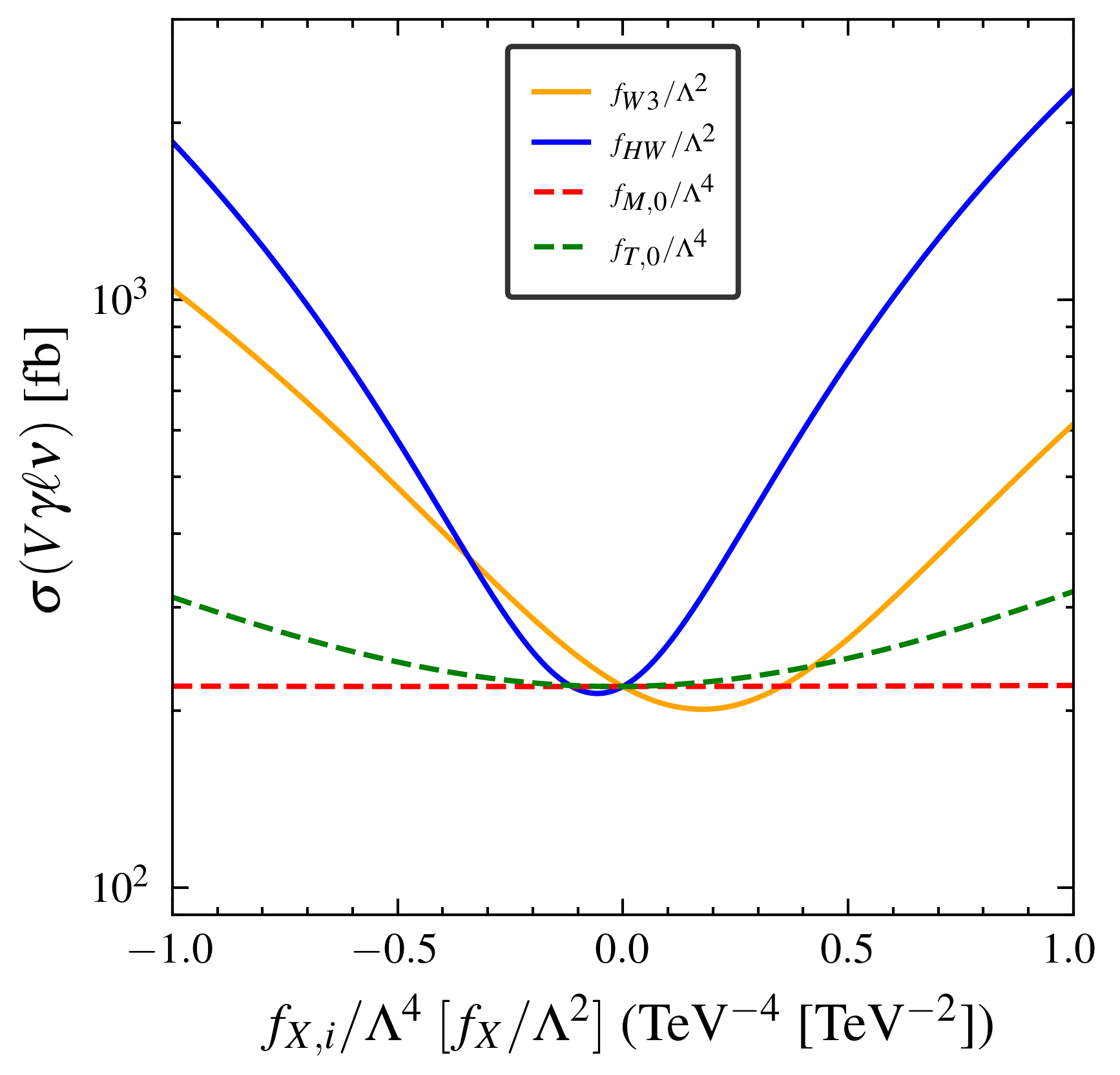}
    \includegraphics[width=0.3\linewidth]{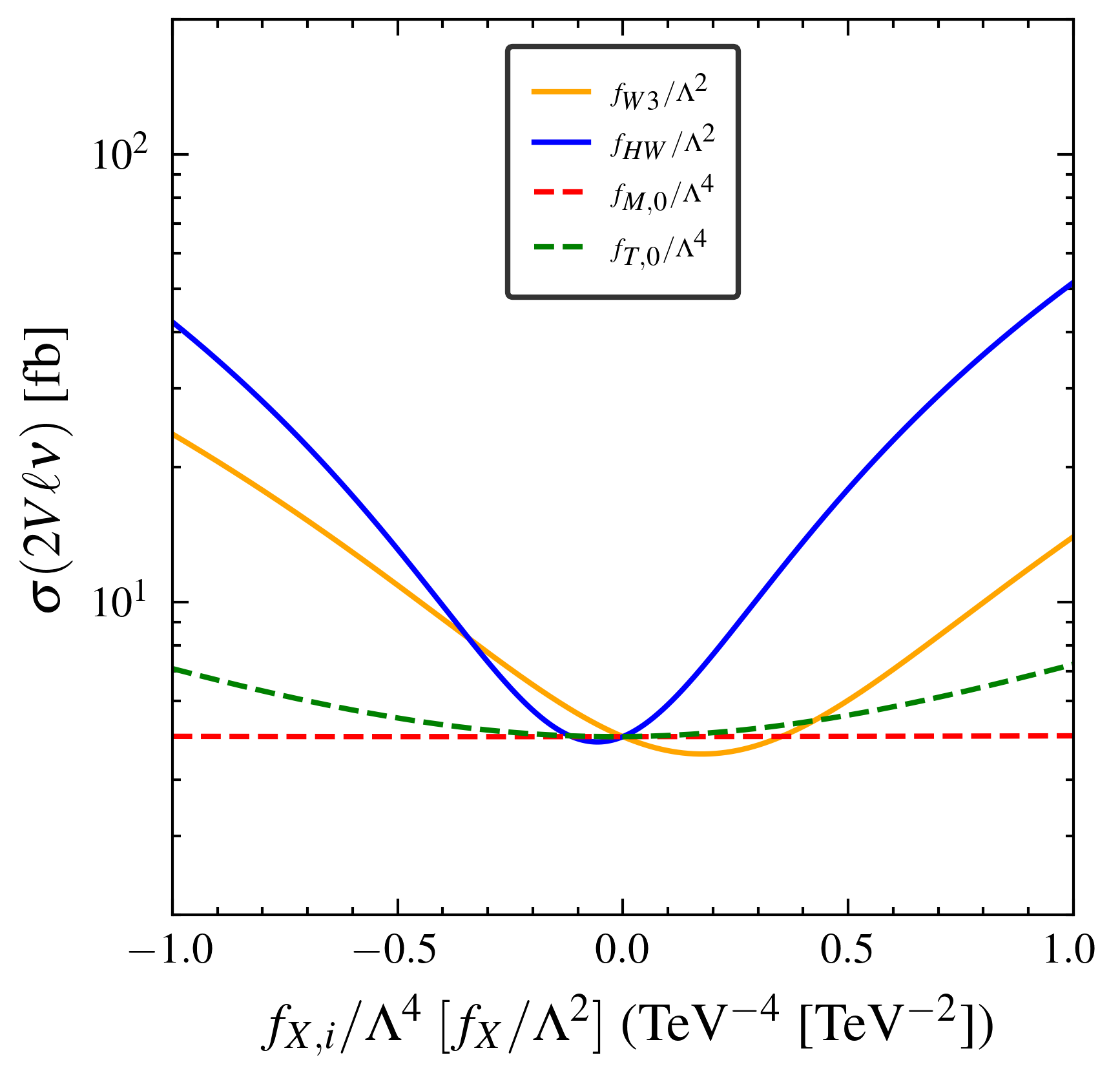} \\
    \includegraphics[width=0.3\linewidth]{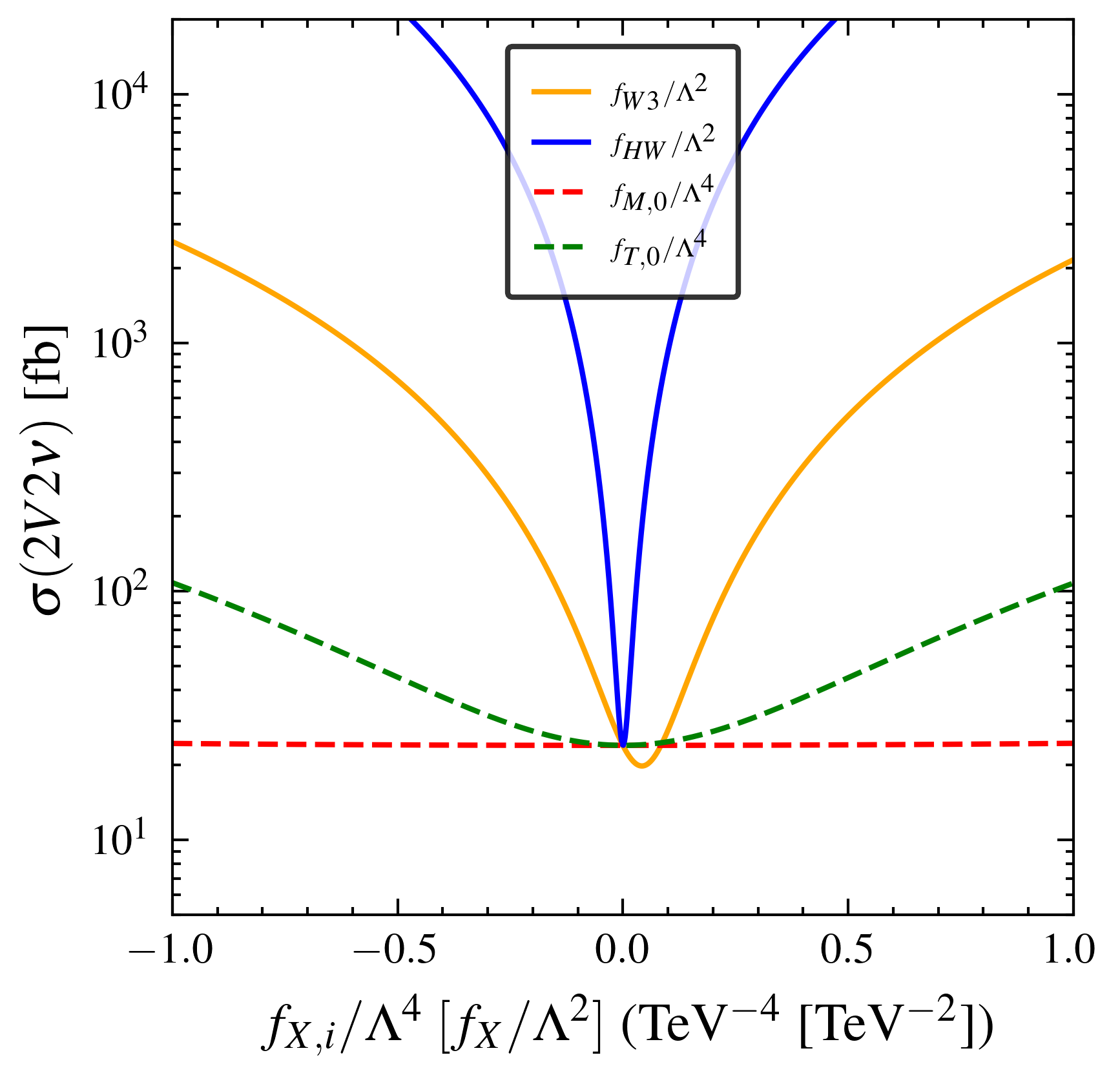}
    \includegraphics[width=0.3\linewidth]{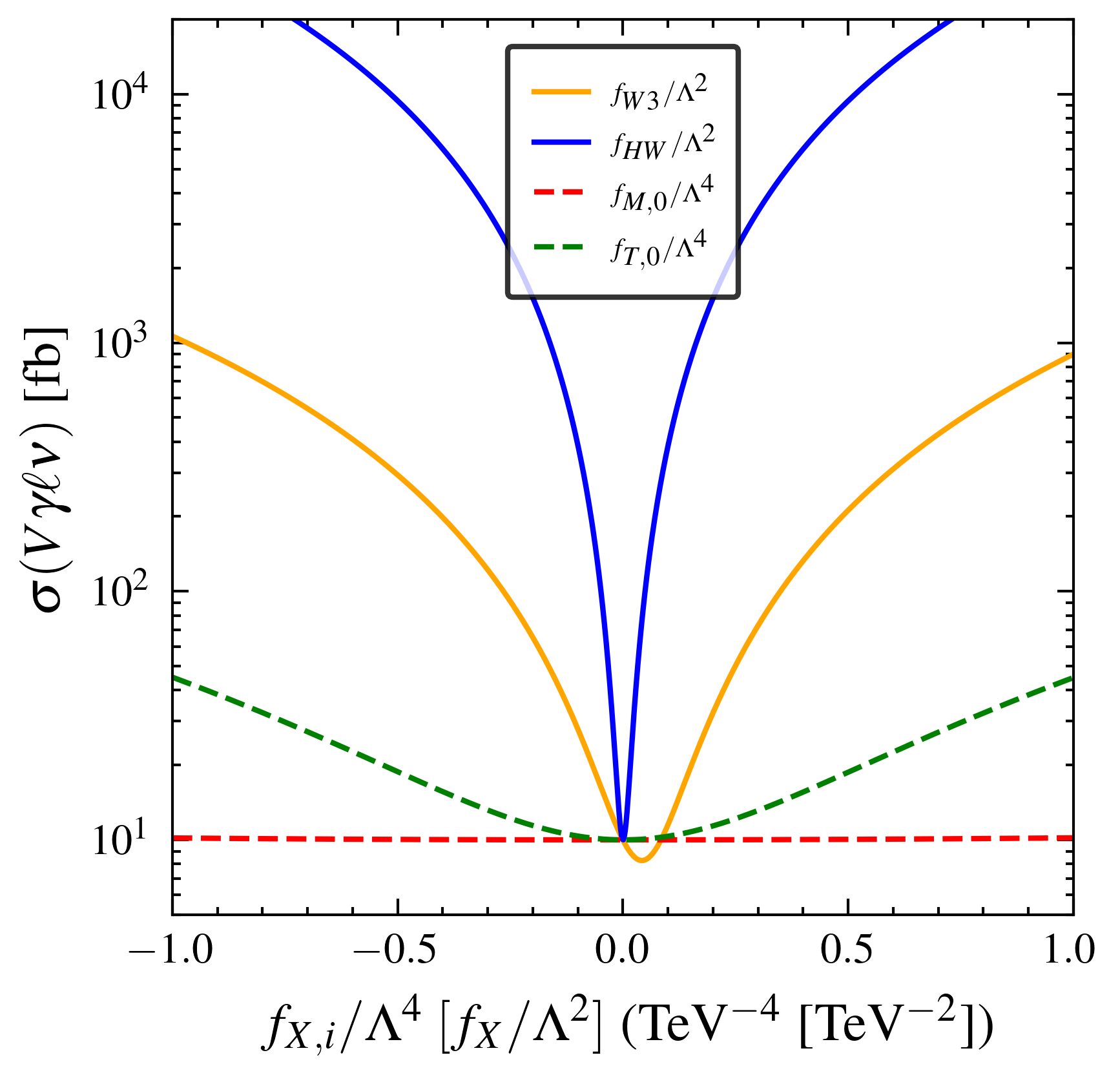}
    \includegraphics[width=0.3\linewidth]{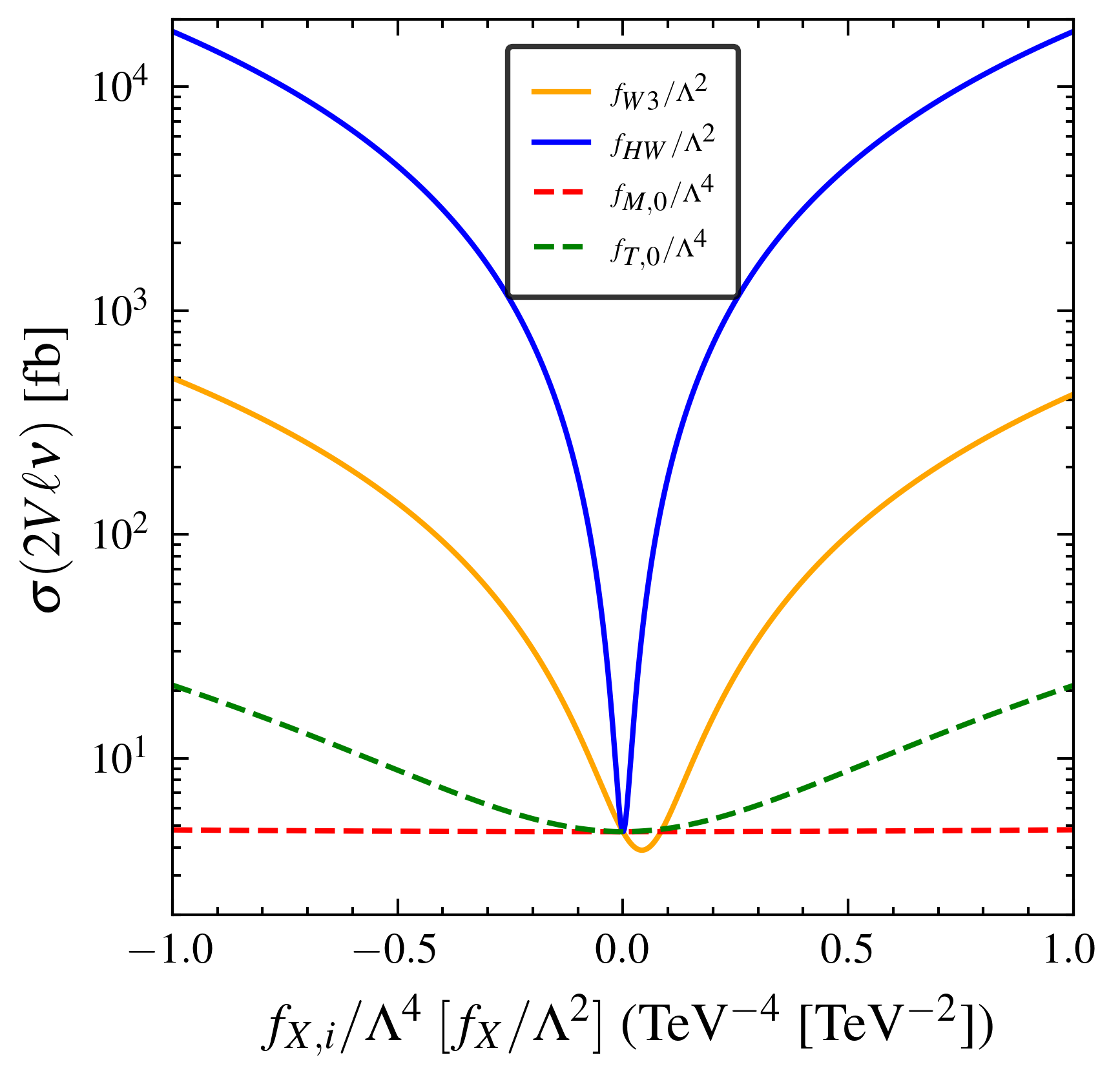}
    \caption{Cross section of different signal regions sensitive to cQGCs ($2V2\nu$, $V\gamma\ell\nu$ and $2V\ell\nu$) corresponding to variation of dimension-6 coefficients ($f_{W3}/\Lambda^2$ and $f_{HW}/\Lambda^2$) and dimension-8 coefficients ($f_{M,0}/\Lambda^4$ and $f_{T,0}/\Lambda^4$), respectively, for $\sqrt{s}=2$ TeV (\textit{top} panel) and $\sqrt{s}=6$ TeV (\textit{bottom} panel) $\mu$TRISTAN configurations.}
    \label{fig:d6eft}
\end{figure*}
To illustrate this point, Fig.~\ref{fig:d6eft} compares the cross sections of the cQGC-sensitive signal regions ($2V2\nu$, $V\gamma\ell\nu$, and $2V\ell\nu$) as functions of representative dimension-6 coefficients ($f_{W3}/\Lambda^2$ and $f_{HW}/\Lambda^2$) and genuine dimension-8 coefficients ($f_{M,0}/\Lambda^4$ and $f_{T,0}/\Lambda^4$) for the $\sqrt{s}=2~{\rm TeV}$ (top panel) and $\sqrt{s}=6~{\rm TeV}$ (bottom panel) $\mu$TRISTAN configurations. As expected, the dimension-6 operators induce significantly larger variations in the cross sections than the dimension-8 operators, owing to both the lower EFT suppression and the simultaneous presence of TGC and QGC contributions. Furthermore, the effects become more pronounced with increasing CM energy, reflecting the energy-growing behaviour of the corresponding amplitudes.

The dimension-6 operators are already subject to constraints from LHC measurements, with typical limits of $f_{W3}/\Lambda^2\sim\mathcal{O}(10^{-1})~{\rm TeV}^{-2}$ and $f_{HW}/\Lambda^2\sim\mathcal{O}(10^{-2})~{\rm TeV}^{-2}$~\cite{Bartocci:2023nvp,Celada:2024mcf}. However, such effects cannot, in general, be neglected. Nevertheless, UV scenarios with suitable symmetry structures can suppress or forbid the dimension-6 operators responsible for TGCs while allowing genuine dimension-8 operators to provide the leading NP effects. In such scenarios, the study of pure dimension-8 contributions to QGCs is well motivated, justifying the framework adopted in this work and in numerous phenomenological and experimental studies discussed in Sec.~\ref{sec:aqgcmodel}.

\bibliography{biblio.bib}

@article{Hamada:2022uyn,
    author = "Hamada, Yu and Kitano, Ryuichiro and Matsudo, Ryutaro and Takaura, Hiromasa",
    title = "{Precision {\ensuremath{\mu}}+{\ensuremath{\mu}}+ and {\ensuremath{\mu}}+e{\ensuremath{-}} elastic scatterings}",
    eprint = "2210.11083",
    archivePrefix = "arXiv",
    primaryClass = "hep-ph",
    reportNumber = "KEK-TH-2462",
    doi = "10.1093/ptep/ptac174",
    journal = "PTEP",
    volume = "2023",
    number = "1",
    pages = "013B07",
    year = "2023"
}

@article{Bossi:2020yne,
    author = "Bossi, Fabio and Ciafaloni, Paolo",
    title = "{Lepton Flavor Violation at muon-electron colliders}",
    eprint = "2003.03997",
    archivePrefix = "arXiv",
    primaryClass = "hep-ph",
    doi = "10.1007/JHEP10(2020)033",
    journal = "JHEP",
    volume = "10",
    pages = "033",
    year = "2020"
}

@article{Chen:2024tqh,
    author = "Chen, Lisong and Iguro, Syuhei and Hamada, Yu",
    title = "{Determining Weak-Mixing Angle at $\mu$TRISTAN}",
    eprint = "2406.04500",
    archivePrefix = "arXiv",
    journal = "arXiv preprint",
    primaryClass = "hep-ph",
    month = "6",
    year = "2024"
}

@article{Das:2022mmh,
    author = "Das, Arindam and Nomura, Takaaki and Shimomura, Takashi",
    title = "{Multi muon/anti-muon signals via productions of gauge and scalar bosons in a $U(1)_{L_\mu - L_\tau }$ model at muonic colliders}",
    eprint = "2212.11674",
    archivePrefix = "arXiv",
    primaryClass = "hep-ph",
    reportNumber = "UME-PP-024, KYUSHU-HET-252",
    doi = "10.1140/epjc/s10052-023-11955-4",
    journal = "Eur. Phys. J. C",
    volume = "83",
    number = "9",
    pages = "786",
    year = "2023"
}

@article{Dev:2023nha,
    author = "Dev, P. S. Bhupal and Heeck, Julian and Thapa, Anil",
    title = "{Neutrino mass models at $\mu $TRISTAN}",
    eprint = "2309.06463",
    archivePrefix = "arXiv",
    primaryClass = "hep-ph",
    reportNumber = "CETUP-2023-005",
    doi = "10.1140/epjc/s10052-024-12496-0",
    journal = "Eur. Phys. J. C",
    volume = "84",
    number = "2",
    pages = "148",
    year = "2024"
}

@article{Fridell:2023gjx,
    author = "Fridell, K{\r{a}}re and Kitano, Ryuichiro and Takai, Ryoto",
    title = "{Lepton flavor physics at {\ensuremath{\mu}}$^{+}${\ensuremath{\mu}}$^{+}$ colliders}",
    eprint = "2304.14020",
    archivePrefix = "arXiv",
    primaryClass = "hep-ph",
    reportNumber = "KEK-TH-2519",
    doi = "10.1007/JHEP06(2023)086",
    journal = "JHEP",
    volume = "06",
    pages = "086",
    year = "2023"
}

@article{Fukuda:2023yui,
    author = "Fukuda, Hajime and Moroi, Takeo and Niki, Atsuya and Wei, Shang-Fu",
    title = "{Search for WIMPs at future {\ensuremath{\mu}}$^{+}${\ensuremath{\mu}}$^{+}$ colliders}",
    eprint = "2310.07162",
    archivePrefix = "arXiv",
    primaryClass = "hep-ph",
    doi = "10.1007/JHEP02(2024)214",
    journal = "JHEP",
    volume = "02",
    pages = "214",
    year = "2024"
}

@article{Lichtenstein:2023iut,
    author = "Lichtenstein, Gabriela and Schmidt, Michael A. and Valencia, German and Volkas, Raymond R.",
    title = "{Complementarity of $\mu$TRISTAN and Belle II in searches for charged-lepton flavour violation}",
    eprint = "2307.11369",
    archivePrefix = "arXiv",
    primaryClass = "hep-ph",
    doi = "10.1016/j.physletb.2023.138144",
    journal = "Phys. Lett. B",
    volume = "845",
    pages = "138144",
    year = "2023"
}

@article{Das:2024gfg,
    author = "Das, Arindam and Orikasa, Yuta",
    title = "{Z' induced forward dominant processes in $\mu$TRISTAN experiment}",
    eprint = "2401.00696",
    archivePrefix = "arXiv",
    primaryClass = "hep-ph",
    doi = "10.1016/j.physletb.2024.138577",
    journal = "Phys. Lett. B",
    volume = "851",
    pages = "138577",
    year = "2024"
}

@article{Ding:2024zaj,
    author = "Ding, Ran and Li, Jingshu and Lu, Meng and You, Zhengyun and Wang, Zijian and Li, Qiang",
    title = "{Study of charged Lepton Flavor Violation in electron muon interactions}",
    eprint = "2405.09417",
    archivePrefix = "arXiv",
    primaryClass = "hep-ex",
    doi = "10.1007/JHEP01(2025)165",
    journal = "JHEP",
    volume = "01",
    pages = "165",
    year = "2025"
}

@article{Calibbi:2024rcm,
    author = "Calibbi, Lorenzo and Li, Tong and Mukherjee, Lopamudra and Yang, Yiming",
    title = "{Probing ALP lepton flavor violation at {\ensuremath{\mu}}TRISTAN}",
    eprint = "2406.13234",
    archivePrefix = "arXiv",
    primaryClass = "hep-ph",
    doi = "10.1103/PhysRevD.110.115009",
    journal = "Phys. Rev. D",
    volume = "110",
    number = "11",
    pages = "115009",
    year = "2024"
}

@article{Das:2024kyk,
    author = "Das, Arindam and Li, Jinmian and Mandal, Sanjoy and Nomura, Takaaki and Zhang, Rao",
    title = "{Testing tree level TeV scale seesaw scenarios in {\ensuremath{\mu}}TRISTAN}",
    eprint = "2410.21956",
    archivePrefix = "arXiv",
    primaryClass = "hep-ph",
    doi = "10.1103/df3g-32t9",
    journal = "Phys. Rev. D",
    volume = "112",
    number = "3",
    pages = "035008",
    year = "2025"
}

@article{Sarkar:2025bgo,
    author = "Sarkar, Abhik",
    title = "{Lepton flavor violating top quark FCNC processes at the {\ensuremath{\mu}}TRISTAN}",
    eprint = "2506.18015",
    archivePrefix = "arXiv",
    primaryClass = "hep-ph",
    doi = "10.1103/jnnt-6t8f",
    journal = "Phys. Rev. D",
    volume = "113",
    number = "9",
    pages = "095010",
    year = "2026"
}

@article{Barducci:2025kuq,
    author = "Barducci, Daniele",
    title = "{Boosting long lived particles searches at {\ensuremath{\mu}}TRISTAN}",
    eprint = "2511.10131",
    archivePrefix = "arXiv",
    primaryClass = "hep-ph",
    doi = "10.1007/JHEP04(2026)084",
    journal = "JHEP",
    volume = "04",
    pages = "084",
    year = "2026"
}

@article{Eboli:2016kko,
    author = "{\'E}boli, O. J. P. and Gonzalez-Garcia, M. C.",
    title = "{Classifying the bosonic quartic couplings}",
    eprint = "1604.03555",
    archivePrefix = "arXiv",
    primaryClass = "hep-ph",
    reportNumber = "YITP-SB-16-09",
    doi = "10.1103/PhysRevD.93.093013",
    journal = "Phys. Rev. D",
    volume = "93",
    number = "9",
    pages = "093013",
    year = "2016"
}

@article{Degrande:2013kka,
    author = "Degrande, Celine",
    title = "{A basis of dimension-eight operators for anomalous neutral triple gauge boson interactions}",
    eprint = "1308.6323",
    archivePrefix = "arXiv",
    primaryClass = "hep-ph",
    doi = "10.1007/JHEP02(2014)101",
    journal = "JHEP",
    volume = "02",
    pages = "101",
    year = "2014"
}

@article{Abbott:2022jqq,
    author = "Abbott, Brad and others",
    title = "{Anomalous production of massive gauge boson pairs at muon colliders}",
    eprint = "2203.08135",
    archivePrefix = "arXiv",
    primaryClass = "hep-ex",
    reportNumber = "FERMILAB-CONF-22-174-PPD",
    doi = "10.1103/PhysRevD.108.093009",
    journal = "Phys. Rev. D",
    volume = "108",
    number = "9",
    pages = "093009",
    year = "2023"
}

@article{Yang:2022fhw,
    author = "Yang, Ji-Chong and Han, Xue-Ying and Qin, Zhi-Bin and Li, Tong and Guo, Yu-Chen",
    title = "{Measuring the anomalous quartic gauge couplings in the W$^{+}$W$^{-}${\textrightarrow} W$^{+}$W$^{-}$ process at muon collider using artificial neural networks}",
    eprint = "2204.10034",
    archivePrefix = "arXiv",
    primaryClass = "hep-ph",
    doi = "10.1007/JHEP09(2022)074",
    journal = "JHEP",
    volume = "09",
    pages = "074",
    year = "2022"
}

@article{Almeida:2020ylr,
    author = "Almeida, Eduardo da Silva and {\'E}boli, O. J. P. and Gonzalez{\textendash}Garcia, M. C.",
    title = "{Unitarity constraints on anomalous quartic couplings}",
    eprint = "2004.05174",
    archivePrefix = "arXiv",
    primaryClass = "hep-ph",
    reportNumber = "YITP-SB-2020-8",
    doi = "10.1103/PhysRevD.101.113003",
    journal = "Phys. Rev. D",
    volume = "101",
    number = "11",
    pages = "113003",
    year = "2020"
}

@article{ATLAS:2014jzl,
    author = "Aad, Georges and others",
    collaboration = "ATLAS",
    title = "{Evidence for Electroweak Production of $W^{\pm}W^{\pm}jj$ in $pp$ Collisions at $\sqrt{s}=8$ TeV with the ATLAS Detector}",
    eprint = "1405.6241",
    archivePrefix = "arXiv",
    primaryClass = "hep-ex",
    reportNumber = "CERN-PH-EP-2014-079",
    doi = "10.1103/PhysRevLett.113.141803",
    journal = "Phys. Rev. Lett.",
    volume = "113",
    number = "14",
    pages = "141803",
    year = "2014"
}

@article{CMS:2014mra,
    author = "Khachatryan, Vardan and others",
    collaboration = "CMS",
    title = "{Study of vector boson scattering and search for new physics in events with two same-sign leptons and two jets}",
    eprint = "1410.6315",
    archivePrefix = "arXiv",
    primaryClass = "hep-ex",
    reportNumber = "CMS-SMP-13-015, CERN-PH-EP-2014-250",
    doi = "10.1103/PhysRevLett.114.051801",
    journal = "Phys. Rev. Lett.",
    volume = "114",
    number = "5",
    pages = "051801",
    year = "2015"
}

@article{CMS:2017fhs,
    author = "Sirunyan, Albert M and others",
    collaboration = "CMS",
    title = "{Observation of electroweak production of same-sign W boson pairs in the two jet and two same-sign lepton final state in proton-proton collisions at $\sqrt{s} = $ 13 TeV}",
    eprint = "1709.05822",
    archivePrefix = "arXiv",
    primaryClass = "hep-ex",
    reportNumber = "CMS-SMP-17-004, CERN-EP-2017-232",
    doi = "10.1103/PhysRevLett.120.081801",
    journal = "Phys. Rev. Lett.",
    volume = "120",
    number = "8",
    pages = "081801",
    year = "2018"
}

@article{ATLAS:2016bkj,
    author = "Aad, Georges and others",
    collaboration = "ATLAS",
    title = "{Measurements of $W^\pm Z$ production cross sections in $pp$ collisions at $\sqrt{s} = 8$ TeV with the ATLAS detector and limits on anomalous gauge boson self-couplings}",
    eprint = "1603.02151",
    archivePrefix = "arXiv",
    primaryClass = "hep-ex",
    reportNumber = "CERN-EP-2016-017",
    doi = "10.1103/PhysRevD.93.092004",
    journal = "Phys. Rev. D",
    volume = "93",
    number = "9",
    pages = "092004",
    year = "2016"
}

@article{ATLAS:2016snd,
    author = "Aaboud, Morad and others",
    collaboration = "ATLAS",
    title = "{Measurement of $W^{\pm}W^{\pm}$ vector-boson scattering and limits on anomalous quartic gauge couplings with the ATLAS detector}",
    eprint = "1611.02428",
    archivePrefix = "arXiv",
    primaryClass = "hep-ex",
    reportNumber = "CERN-EP-2016-167",
    doi = "10.1103/PhysRevD.96.012007",
    journal = "Phys. Rev. D",
    volume = "96",
    number = "1",
    pages = "012007",
    year = "2017"
}

@article{CMS:2017rin,
    author = "Khachatryan, Vardan and others",
    collaboration = "CMS",
    title = "{Measurement of the cross section for electroweak production of Z$\gamma$ in association with two jets and constraints on anomalous quartic gauge couplings in proton{\textendash}proton collisions at $\sqrt{s} = 8$ TeV}",
    eprint = "1702.03025",
    archivePrefix = "arXiv",
    primaryClass = "hep-ex",
    reportNumber = "CMS-SMP-14-018, CERN-EP-2016-308",
    doi = "10.1016/j.physletb.2017.04.071",
    journal = "Phys. Lett. B",
    volume = "770",
    pages = "380--402",
    year = "2017"
}

@article{CMS:2017zmo,
    author = "Sirunyan, Albert M and others",
    collaboration = "CMS",
    title = "{Measurement of vector boson scattering and constraints on anomalous quartic couplings from events with four leptons and two jets in proton{\textendash}proton collisions at $\sqrt{s}=$ 13 TeV}",
    eprint = "1708.02812",
    archivePrefix = "arXiv",
    primaryClass = "hep-ex",
    reportNumber = "CMS-SMP-17-006, CERN-EP-2017-177",
    doi = "10.1016/j.physletb.2017.10.020",
    journal = "Phys. Lett. B",
    volume = "774",
    pages = "682--705",
    year = "2017"
}

@article{CMS:2019uys,
    author = "Sirunyan, Albert M and others",
    collaboration = "CMS",
    title = "{Measurement of electroweak WZ boson production and search for new physics in WZ + two jets events in pp collisions at $\sqrt{s} =$ 13TeV}",
    eprint = "1901.04060",
    archivePrefix = "arXiv",
    primaryClass = "hep-ex",
    reportNumber = "CMS-SMP-18-001, CERN-EP-2018-333",
    doi = "10.1016/j.physletb.2019.05.042",
    journal = "Phys. Lett. B",
    volume = "795",
    pages = "281--307",
    year = "2019"
}

@article{CMS:2016gct,
    author = "Khachatryan, Vardan and others",
    collaboration = "CMS",
    title = "{Measurement of electroweak-induced production of W$\gamma$ with two jets in pp collisions at $ \sqrt{s}=8 $ TeV and constraints on anomalous quartic gauge couplings}",
    eprint = "1612.09256",
    archivePrefix = "arXiv",
    primaryClass = "hep-ex",
    reportNumber = "CMS-SMP-14-011, CERN-EP-2016-289",
    doi = "10.1007/JHEP06(2017)106",
    journal = "JHEP",
    volume = "06",
    pages = "106",
    year = "2017"
}

@article{ATLAS:2017vqm,
    author = "Aaboud, Morad and others",
    collaboration = "ATLAS",
    title = "{Studies of $Z\gamma$ production in association with a high-mass dijet system in $pp$ collisions at $\sqrt{s}=$ 8 TeV with the ATLAS detector}",
    eprint = "1705.01966",
    archivePrefix = "arXiv",
    primaryClass = "hep-ex",
    reportNumber = "CERN-EP-2017-046",
    doi = "10.1007/JHEP07(2017)107",
    journal = "JHEP",
    volume = "07",
    pages = "107",
    year = "2017"
}

@article{CMS:2019qfk,
    author = "Sirunyan, Albert M and others",
    collaboration = "CMS",
    title = "{Search for anomalous electroweak production of vector boson pairs in association with two jets in proton-proton collisions at 13 TeV}",
    eprint = "1905.07445",
    archivePrefix = "arXiv",
    primaryClass = "hep-ex",
    reportNumber = "CMS-SMP-18-006, CERN-EP-2019-089",
    doi = "10.1016/j.physletb.2019.134985",
    journal = "Phys. Lett. B",
    volume = "798",
    pages = "134985",
    year = "2019"
}

@article{ATLAS:2016nmw,
    author = "Aaboud, Morad and others",
    collaboration = "ATLAS",
    title = "{Search for anomalous electroweak production of $WW/WZ$ in association with a high-mass dijet system in $pp$ collisions at $\sqrt{s}=8$ TeV with the ATLAS detector}",
    eprint = "1609.05122",
    archivePrefix = "arXiv",
    primaryClass = "hep-ex",
    reportNumber = "CERN-EP-2016-171",
    doi = "10.1103/PhysRevD.95.032001",
    journal = "Phys. Rev. D",
    volume = "95",
    number = "3",
    pages = "032001",
    year = "2017"
}

@article{CMS:2020gfh,
    author = "Sirunyan, Albert M and others",
    collaboration = "CMS",
    title = "{Measurements of production cross sections of WZ and same-sign WW boson pairs in association with two jets in proton-proton collisions at $\sqrt{s} =$ 13 TeV}",
    eprint = "2005.01173",
    archivePrefix = "arXiv",
    primaryClass = "hep-ex",
    reportNumber = "CMS-SMP-19-012, CERN-EP-2020-064",
    doi = "10.1016/j.physletb.2020.135710",
    journal = "Phys. Lett. B",
    volume = "809",
    pages = "135710",
    year = "2020"
}

@article{CMS:2020ypo,
    author = "Sirunyan, Albert M and others",
    collaboration = "CMS",
    title = "{Observation of electroweak production of W$\gamma$ with two jets in proton-proton collisions at $\sqrt {s}$ = 13 TeV}",
    eprint = "2008.10521",
    archivePrefix = "arXiv",
    primaryClass = "hep-ex",
    reportNumber = "CMS-SMP-19-008, CERN-EP-2020-143",
    doi = "10.1016/j.physletb.2020.135988",
    journal = "Phys. Lett. B",
    volume = "811",
    pages = "135988",
    year = "2020"
}

@article{CMS:2021gme,
    author = "Tumasyan, Armen and others",
    collaboration = "CMS",
    title = "{Measurement of the electroweak production of Z$\gamma$ and two jets in proton-proton collisions at $\sqrt{s} =$ 13 TeV and constraints on anomalous quartic gauge couplings}",
    eprint = "2106.11082",
    archivePrefix = "arXiv",
    primaryClass = "hep-ex",
    reportNumber = "CMS-SMP-20-016, CERN-EP-2021-095",
    doi = "10.1103/PhysRevD.104.072001",
    journal = "Phys. Rev. D",
    volume = "104",
    pages = "072001",
    year = "2021"
}

@article{CMS:2020fqz,
    author = "Sirunyan, Albert M and others",
    collaboration = "CMS",
    title = "{Evidence for electroweak production of four charged leptons and two jets in proton-proton collisions at $\sqrt {s}$ = 13 TeV}",
    eprint = "2008.07013",
    archivePrefix = "arXiv",
    primaryClass = "hep-ex",
    reportNumber = "CMS-SMP-20-001, CERN-EP-2020-127",
    doi = "10.1016/j.physletb.2020.135992",
    journal = "Phys. Lett. B",
    volume = "812",
    pages = "135992",
    year = "2021"
}

@article{Gutierrez-Rodriguez:2025wcy,
    author = "Gutierrez-Rodr{\i}guez, A. and Cetinkaya, V. and Koksal, M. and Gurkanli, E. and Ari, V. and Hernandez-Ru{\i}z, M. A.",
    title = "{The future muon collider for the research of the anomalous neutral quartic $Z\gamma\gamma\gamma$, $ZZ\gamma\gamma$, and $ZZZ\gamma$ couplings}",
    eprint = "2312.06329",
    archivePrefix = "arXiv",
    primaryClass = "hep-ph",
    doi = "10.1140/epjp/s13360-025-06400-2",
    journal = "Eur. Phys. J. Plus",
    volume = "140",
    pages = "5",
    year = "2025"
}

@article{ATLAS:2022nru,
    author = "Aad, Georges and others",
    collaboration = "ATLAS",
    title = "{Measurement of electroweak $ Z\left(\nu \overline{\nu}\right)\gamma jj $ production and limits on anomalous quartic gauge couplings in pp collisions at $ \sqrt{s} $ = 13 TeV with the ATLAS detector}",
    eprint = "2208.12741",
    archivePrefix = "arXiv",
    primaryClass = "hep-ex",
    reportNumber = "CERN-EP-2022-138",
    doi = "10.1007/JHEP06(2023)082",
    journal = "JHEP",
    volume = "06",
    pages = "082",
    year = "2023"
}

@article{ATLAS:2023sua,
    author = "Aad, Georges and others",
    collaboration = "ATLAS",
    title = "{Measurement and interpretation of same-sign W boson pair production in association with two jets in pp collisions at $ \sqrt{s} $ = 13 TeV with the ATLAS detector}",
    eprint = "2312.00420",
    archivePrefix = "arXiv",
    primaryClass = "hep-ex",
    reportNumber = "CERN-EP-2023-221",
    doi = "10.1007/JHEP04(2024)026",
    journal = "JHEP",
    volume = "04",
    pages = "026",
    year = "2024"
}

@article{CMS:2025dbm,
    author = "Hayrapetyan, Aram and others",
    collaboration = "CMS",
    title = "{Vector boson scattering and anomalous quartic couplings in final states with {\ensuremath{\ell}}{\ensuremath{\nu}}qq or {\ensuremath{\ell}}{\ensuremath{\ell}}qq plus jets using proton-proton collisions at $ \sqrt{s}=13 $ TeV}",
    eprint = "2510.00118",
    archivePrefix = "arXiv",
    primaryClass = "hep-ex",
    reportNumber = "CMS-SMP-22-011, CERN-EP-2025-147",
    doi = "10.1007/JHEP03(2026)022",
    journal = "JHEP",
    volume = "03",
    pages = "022",
    year = "2026"
}

@article{Hamada:2022mua,
    author = "Hamada, Yu and Kitano, Ryuichiro and Matsudo, Ryutaro and Takaura, Hiromasa and Yoshida, Mitsuhiro",
    title = "{$\mu$TRISTAN}",
    eprint = "2201.06664",
    archivePrefix = "arXiv",
    primaryClass = "hep-ph",
    reportNumber = "KEK-TH-2385",
    doi = "10.1093/ptep/ptac059",
    journal = "PTEP",
    volume = "2022",
    number = "5",
    pages = "053B02",
    year = "2022"
}

@article{Passarino:1990hk,
    author = "Passarino, G.",
    title = "{W W scattering and perturbative unitarity}",
    doi = "10.1016/0550-3213(90)90593-3",
    journal = "Nucl. Phys. B",
    volume = "343",
    pages = "31--59",
    year = "1990"
}

@article{Cornwall:1974km,
    author = "Cornwall, John M. and Levin, David N. and Tiktopoulos, George",
    title = "{Derivation of Gauge Invariance from High-Energy Unitarity Bounds on the s Matrix}",
    reportNumber = "UCLA-74-TEP-2",
    doi = "10.1103/PhysRevD.10.1145",
    journal = "Phys. Rev. D",
    volume = "10",
    pages = "1145",
    year = "1974",
    note = "[Erratum: Phys.Rev.D 11, 972 (1975)]"
}

@article{Zhang:2018shp,
    author = "Zhang, Cen and Zhou, Shuang-Yong",
    title = "{Positivity bounds on vector boson scattering at the LHC}",
    eprint = "1808.00010",
    archivePrefix = "arXiv",
    primaryClass = "hep-ph",
    reportNumber = "USTC-ICTS-18-13",
    doi = "10.1103/PhysRevD.100.095003",
    journal = "Phys. Rev. D",
    volume = "100",
    number = "9",
    pages = "095003",
    year = "2019"
}

@article{Bi:2019phv,
    author = "Bi, Qi and Zhang, Cen and Zhou, Shuang-Yong",
    title = "{Positivity constraints on aQGC: carving out the physical parameter space}",
    eprint = "1902.08977",
    archivePrefix = "arXiv",
    primaryClass = "hep-ph",
    reportNumber = "USTC-ICTS-19-01",
    doi = "10.1007/JHEP06(2019)137",
    journal = "JHEP",
    volume = "06",
    pages = "137",
    year = "2019"
}

@article{Yamashita:2020gtt,
    author = "Yamashita, Kimiko and Zhang, Cen and Zhou, Shuang-Yong",
    title = "{Elastic positivity vs extremal positivity bounds in SMEFT: a case study in transversal electroweak gauge-boson scatterings}",
    eprint = "2009.04490",
    archivePrefix = "arXiv",
    primaryClass = "hep-ph",
    reportNumber = "USTC-ICTS/PCFT-20-29",
    doi = "10.1007/JHEP01(2021)095",
    journal = "JHEP",
    volume = "01",
    pages = "095",
    year = "2021"
}

@article{Alloul:2013bka,
    author = "Alloul, Adam and Christensen, Neil D. and Degrande, C{\'e}line and Duhr, Claude and Fuks, Benjamin",
    title = "{FeynRules  2.0 - A complete toolbox for tree-level phenomenology}",
    eprint = "1310.1921",
    archivePrefix = "arXiv",
    primaryClass = "hep-ph",
    reportNumber = "CERN-PH-TH-2013-239, MCNET-13-14, IPPP-13-71, DCPT-13-142, PITT-PACC-1308",
    doi = "10.1016/j.cpc.2014.04.012",
    journal = "Comput. Phys. Commun.",
    volume = "185",
    pages = "2250--2300",
    year = "2014"
}

@article{Degrande:2011ua,
    author = "Degrande, Celine and Duhr, Claude and Fuks, Benjamin and Grellscheid, David and Mattelaer, Olivier and Reiter, Thomas",
    title = "{UFO - The Universal FeynRules Output}",
    eprint = "1108.2040",
    archivePrefix = "arXiv",
    primaryClass = "hep-ph",
    reportNumber = "CP3-11-25, IPHC-PHENO-11-04, IPPP-11-39, DCPT-11-78, MPP-2011-68",
    doi = "10.1016/j.cpc.2012.01.022",
    journal = "Comput. Phys. Commun.",
    volume = "183",
    pages = "1201--1214",
    year = "2012"
}

@article{Alwall:2011uj,
    author = "Alwall, Johan and Herquet, Michel and Maltoni, Fabio and Mattelaer, Olivier and Stelzer, Tim",
    title = "{MadGraph 5 : Going Beyond}",
    eprint = "1106.0522",
    archivePrefix = "arXiv",
    primaryClass = "hep-ph",
    reportNumber = "FERMILAB-PUB-11-448-T",
    doi = "10.1007/JHEP06(2011)128",
    journal = "JHEP",
    volume = "06",
    pages = "128",
    year = "2011"
}

@article{Bierlich:2022pfr,
    author = "Bierlich, Christian and others",
    title = "{A comprehensive guide to the physics and usage of PYTHIA 8.3}",
    eprint = "2203.11601",
    archivePrefix = "arXiv",
    primaryClass = "hep-ph",
    reportNumber = "LU-TP 22-16, MCNET-22-04, FERMILAB-PUB-22-227-SCD",
    doi = "10.21468/SciPostPhysCodeb.8",
    journal = "SciPost Phys. Codeb.",
    volume = "2022",
    pages = "8",
    year = "2022"
}

@article{deFavereau:2013fsa,
    author = "de Favereau, J. and Delaere, C. and Demin, P. and Giammanco, A. and Lema{\^\i}tre, V. and Mertens, A. and Selvaggi, M.",
    collaboration = "DELPHES 3",
    title = "{DELPHES 3, A modular framework for fast simulation of a generic collider experiment}",
    eprint = "1307.6346",
    archivePrefix = "arXiv",
    primaryClass = "hep-ex",
    doi = "10.1007/JHEP02(2014)057",
    journal = "JHEP",
    volume = "02",
    pages = "057",
    year = "2014"
}

@article{Cacciari:2011ma,
    author = "Cacciari, Matteo and Salam, Gavin P. and Soyez, Gregory",
    title = "{FastJet User Manual}",
    eprint = "1111.6097",
    archivePrefix = "arXiv",
    primaryClass = "hep-ph",
    reportNumber = "CERN-PH-TH-2011-297",
    doi = "10.1140/epjc/s10052-012-1896-2",
    journal = "Eur. Phys. J. C",
    volume = "72",
    pages = "1896",
    year = "2012"
}

@article{ParticleDataGroup:2024cfk,
    author = "Navas, S. and others",
    collaboration = "Particle Data Group",
    title = "{Review of particle physics}",
    doi = "10.1103/PhysRevD.110.030001",
    journal = "Phys. Rev. D",
    volume = "110",
    number = "3",
    pages = "030001",
    year = "2024"
}

@article{Wilks:1938dza,
    author = "Wilks, S. S.",
    title = "{The Large-Sample Distribution of the Likelihood Ratio for Testing Composite Hypotheses}",
    doi = "10.1214/aoms/1177732360",
    journal = "Annals Math. Statist.",
    volume = "9",
    number = "1",
    pages = "60--62",
    year = "1938"
}

@article{Fichet:2013ola,
    author = "Fichet, Sylvain and von Gersdorff, Gero",
    title = "{Anomalous gauge couplings from composite Higgs and warped extra dimensions}",
    eprint = "1311.6815",
    archivePrefix = "arXiv",
    primaryClass = "hep-ph",
    doi = "10.1007/JHEP03(2014)102",
    journal = "JHEP",
    volume = "03",
    pages = "102",
    year = "2014"
}

@article{Gupta:2011be,
    author = "Gupta, Rick S.",
    title = "{Probing Quartic Neutral Gauge Boson Couplings using diffractive photon fusion at the LHC}",
    eprint = "1111.3354",
    archivePrefix = "arXiv",
    primaryClass = "hep-ph",
    reportNumber = "CERN-PH-TH-2011-284, MCTP-11-40",
    doi = "10.1103/PhysRevD.85.014006",
    journal = "Phys. Rev. D",
    volume = "85",
    pages = "014006",
    year = "2012"
}

@article{Bartocci:2023nvp,
    author = {Bartocci, Riccardo and Biek{\"o}tter, Anke and Hurth, Tobias},
    title = "{A global analysis of the SMEFT under the minimal MFV assumption}",
    eprint = "2311.04963",
    archivePrefix = "arXiv",
    primaryClass = "hep-ph",
    reportNumber = "MITP/23-063",
    doi = "10.1007/JHEP05(2024)074",
    journal = "JHEP",
    volume = "05",
    pages = "074",
    year = "2024"
}

@article{Celada:2024mcf,
    author = "Celada, Eugenia and Giani, Tommaso and ter Hoeve, Jaco and Mantani, Luca and Rojo, Juan and Rossia, Alejo N. and Thomas, Marion O. A. and Vryonidou, Eleni",
    title = "{Mapping the SMEFT at high-energy colliders: from LEP and the (HL-)LHC to the FCC-ee}",
    eprint = "2404.12809",
    archivePrefix = "arXiv",
    primaryClass = "hep-ph",
    doi = "10.1007/JHEP09(2024)091",
    journal = "JHEP",
    volume = "09",
    pages = "091",
    year = "2024"
}
\end{document}